\documentclass[fleqn,usenatbib]{mnras}

\usepackage{newtxtext,newtxmath}

\usepackage[T1]{fontenc}
\usepackage{ae,aecompl}

\usepackage{graphicx}	
\usepackage{amsmath}	
\usepackage{xcolor}

\title[Star clusters in the Outskirts of the Large Magellanic Cloud]{A Search for star clusters in the outskirts of the Large Magellanic Cloud: indication of clusters in the age gap.\thanks{This work is based on INAF-VST guaranteed observing time under ESO programmes 098.D-0587; 0100.D-0565}}

\author[Gatto et al.]{
M. Gatto$^{1,2}$,\thanks{e-mail: massimiliano.gatto@inaf.it}
V. Ripepi$^{1}$, M. Bellazzini$^{3}$, M. Cignoni$^{4}$, M.-R. L. Cioni$^{5}$,\\ 
\newauthor{M. Dall'Ora$^{1}$, G. Longo$^{2}$, M. Marconi$^{1}$, P. Schipani$^{1}$, M. Tosi$^{3}$}
\\
\\
$^{1}$ INAF-Osservatorio Astronomico di Capodimonte, Via Moiariello
 16, 80131, Naples, Italy \\
 $^{2}$ Dept. of Physics, University of Naples Federico II, C.U. Monte Sant'Angelo, Via Cinthia, 80126, Naples, Italy\\
$^{3}$ INAF-Osservatorio di Astrofisica e Scienza dello Spazio, Via Gobetti 93/3, I-40129 Bologna, Italy \\
$^{4}$ Physics Departement, University of Pisa, Largo Bruno Pontecorvo, 3, I-56127 Pisa, Italy\\
$^5$ Leibniz-Institut f\"ur Astrophysik Potsdam, An der Sternwarte 16, D-14482 Potsdam, Germany
}

\date{Accepted XXX. Received YYY; in original form ZZZ}

\pubyear{}

\begin{document}
\label{firstpage}
\pagerange{\pageref{firstpage}--\pageref{lastpage}}
\maketitle

\begin{abstract}
The YMCA ({\it Yes, Magellanic Clouds Again}) and STEP ({\it The SMC in Time: Evolution of a Prototype interacting late-type dwarf galaxy}) projects are deep $g$,$i$ photometric surveys carried out with the VLT Survey Telescope (VST) and devoted to study the outskirts of the Magellanic System. A main goal of YMCA and STEP is to identify candidate stellar clusters and complete their census out to the outermost regions of the Magellanic Clouds. We adopted a specific over-density search technique coupled with a visual inspection of the color magnitude diagrams (CMDs) to select the best candidates and estimate their ages. To date, we analysed a region of 23 sq. deg. in the outskirts of the Large Magellanic Cloud, detecting 85 candidate cluster candidates, 16 of which have estimated ages falling in the so called ``age gap". We use these objects together with literature data to gain insight into the formation and interaction history of the Magellanic Clouds.
\end{abstract}




\section{Introduction}

The Magellanic Clouds (MCs) are the nearest example of a pair of interacting galaxies. Due to their relatively small distances, about 50 kpc for the Large Magellanic Cloud \citep[LMC,][]{degrijs-wicker-bono-2014}, and a little more than 60 kpc for the Small Magellanic Cloud \citep[SMC,][]{degrijs&bono-2015}, this system is an ideal laboratory where to test theories of merger and galaxy evolution.
Moreover, the MCs are interacting also with the Milky Way (MW), thus representing a primary benchmark to understand the formation and evolution, via accretions, of the MW galaxy halo.\\
The interaction signatures are striking, the most evident being the Magellanic Stream (MS), an extended cloud of HI gas that covers about 180 degrees ($\sim$ 180 kpc at the MS distance) around the Galactic South pole of the Milky Way \citep{Putman2003, Bruns2005}. Other interaction footprints, are the Magellanic Bridge (MB) \citep{Kerr1957}, and the Wing of the SMC, showing the disturbed geometry of the galaxy, likely due to a direct collision or tidal interactions with the LMC \citep{Zaritsky&harris2004, Cioni2009, Gordon2009, Besla2010, Besla2012, Diaz&Bekki2012}.
Even though the MCs were traditionally believed to have been MW satellites for a Hubble time \citep{Murai&Fujimoto1980, Lin&Linden-bell1982,Gardiner1994,Heller&Rohlfs1994,Moore&Davis1994,Lin1995,Gardiner&Noguchi1996,Bekki&Chiba2005,Yoshizawa2003,Connors2004,Connors2006,Mastropietro2005}, nowadays a large consensus exists on the idea that the LMC and the SMC are in their first passage around the MW. This is supported by recent and more precise measurements of their proper motion \citep{Kallivayalil2006a, Kallivayalil2006b, Kallivayalil2013}.
Moreover, based on these new findings, also the idea of the MCs as interacting binaries for a Hubble time has been challenged \citep[e.g.][]{Besla2012,Diaz&Bekki2012}.\\
In recent years many authors tried to reconstruct the evolutionary history of the MCs, either by studying their star formation history \citep[SFH, e.g.][]{Harris2004, Harris2009, Weisz2013, Cignoni2013} or by investigating the age distribution of their star clusters \citep{Pietrzynski2000a,Glatt2010,Baumgardt2013,Piatti2015a,Piatti2016,Piatti2018,Nayak2016,Nayak2018,Pieres2016}.
Even though great advances in the knowledge of the MCs recent past has been achieved, supporting a scenario in which the LMC and the SMC became an interacting pair only a few Gyr ago \citep{Besla2010,Besla2012,Diaz&Bekki2012}, some important questions still remain unanswered.\\

As for the LMC, the almost total absence of star clusters (SCs) in the so called ``age gap" - i.e. an interval of ages ranging from $\sim 4$ to $\sim 10$ Gyr\footnote{In addition, the LMC has 15 ancient globular clusters similar to those present in the Milky Way, with ages older than 10 Gyr.} - which was first noticed by \citet{DaCosta1991}, has not been clearly explained yet.
Since this gap is not present in the LMC star field population \citep{Tosi2004,Carrera2011,Piatti2013}, the SC formation, and the SFH seem to be decoupled.
These occurrences make the cluster formation history of the LMC peculiar with respect to the Galactic and SMC counterparts.\\
Within this framework, a complete catalog of SCs with accurate age estimates is fundamental to unveil the whole evolutionary history of the LMC and to understand how the SMC and the MW could have influenced it.\\
So far, the most complete catalogue of Magellanic SCs is that by \citet{Bica2008}, which consists of several thousands of clusters and young associations.
With the advent of deeper surveys of higher spatial resolution, many works have focused on the search of previously unrecognised SCs, thus increasing the total number of the LMC cluster system \citep{Sitek2016,Sitek2017,Piatti2017a,Piatti2018}.
In some of the observed fields the number of the local SCs has been raised by 55\% \citep{Piatti2016} implying that the catalog of LMC SCs is still far from complete.
Moreover, the majority of the above quoted surveys covered only the main body of the LMC, leaving the area beyond $\sim$ 4 degrees from the centre almost unexplored.
In fact, the quest for SCs in the outskirts of the LMC is limited to few works. For example, \citet{Pieres2016} exploited data from the Dark Energy Survey (DES) to perform a SC search in the Northern part of the LMC. This led to visually identify 255 clusters, among which 109 are new candidates, out to a distance of about 10 kpc from the LMC centre, in cylindrical coordinates.
\citet{Sitek2016} identified 226 new SC candidates in the outer disk of the LMC by using the Optical Gravitational Lensing Experiment \citep[OGLE-IV;][]{Udalski2015}, and more recently, \citet{Piatti2017a} found 24 new SCs in the MCs periphery through The Survey of Magellanic Stellar History \citep[SMASH;][]{Nidever2015}.

It is worth noticing that the periphery of a galaxy is important to constrain theories of galaxy evolution because it more easily preserves the signatures of recent interactions with neighbouring systems. This is because in the outskirts of a galaxy, the dynamical timescale is longer than in the inner regions \citep[e.g.][]{Bullock2005}.

\begin{figure*}
    \centering
    \includegraphics[width=\textwidth]{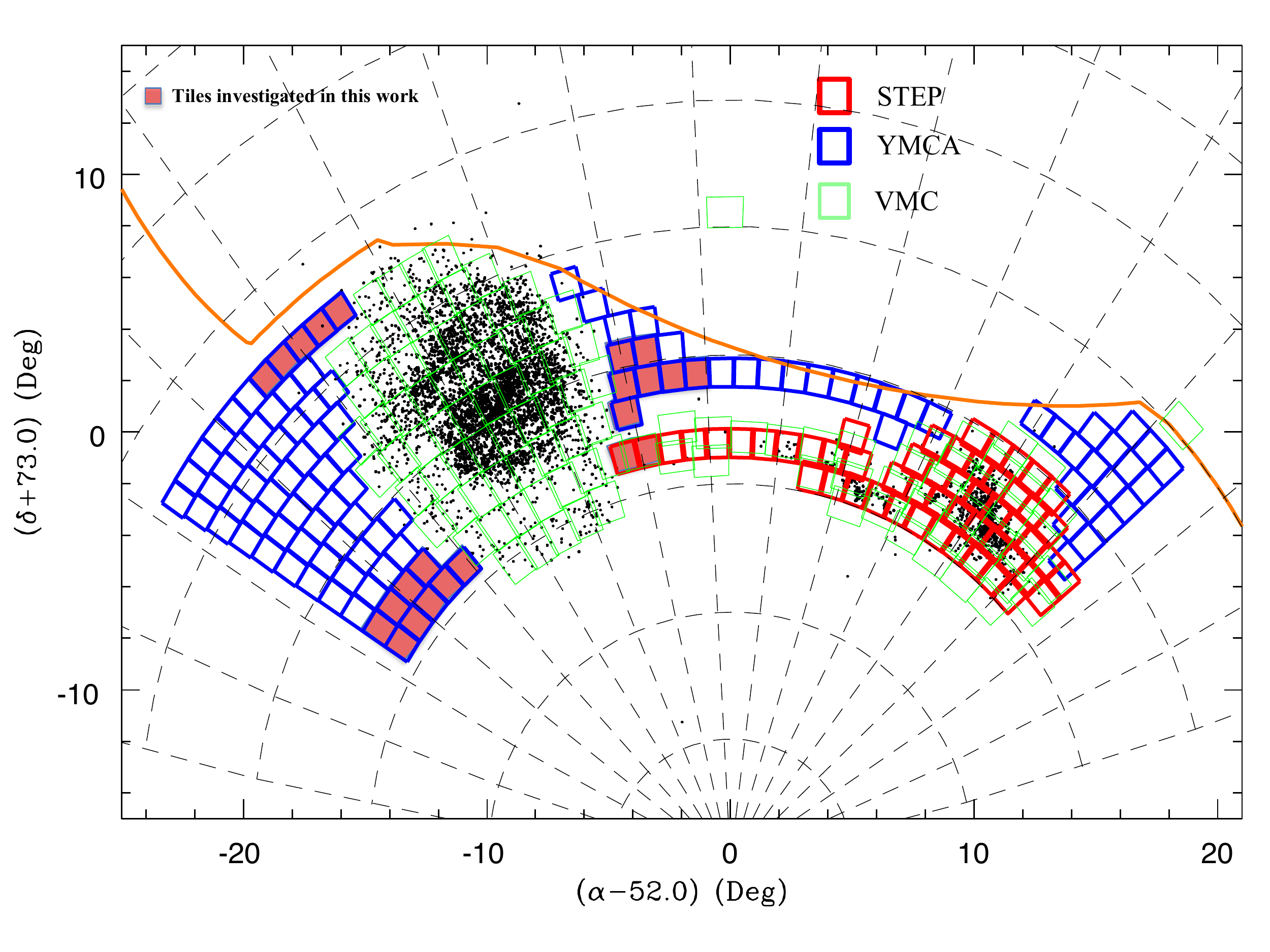}
    \caption{Footprint of the STEP and YMCA surveys (see labels) in a zenithal equidistant projection along with all objects present in \citet{Bica2008} (black dots). For comparison, we overdraw the VMC \citep{Cioni2011} regions in green, whereas the Dark Energy Survey (DES) surveyed area lays Northward of the orange line. The 23 tiles analysed in this work are filled in red.}
    \label{fig:sampled_fields}
\end{figure*}

This work aims at answering some still debated aspects of the MCs evolution, by searching for new SCs in the periphery of the LMC through two surveys: {\it The SMC in Time: Evolution of a Prototype interacting late-type dwarf galaxy} (STEP: PI V. Ripepi) and {\it Yes, Magellanic Clouds Again} (YMCA: PI V. Ripepi).
The STEP survey has been presented in \citet{Ripepi2014}, whereas the YMCA survey will be the subject of a forthcoming paper (Ripepi et al. 2020, in preparation). 
Both surveys reach 1.5-2 magnitudes ($\sim$ 24 mag in the $g$-band; \citealt{Ripepi2014}, hereafter R14) below the main sequence turn-off of the oldest stellar population (> 10 Gyr, which in the LMC is $\sim 22.5$ mag in the $g$-band) thus allowing us to detect even the oldest LMC SCs.
On the contrary, some of the recent surveys which looked for SCs could not reveal clusters older than 1 Gyr \citep[e.g.][]{Pietrzynski2000a, Glatt2010, Nayak2016}, thus lacking coverage for an important period of the LMC evolution, in particular the age gap.\\
Finally, as far as we know, the regions analysed in this work (see R14 and Fig.~\ref{fig:sampled_fields}) fall in areas of the LMC outskirts that have never been observed to this depth, and indeed, just a few of our candidate clusters were previously known (see next sections). \\

In this context, we aim at publishing the first census of SCs in the outer regions of the LMC, with estimated ages, reddening and metallicity, complete down to the oldest clusters. In this paper we present the first results of this project, analysing 23 tiles in the outskirts of the LMC. These tiles sample three different regions around the galaxy, at Northeast, Southeast, and West-Southwest (see Fig. \ref{fig:sampled_fields}), thus covering a range of projected distances between 4.4 kpc and 10.4 kpc.
The analysis of fields located in opposite directions with respect to the LMC centre will allow us to understand if the whole galaxy, at least in its outskirts, shares the same evolutionary history.\\

The paper is organized as follow: in Sect. 2 we briefly describe the survey and the data reduction; in Sect. 3 we focus on the procedure used to detect new SCs; in Sect. 4 we assess the accuracy of the detection procedure by simulating artificial SCs; in Sect. 5 we describe the methods employed to derive SC parameters; Sect. 6 and 7 are devoted to the discussion of the results and the comparison with the literature, respectively. A brief summary concludes the paper.

{}

\section{Observations and data reduction}
\label{sect:observations}
\begin{figure*}
    \centering
    \includegraphics[width=\textwidth]{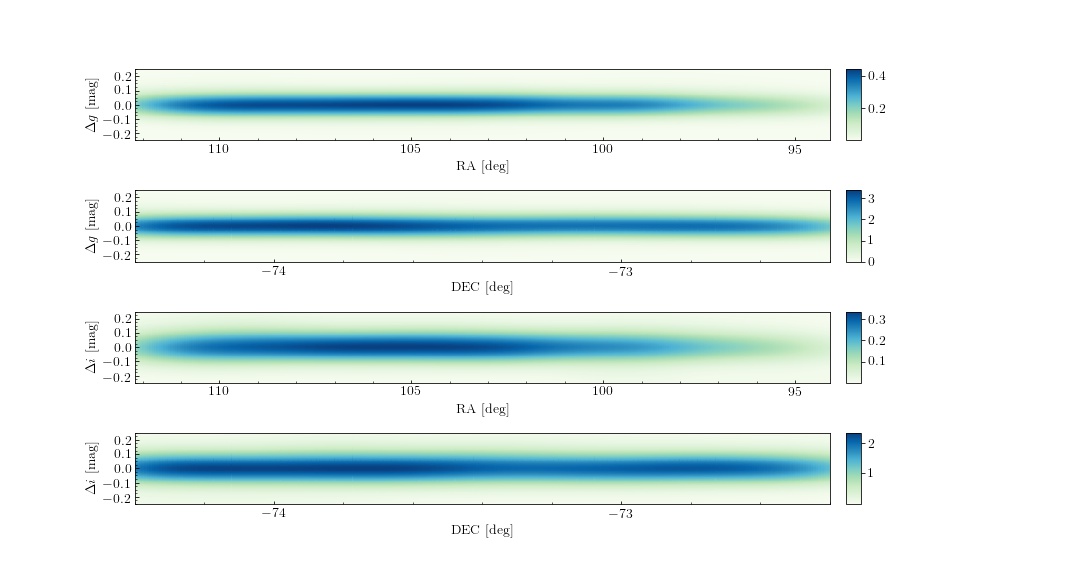}
    \caption{Comparison of present calibrated photometry and the APASS one vs R.A. and Dec. $\Delta g$ and $\Delta i$ are in the direction "this work"-APASS. The data has been smoothed by means of a KDE. The colorbars report the density of data points.}
    \label{fig:deltaMag}
\end{figure*}

The observations used in this work are part of the STEP and YMCA surveys. Both have been carried out with the VLT Survey Telescope \citep[VST,][]{Capaccioli2011} using part of the Guaranteed Time Observations (GTO) allocated by the European Southern Observatory (ESO) to the Istituto Nazionale di Astrofisica (INAF). The telescope is equipped with OmegaCAM, a camera with a field-of-view of 1 deg$^2$ built by a consortium of European Institutes \citep{Kuijken2011}. The camera is a mosaic of 32-CCD, $16k\times16k$ detectors with a pixel scale of 0.214 arcsec/pixel.
The footprint of the surveyed area of STEP, as well as full details about the observing strategy, can be found in R14. In this work we use only the tiles 3\_20 and 3\_21 of STEP whose coordinates can be found in Table 2 of R14.\\
The YMCA survey will be described in detail in a future paper (Ripepi et al. 2020, in preparation), here we recall only its main characteristics. The YMCA footprint is shown in Fig~\ref{fig:sampled_fields} (blue boxes), where it is compared with the footprints of the STEP and VISTA survey of the Magellanic Clouds system \citep[VMC,][red and green boxes, respectively]{Cioni2011}. The tiles filled in red are those used in this work; their features can be found in Table~\ref{tab:log}. The survey strategy is very similar to STEP. It is conducted in the $g,i$ bands, but adopting slightly shorter exposure times due to the smaller crowding in the external MC regions. In particular, while STEP exposure times were 3000s and 2600s for the $g,i$ bands, respectively, in YMCA they were 1900s and 1500s. The dithering procedure to cover the gaps between the CCDs is the same as in STEP, the relative number of images is given in the second row in Table 3 of R14. 
The data reduction of the analysed images was conducted as in R14. In particular, the pre-reduction, astrometry and stacking of the different dithered frames to provide single mosaic images have been carried out with the VST--Tube imaging pipeline \citep{Grado2012}, while the Point Spread Function (PSF) photometry was obtained using the standard packages DAOPHOT IV/ALLSTAR \citep[][]{Stetson1987,Stetson1992}. An important difference with respect to R14 concerns the absolute photometric calibration. In this work, the PSF photometry in each tile was calibrated adopting local standard stars provided by the AAVSO Photometric All-Sky Survey (APASS).\\ 
For each tile, the different steps were the following: 
i) we cross-matched the PSF photometric catalog with the APASS database using a search radius of 0.5\arcsec to reduce the number of wrong matches (the APASS instrument's pixel-size is 2.57\arcsec), retaining only APASS observations with Signal to Noise (S/N) ratio larger than 10; ii) we searched and corrected possible residual spatial variations of the photometric zero points (i.e. we searched for trends in photometry vs RA/Dec); iii) we corrected for the colour dependence of the zero points in $g$ and $i$. At the end of this procedure, we obtained an average accuracy of the order of 0.02 and 0.03 mag in $g$ and $i$, respectively. Figure~\ref{fig:deltaMag} shows the typical plot used to check the absolute photometric calibration in the cases of South-East tiles 1\_27-2\_33: the lack of any spatial trend and the reasonably low dispersion of the residuals are evident. Similar results have been obtained for the other tiles with an analogous level of crowding.\\
Finally, the photometric catalogues have been purged from extended/spurious objects by retaining only objects with -0.6$\leq SHARPNESS \leq$0.7, where $SHARPNESS$ is an output parameter of the DAOPHOT package useful to detect both extended objects and spurious detection due to bad pixels. No cuts have been applied to the $CHI$ parameter to avoid eliminating bright stars.   

{}

\begin{table}
 \caption{Log of Observations. The different columns show: name of the tile, its centre, date of observation, average FWHM over the images (S$_g$ and $S_i)$}
 \label{tab:log}
 \begin{tabular}{lccccc}
  \hline
  Tile & R.A. & Dec & Date & S$_g$ & S$_i$\\
      &  hms & dms &      & \arcsec & \arcsec \\
  \hline
     11\_41    & 06:00:26.98 & -62:57:31.7 & 2017-12-09 & 0.98 & 0.96\\
     11\_42    & 06:13:47.98 & -62:57:31.7 & 2018-01-08 & 1.15 & 0.77\\
     11\_43    & 06:13:47.98 & -62:57:31.7 & 2018-01-14 & 1.05 & 0.72\\
     11\_44    & 06:31:35.99 & -62:57:31.7 & 2018-01-19 & 0.93 & 0.85\\
     11\_45    & 06:31:35.99 & -62:57:31.7 & 2018-01-19 & 0.97 & 0.82\\
    1\_27    & 06:23:43.30 & -73:59:14.1 & 2016-10-11 & 1.24 & 1.01\\
    1\_28    & 06:38:11.88 & -73:59:14.1 & 2016-10-12 & 1.34 & 1.00\\
    1\_29    & 06:52:40.49 & -73:59:14.1 & 2016-10-12 & 1.51 & 0.96\\
    1\_30    & 07:07:09.10 & -73:59:14.1 & 2016-10-23 & 1.30 & 0.87\\
    1\_31    & 07:21:37.68 & -73:59:14.1 & 2016-11-19 & 1.34 & 0.82\\
    2\_30    & 06:41:10.37 & -72:53:04.1 & 2017-10-12 & 1.30 & 1.06\\
    2\_31    & 06:54:46.16 & -72:53:04.1 & 2017-10-13 & 1.17 & 1.17\\
    2\_32    & 07:08:21.96 & -72:53:04.1 & 2017-10-12 & 1.30 & 1.01\\
    2\_33    & 07:21:57.77 & -72:53:04.1 & 2016-12-18 & 1.05 & 0.73\\
    3\_21    & 04:22:54.42 & -71:46:54.4 & 2017-10-12 & 1.09 & 1.09\\
    4\_19    & 03:50:36.64 & -70:40:44.2 & 2017-10-12 & 1.19 & 0.98\\
    4\_20    & 03:50:36.64 & -70:40:44.2 & 2017-10-13 & 1.13 & 0.99\\
    4\_21    & 04:14:52.86 & -70:40:44.2 & 2017-12-23 & 1.08 & 0.97\\
    4\_22    & 04:14:52.86 & -70:40:44.2 & 2017-12-25 & 1.12 & 0.99\\
    5\_22    & 04:13:31.25 & -69:34:33.9 & 2017-12-26 & 1.13 & 1.08\\
    5\_23    & 04:13:31.25 & -69:34:33.9 & 2018-01-11 & 1.25 & 1.09\\
  \hline
 \end{tabular}
\end{table}

\section{Cluster detection}
\label{ch:algorithm}

As stated above, our aim is to obtain a SC catalog as complete as possible in the outskirts of the LMC. To achieve this goal, we need to go beyond the simple visual inspection of the images. Indeed, visual methods are not very effective in detecting the less luminous and less dense SCs, that constitute a significant fraction of a galaxy's SC population, since the SC luminosity function steeply increases towards the faint end \citep[e.g.][]{Degrijs2003}. Furthermore, using an automated method allows us to estimate the completeness of the procedure on more objective grounds (see \S\ref{sec:Efficiency of the algorithm}).

Therefore, we adopted the procedure introduced by \citet[][]{Zaritsky1997} and successfully developed by other authors \citep[e.g.][]{Piatti2016,Piatti2018,Sitek2016,Ivanov2017}, to automatically search for new SCs among the over-densities in the space of positions.\\
In this section we describe in detail each step of the semi-automated procedure adopted to find SCs in YMCA and STEP images.

\subsection{Identification of SC candidates: search for over-densities with the KDE method}
\label{overdensity}

\begin{figure}
    \centering
    \includegraphics[width=8.5cm]{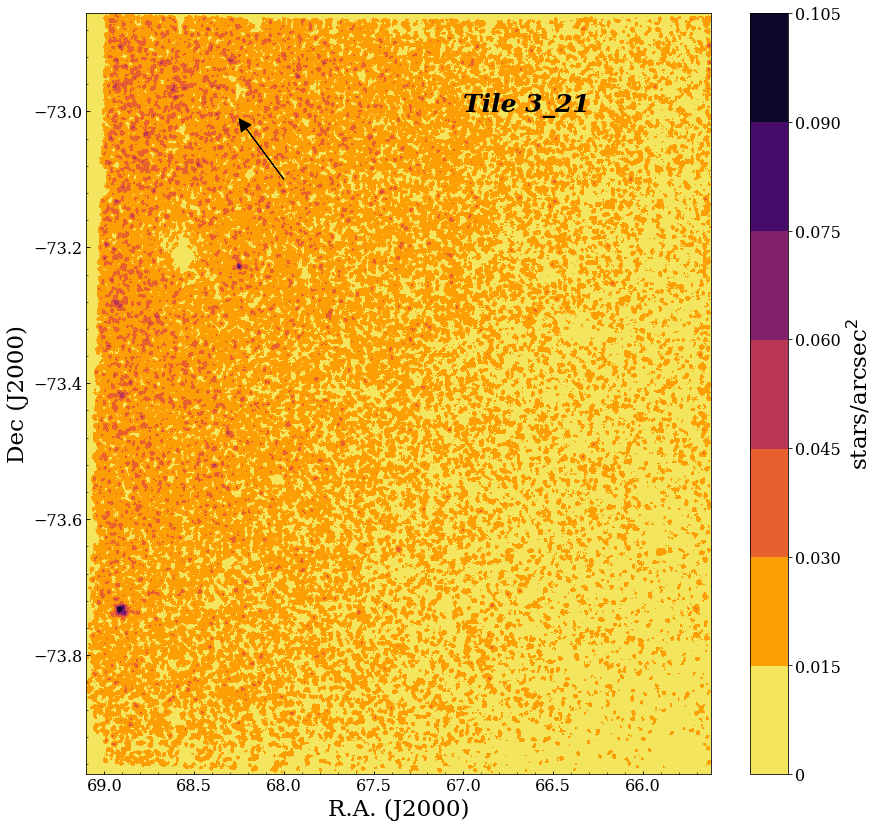}
    \caption{Example of the density surface map generated by the KDE, with density increasing from lighter to darker colors. This image corresponds to the tile 3\textunderscore21 of the STEP survey. The black arrow indicates the direction of the LMC centre. The dark spot at the bottom left of the figure corresponds to the known cluster SL63.}
    \label{fig:kde}
\end{figure}

The first step of the procedure is to find regions in R.A., Dec where the local density is significantly above the background level, by counting the local number of stars and comparing it with the mean estimated density of the field stellar population.
To do this efficiently, we adopted a two dimensional kernel density estimator (KDE)\footnote{We used the version available in the scikit-learn package \citep{scikit-learn}} to generate a surface density map and thus look for over-densities in the R.A., Dec space.
The KDE is a non-parametric technique utilized to estimate the probability density function of a random variable smoothing data through a kernel function, avoiding histogram troubles like the choice of the bin size or of the bin phase \citep{Rosenblatt1956}.
The only parameter that must be set in KDE is the bandwidth of the kernel function, and it should be of the same size of the smallest objects that need to be detected \citep{Piatti2018}.
In order to detect even the tiniest SC present in the data, we run the KDE with a bandwidth of 0.2\arcmin, comparable with the size of the smallest SCs around the LMC \citep[see][]{Bica2008}. 
To improve our ability in detecting over-densities, the KDE analysis was carried out adopting two different kernel functions, namely the gaussian and tophat ones. 
The KDE analysis was carried out on a tile by tile basis, sub-dividing the star catalog in squares (pixels) with size 4\arcsec$\times$4\arcsec and successively computing the density value in every single pixel with the KDE.\footnote{As already mentioned, the size of the smallest SCs around the LMC is $\sim$ 0.2\arcmin; hence such a choice of pixel size allows us to sample the candidate over-densities with at least three pixels.}\\
An example is shown in Fig.~\ref{fig:kde} that shows the density surface map related to the STEP tile 3\textunderscore{21} (Southwest LMC). In the figure the density increases from lighter to darker colours, revealing the presence of a stellar density gradient towards the LMC centre, whose direction is indicated by the black arrow. The presence of such a significant gradient in the background (also detected in several other tiles) suggested us to use as threshold an estimate of the local background density rather than an average value measured over the whole tile. 
In order to define such threshold (local over-density), we compared each pixel value with a local estimated mean. For each selected pixel, this local mean was measured averaging the values of all the pixels external to a box with a size of 1.5\arcmin and internal to a box with a size of  2.5\arcmin.
The size of the inner box prevents the chance that the presence of a SC could raise the local mean, thus decreasing the signal-to-noise ratio.
The outer box is large enough to ensure a statistically significant sampling and is small enough to guarantee that we were probing the local density. 
The estimated background density allows us to measure the signal-to-noise ratio, or significance of each pixel, defined as:

\begin{equation}
\label{eq:significance}
S = \frac{m - \mu}{\sigma}   
\end{equation}

\noindent where \emph{m} is the pixel density, $\mu$ is the background density and $\sigma$ its standard deviation.

{}

The definition of a significance threshold, or $S \geq S_{\rm th}$, above which we can define an over-density, is an important step since its value determines both the lowest cluster density that the algorithm is able to detect and the number of false positives that it could yield.
To better constrain its value, we relied on Montecarlo simulations, finding that it is important to use a threshold dependent on the local field stellar density (full details in Appendix~\ref{appendix:appendix_a}).
We selected and assembled all adjacent pixels that are above the threshold, discarding all groups with a number of pixels lower than four (corresponding to a dimension of 8\arcsec$\times$8\arcsec) to remove likely spurious over-densities, originating from stochastic fluctuations of the stellar field. The mentioned Montecarlo simulations allowed us to estimate that such a choice would decrease the spurious over-densities by about 70\%.\\
At the end of this procedure we have a list of over-densities, that need to be further inspected since an agglomerate of stars is not sufficient to define a real SC. The next step was then to estimate the centre and radius for each over-density, as described in the next section. 

\subsection{Center and radius estimation}
\label{sec:center and radius estimation}

\begin{figure}
    \centering
    \includegraphics[width=0.45\textwidth]{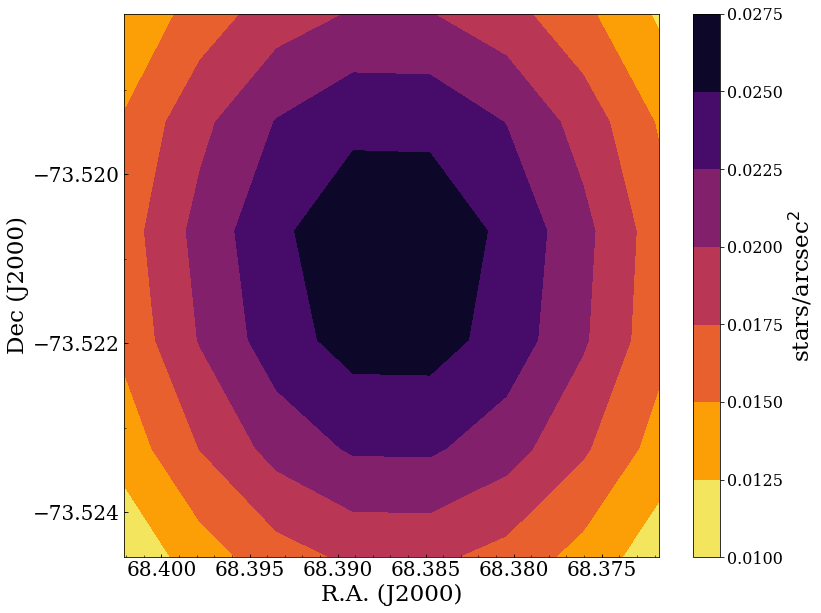}\\
    \includegraphics[width=0.45\textwidth]{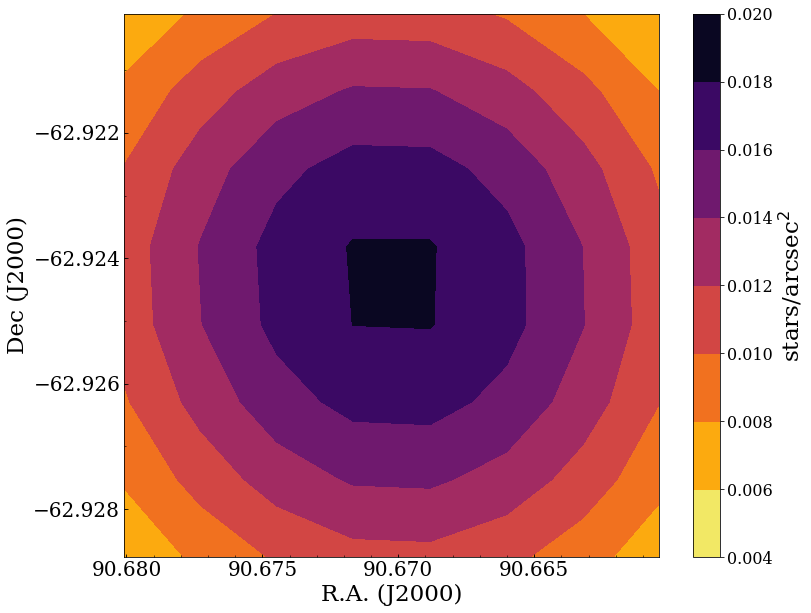}
    \caption{Example of the 2-D KDE in the region of the two newly discovered SCs: STEP-0025 (top) and YMCA-0032 (bottom).}
    \label{fig:kde2D}
\end{figure}

The estimate of the over-density's centre and radius is crucial since the former could have an important effect on the radial density profile, hence in the radius value, and the latter could influence the procedure (described in the next section) we adopted to disentangle real SCs from false positives.
We used an automated method to infer the coordinates of the centre of the over-density, which consists in running another KDE in the SC region, looking for the pixel with the highest stellar density value, iterating the procedure until convergence is obtained. In particular, we picked as trial SC centre the pixel with the maximum value from the first KDE run, and considered as effective SC area a circle with radius twice as long as the greatest distance among all pixels belonging to the same over-density. Then, we performed another bi-dimensional KDE in this area, taking the new highest pixel value as new trial center. We repeated this process until two subsequent centres differed less than 1.5\arcsec, or after ten steps, if the method did not converge, retaining the last estimate as the good one. However, more than 90\% of the over-densities converged before the last step was reached.
Figure~\ref{fig:kde2D} displays two examples of centre determination in the region of a typical over-density (candidate SC STEP-0025 and YMCA-0032).\\
\par As the previous procedure is very effective but not perfect, in some cases a visual inspection of the results revealed a clear offset of the estimated centre with respect to the real one. 
This could happen for example, if another agglomerate of stars is located nearby, or because of border effects, when the over-density is close to the tile edge, or in the case of binary SCs. In all these cases, we corrected manually the centre values.

{}

The radius plays an important role in assigning a membership probability to stars within the SC region (see  \S\ref{sec:cleaning procedure}).
It is usually estimated measuring the distance from the centre at which the radial density profile (RDP, i.e. number of stars per unit area) stabilizes around the background density level \citep{Bonatto2009,Pavani2011,Perren2017}.
To measure the RDP for each over-density, we calculated the number of stars located in concentric shells around the SC centre, separated in bins of 0.05\arcmin, starting from the center, up to a distance of 2.0\arcmin, and divided this value for the area of each shell.
We did not apply any magnitude cut since most of our newly detected SCs did not show a significant crowding level.
To estimate the background, i.e. the expected number density of field stars, we considered four shells with radius 0.5\arcmin in a region between 2\arcmin and 4\arcmin from the candidate SC centre and measured the number of stars divided by the area of each shell, then we took the mean as the background\footnote{In \citet{Bica2008}'s catalogue only $\sim$3\% of the SCs have a radius $\geq$ 2\arcmin, however, they are easily detectable by eye and we are pretty confident that such SCs are not present in the analysed tiles.}.
\begin{figure}
    \centering
    \includegraphics[width=0.45\textwidth]{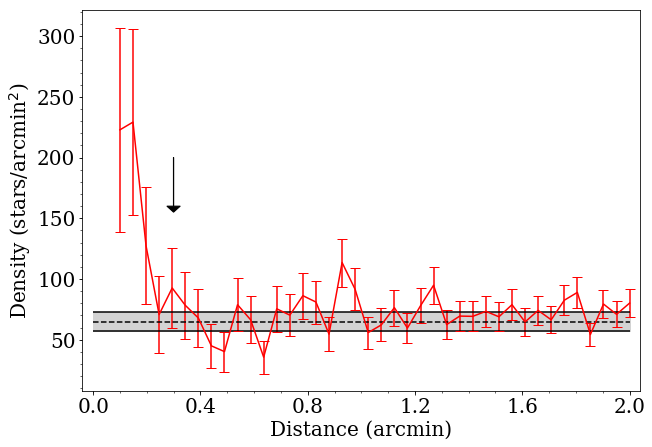}\\
    \includegraphics[width=0.45\textwidth]{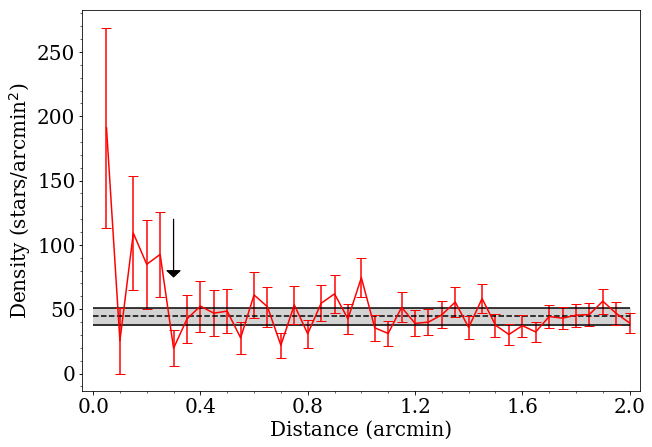}
    \caption{Radial density profile of the SCs STEP-0025 (top) and YMCA-0032 (bottom). The dashed line is the estimated background mean, and the two black solid lines represent the 1$\sigma$ deviation. The red solid line sets the density of stars of the SC as a function of the distance to its centre. All errors are Poissonian. The black arrow indicates the estimated radius.}
    \label{fig:RDP}
\end{figure}{}
To illustrate the procedure, Fig.~\ref{fig:RDP} displays the RDP of the candidate clusters STEP-0025 and YMCA-0032 (red dots/solid red line). It can be seen that the RDP steadily decreases from the centre until it reaches and settles around the estimated background density of field stars (dashed line). The errors on the RDP (red vertical lines) and on the background density (solid black lines/grey area) are calculated assuming a Poissonian noise (the square root of the value).
In the last step, the procedure uses both the RDP and the background density (with their errors) to estimate the proper radius, comparing at each distance from the centre the quantities $N_{\rm RDP} \pm \sqrt{N_{\rm RDP}}$ and $\mu_{bkg} \pm \sqrt{\mu_{bkg}}$, where $N_{\rm RDP}$ is the value of the RDP at a certain distance from the centre, and $\mu_{bkg}$ is the background density value. The right SC radius is estimated when the two quantities are congruent. However, this step is not straightforward because of the fluctuations of the RDP when it approaches the background density value. 
To overcome this problem we tested several criteria to define a stabilization condition, making this procedure more robust against stochastic fluctuations of the RDP. In particular we considered the RDP and the background level congruent if the condition $N_{\rm RDP} - \sqrt{N_{\rm RDP}} \leq \mu_{bkg}$ is satisfied two consecutive times or if $N_{\rm RDP} - \sqrt{N_{\rm RDP}} \leq \mu_{bkg} + \sqrt{\mu_{bkg}}$ is reached three times on four adjacent concentric shells.


\subsection{Cluster selection through the colour-magnitude diagram}
\label{sec:cleaning procedure}


Once a list of over-densities is obtained as outlined above, the following fundamental step is to remove spurious objects, i.e. groups of stars clumping due to projection effects (asterisms).  
Obviously, this is crucial to remove the contaminants, since they can alter the statistical properties of the SC population. 
\citet{Piatti2018} in their paper devoted to the analysis of SCs located along the minor axis of the LMC, concluded that about 30\% of the SCs in their surveyed regions belonging to Bica's catalog are not real physical systems, demonstrating that in the absence of a rigorous procedure for removing false SCs, the degree of contamination can be significantly high.\\
In spite of several studies on the proper motion of MC stars \citep[e.g.][]{Kallivayalil2018,Zivick2019} and satellites, accurate proper motions and radial velocities of the MC SCs are not available yet, hence the only additional information available other than the positions are the magnitudes and colors of the stars belonging to the candidate SC. We can exploit the fact that, contrary to asterisms, the stars belonging to actual SCs  show precise sequences in the Colour-Magnitude Diagram (CMD). Indeed, stellar evolution theory predicts that the members of a coeval system, like a SC, evolve along well defined sequences in the CMD and, once corrected for the distance modulus and the reddening, they are expected to lay around an isochrone for a given age and metallicity. Thus the analysis of the over-densities CMD allows us not only to remove spurious objects, but also to directly estimate the main parameters of the stellar population of the SC, namely its reddening, age and metallicity.\\
However, disentangling true from false SCs on the basis of the CMD is not a straightforward task. For instance, due to the 3D geometry of the investigated galaxy, also field stars can be projected in the region occupied by genuine cluster stars. 
Therefore, a cleaning procedure of the SC CMD is necessary to confirm the SC nature of the over-density and to estimate correctly the main SC parameters.
To this aim we used the procedure developed by \citet[][]{Piatti2012} and commonly used in the literature \citep[e.g.][]{Piatti2014,Piatti2015b,Piatti2016,Ivanov2017}. 
In the following, we briefly describe the main steps of this method and the interested reader can refer to the original paper for full details.
In a nutshell, the procedure consists in cleaning the SC candidate CMD by using for comparison four distinct CMDs representing field stars, located along four different directions (North, South, East and West) with respect to the SC centre. The distance between the comparison fields and the SC is chosen to be large enough to avoid the inclusion of cluster stars but also sufficiently short to sample the local stellar properties, i.e. stellar density, luminosity and colour distribution.
Following \citet{Piatti2014}, in order to increase the statistics, the region around the cluster to be cleaned covers an area described by a radius $R_{\rm ca}$=3$R_{\rm cl}$, where $R_{\rm cl}$ is the SC radius. Each of the four comparison fields is built using a radius $R_{\rm field}$=$R_{\rm ca}$ and the coordinates of their centres are positioned at a distance calculated as follows:

\begin{equation}
    d = 2R_{\rm ca} + R_{\rm cl}
\end{equation}{}

\begin{figure}
    \centering
    \includegraphics[scale=0.38]{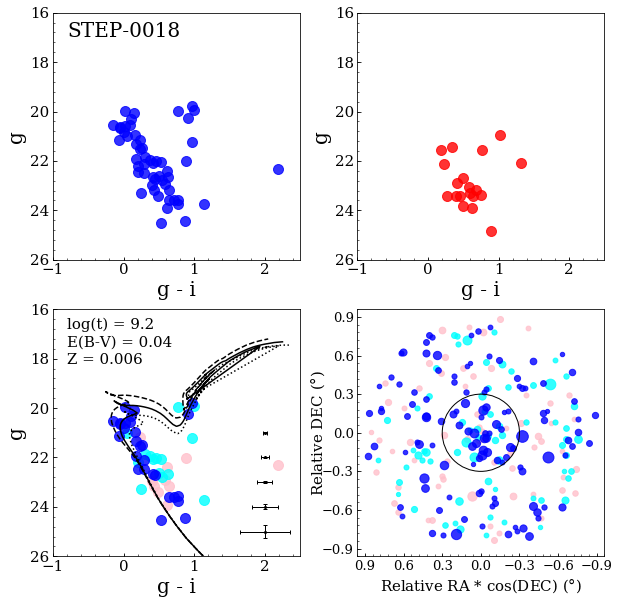}
    \caption{Example of the cleaning procedure for the SC STEP-0018. \emph{Top:} SC CMD (\emph{left side}) and CMD of a field (\emph{right side}) taken at 7$\times R_{\rm cl}$ and with an area equal to the cluster one.
    \emph{Bottom left:} cluster CMD after the cleaning procedure with stars colored by their membership probability. Blue, cyan and pink points are stars with $P \geq 75\%$, $P \geq 50\%$, $P \leq 25\%$, respectively.
    The black solid line marks the best fitting isochrone, obtained with the values listed in the top left corner of the figure, while dashed and dotted lines show isochrones with ages $\pm 0.1$ in log(t) with respect to the best fitting one. The photometric errors are also displayed.
    \emph{Bottom right:} relative positions for all stars within 3$\times R_{\rm cl}$, with the origin at the SC center, and the size proportional to their luminosity. The black circle indicates the cluster radius.}
    \label{fig:cleaned_cmd}
\end{figure}

\indent
In each of the four fields, the local CMD density is modeled by cells with different sizes: smaller boxes being generated in denser CMD areas, and larger ones in less dense CMD regions. This is because different parts of the CMD have a different density. In particular, the main sequence (MS) is more populated than other CMD regions, like the sub-giant branch (SGB) or the red-giant branch (RGB), because the lifetime of any stars in the MS phase is at least ten times longer than in any other phase.
Starting with a box centered on each of the star field and with sides (mag, col) = (2.0, 0.5), the shape of the box is varied according to the local CMD density, by reducing it until it reaches the closest star in magnitude and colour, separately. At the end, there is a box for each star, with its size depending on the local crowding in the CMD \citep[see][their Fig. 12]{Piatti2012}
Then the comparison field CMD is overlapped to the cluster CMD, and for each of the boxes we delete the cluster star closest to its centre, considering just the stars within the box.
This operation is repeated four times, one for each field. At the end of the process, for each candidate SC, the number of times a star has been subtracted can be used to derive a membership probability \emph{P} for all stars within the SC radius. Stars that have been eliminated once or never have $P \geq 75 \%$ of probability to belong to the cluster, stars with two subtractions have the same probability to belong to the cluster or to the field stellar population, while stars with $P \leq 25 \%$ are likely field stars.
Removing the contamination from field stars ($P < 50 \%$) allowed us to discard all over-densities whose remaining stars did not follow a SC isochrone on the CMD. 
In Fig.~\ref{fig:cleaned_cmd} there is an example of such a procedure for the candidate cluster STEP-0018. The top panels show the CMD of the SC (\emph{left side}) and the CMD of a representative field (\emph{right side}).
In the bottom-left panel of the figure we show the CMD after the cleaning procedure, color-coded by the membership of the stars (blue, cyan and pink for $P \geq 75\%$, $P = 50\%$ and $P \leq 25\%$, respectively) with the best isochrone over-plotted as a solid line. The plot in the bottom-right panel shows the star positions with respect to the centre of the SC in the cleaned area (three times the estimated cluster radius) and the solid circle indicates the radius of the SC. 
{}
It is worth noticing that following to the original procedure outlined by \citet[][]{Piatti2012}, an area as large as nine times as that of the cluster is chosen to enlarge the statistics and to improve the performance of the cleaning process.
This procedure does not take into account the distance from a star to the cluster centre, but only its position on the CMD. 
This is useful in the case of poorly populated objects as our candidate SCs.
Bearing in mind all these considerations, the presence of some residuals (stars with P $\geq$ 75\%) beyond the cluster radius it is expected, actually.

We produced figures like Fig.~\ref{fig:cleaned_cmd} for all the over-densities found with the KDE procedure (more than 3000 candidates over the 23 tiles analyzed in this work). A visual inspection of these CMDs allowed us to remove the spurious objects, that resulted to be the large majority of the candidate SCs. Indeed, at the end of this procedure, we were left with 104 candidate SCs. 
After a closer inspection of the remaining CMDs, we removed from the SC candidate list all those (namely, 19) that did not show clear MSs, and/or show MS for stars with P < 50\% more populated and better delineated than those for stars with P $\geq$ 75\%, and/or show spatial distributions of stars with P < 50\% more concentrated than those for stars 
with P $\geq$ 75\%. At the end of this careful analysis we had a list composed by 85 candidate SCs.

Identification, centre coordinates and radii of these objects are listed in the first four columns of Table~\ref{tab:clusters}, ordered according their right ascension.  

Since the above outlined procedure contains a certain degree of subjectivity, we also provide a statistical parameter to quantitatively assess the goodness of each SC. Thus, we define
\begin{equation}
G = \frac{N_{\rm cl} - N_{\rm bkg}}{\sqrt{N_{\rm bkg}}} 
\end{equation}{}
\noindent
where $N_{\rm cl}$ is the number of stars within the SC radius and $N_{\rm bkg}$ is the average number of field stars within an area equal to that defined by the radius of the SC. 
This mean was obtained through a measure of the star density in four circular regions with radius  $0.5\arcmin$ placed around the SC in a region comprised between $2\arcmin$ and $4\arcmin$ from its centre. Finally, this density was normalized to the SC area.
The $G$ value, which ranges from 1.32 to 48.75 and it has a median of 3.48, is listed in column (10) of Table~\ref{tab:clusters}, along with the total number of stars within the SC radius (column 11). 
\\
Finally, before discussing the results, it is worth testing the reliability of our SC detection method by means of artificial cluster simulations. This kind of analysis allows us also to estimate the completeness of the cluster catalog provided in this work.

\section{Testing the SC detection method with artificial clusters}
\label{ch:efficiency_algorithm}

In order to quantify the completeness of the catalog, it is important to evaluate the accuracy of the search algorithm in finding targets and to measure the detection limit of the method.
Since our method is based on the measurement of the significance (\emph{S}) of each pixel, and a "good" pixel must exceed a given threshold ($S  \geq S_{\rm th}$), it follows that every parameter which plays a role in the measure of the significance has also an influence in the detection of a SC.
Following this argument, both the concentration of the SC and the stellar field density have an impact on the SC detection algorithm, since varying the former changes the signal, while different values of the latter alter the noise.
For example, decreasing the compactness of the SC or increasing the field stars density (or both), makes it harder to reveal a SC, since both contribute to reduce the S/N ratio. Consequently, if the pixel significance drops below our selected threshold, the SC will be lost, as could happen for sparse SCs embedded in dense fields. Therefore, a quantification of the impact of the threshold value is required to estimate the completeness.

To measure the efficiency of the algorithm in detecting SCs and the completeness, we generated artificial clusters with different values of density, and we overlaid them on simulated stellar field populations with a different density as well. In this way, we could test the sensitivity of our algorithm to changes in both SCs and stellar field densities.
In the next sub-sections we describe in detail the procedure used to simulate artificial clusters and the results of the tests.

\subsection{Generation of artificial clusters}

To simulate the stellar field, we used the same suite of Montecarlo simulations described in the appendix \ref{appendix:appendix_a}. We generated 2000 stellar fields with 100 different density values (20 for each field density value ranging from 1.7 to 89.7 stars/arcmin$^2$), bracketing the stellar density values observed in the analyzed regions.
More in detail, for each simulated field we generated artificial SCs with four different values of density, namely $\rho_{\rm cl} = 30, 50, 70, 90$ stars/arcmin$^2$. The chosen values encompass the whole range of observed SCs compactness, from sparse to very dense. 
To generate the artificial clusters, we randomly selected the centre positions of each of the four SCs, preventing overlap, by imposing that each centre lays more than 5\arcmin  from all the others. We then randomly extracted the number of stars belonging to the SC, using a Gaussian with $\mu = 30$ and $\sigma = 20$. In this way we not only sampled different SC densities but also diverse degrees of compactness, since the SC radius was adjusted to match the proper value of density.
Once SC centres and radii are generated, we assign random positions to the cluster stars, using a Gaussian profile with mean the position of the SC centre and standard deviation the half of its radius.
As last step we removed all simulated stars closer than 0.8\arcsec to each other in order to reduce crowding. 
Finally, the 10000 simulated SCs were analyzed following the same procedure outlined in Sect.~\ref{ch:algorithm}. The results of this test are presented in the next section.

\subsection{Efficiency of the algorithm}
\label{sec:Efficiency of the algorithm}

\begin{figure*}
    \centering
    \includegraphics[scale=0.47]{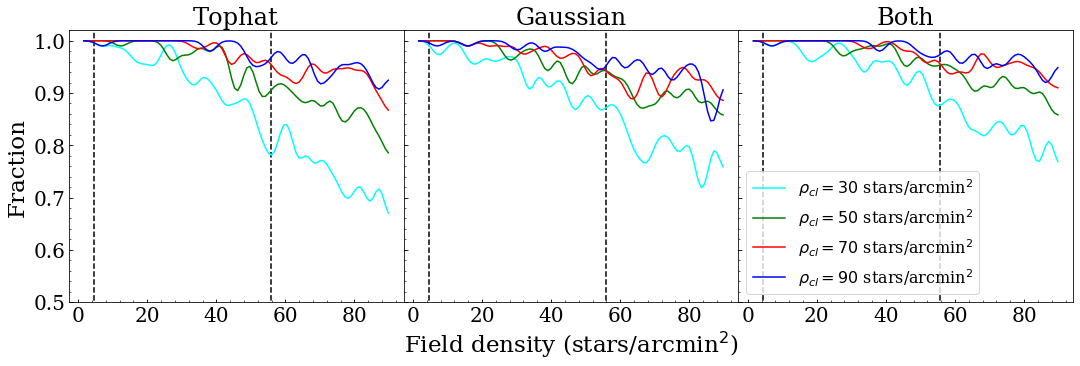}
    \caption{Recovery fraction of the four artificial clusters as a function of the stellar field density. Vertical dashed lines represent limits of field density in our tiles, estimated by dividing the number of stars within a tile by its area. Fractional values have been smoothed with a gaussian filter with $\sigma$ = 2 stars/arcmin$^2$ to make less noisy images.}
    \label{fig:recovery fraction}
\end{figure*}{}

Figure~\ref{fig:recovery fraction} shows the recovery fraction, i.e. the number of positive detection of artificial SCs over the 20 simulated stellar fields (for each of the 100 field density steps), as a function of the different SC (solid lines with different colours) and field densities.
In the left and centre panels of the figure, the recovery fraction is displayed for two different kernels (\textit{tophat} and \textit{gaussian}, respectively), giving insights on how the algorithm works in the two cases.
In the right panel both kernel functions are considered. 
As a reference, the vertical dashed lines indicate the lowest and highest density values measured on the tiles analyzed in this work.
An inspection of the figure reveals that at any SC density, the fraction of positive detections decreases towards higher field density values, as expected. Similarly, less concentrate SCs have a lower fraction of positive detection and are harder to reveal. 
This is true for both kernel functions.
However, the tophat has a slightly higher detection fraction towards lower field densities and denser SCs ($\rho_{\rm cl} = 70 - 90$ stars/arcmin$^2$) while the gaussian kernel performs better on sparse SCs and in more crowded stellar fields.
Therefore, running the algorithm with both kernels yields a great gain in the SC detection (even if the computational time is longer), since in this way the recovery fraction settles above 90\% even for the sparser SCs embedded in the highest observed field densities (right panel).\\
We note that the impact of the field density on the recovery of artificial SCs with $\rho_{\rm cl} = 70 - 90$ stars/arcmin$^2$ (red and blue lines in the figure) is minimal, since the recovery rate is $\sim$100\% till $\sim$50 stars/arcmin$^2$, and remains above $\geq$ 95\% even at the highest field densities present in our images.
In artificial SCs with $\rho = 50$ stars/arcmin$^2$ (green line) the recovery fraction remains close to 100\% up to $\sim$ 40 stars/arcmin$^2$ and $\sim$ 90\% at the upper field observed density limit. Similarly, the artificial SCs with $\rho = 30$ stars/arcmin$^2$ (cyan line) that represent the actual lower limit of the real objects found in this work, follow the same trend, but the recovery fraction drops at $\sim$90\% at high stellar field density.
These results suggest that our method likely produces a catalog of SCs with a very high level of completeness, indeed all of our simulated SCs have more than 90\% probability of detection in our observed tiles.

Once we tested the ability of the algorithm to detect SCs as over-densities in the position space, we investigated also its accuracy on the estimation of the cluster centre and radius, since this measure could influence the cleaning procedure (see also \S\ref{sec:cleaning procedure}).
To this aim we repeated the same procedure described above, generating and inserting in simulated stellar fields other four artificial clusters with four different $\rho_{\rm cl}$, but this time keeping constant also their radius and their total number of stars. In brief, we created the same four clusters and we randomly placed them on each simulated stellar field.

\begin{figure}
    \centering
    \includegraphics[width=0.48\textwidth]{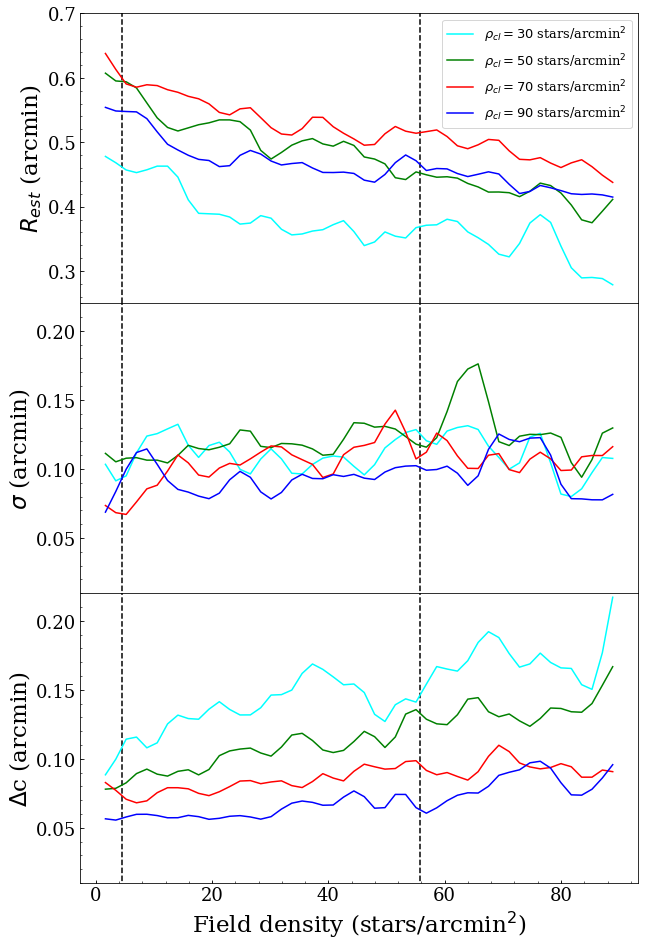}
    \caption{\emph{Top:} radius ($R_{\rm est}$) resulting from the average of 20 simulations per stellar field density vs stellar field density. Different colours identify simulated SCs with different $\rho_{\rm cl}$. \emph{Centre:} standard deviation of the estimated radii. 
    \emph{Bottom:} as above but for the distance separation between the simulated SC centre and its estimate with the procedure described in \S\ref{sec:center and radius estimation}. All values have been smoothed with a Gaussian filter having $\sigma$ = 2 stars/arcmin$^2$ to make less noisy images.}
    \label{fig:artificial cluster radius}
\end{figure}

Figure~\ref{fig:artificial cluster radius} shows the trend with field density of the estimated mean radius (top panel) and its standard deviation (centre panel) for the SCs simulated with four different densities. 
As the field density increases, the measured radius becomes smaller for all artificial SCs. However, this difference is within the uncertainties in the stellar field density interval observed (vertical dashed lines). Indeed, the standard deviation settles around 0.1\arcmin, thus defining our error on the estimated radius. 

{}

The bottom panel of Figure~\ref{fig:artificial cluster radius} displays the distance separation between the simulated SC centre and its estimate with the procedure described in \S\ref{sec:center and radius estimation}, averaged over the 20 simulations per each stellar field density, as a function of the field density.
An inspection of this figure shows that the measurement of SC centres depends on both the background field density and SC density, improving towards SCs with high $\rho_{\rm cl}$, and less populate stellar fields, even if this effect is barely visible for SCs with $\rho_{\rm cl} = 70 - 90$ stars/arcmin$^2$.


\section{Cluster parameter estimation}

In this section we describe the methods used to estimate the main SC parameters such as age, metallicity, reddening, absolute magnitude, and their radial density profiles.

\subsection{Isochrone fitting}
\label{sec:isochrone fitting}

The age of a SC is estimated by identifying the isochrone that best matches the CMD cluster stars. This isochrone fit can be carried out visually or using an automated method.
Both methods have been used in the literature, and each of them has its pros and cons. Some authors preferred automated method \citep[e.g.][]{Nayak2016, Nayak2018} to estimate age, reddening, distance modulus and metallicity of the SCs. Their procedure is definitely more efficient when the SC sample is large and it enables to better quantify the errors on the SC parameters \citep[e.g.][]{Nayak2016, Nayak2018}.
The visual fitting is more subjective and less efficient since it requires to analyze each SC singularly and often it is necessary to fix some physical quantities, like the distance modulus or the metal content, in order to reduce the space of parameters. \citep{Glatt2010, Piatti2014, Piatti2015a, Piatti2015b, Piatti2016}. However, this procedure has to be preferred in case of sparse, poorly populated SCs, where significant statistical fluctuations are present, and the inclusion or exclusion of a few stars can make a difference \citep[][]{Lancon2000}.
For instance, some residual field stars could still be present with high membership probability in the CMD, even after the cleaning procedure described in the previous section, and therefore in these cases the visual fitting avoids that residuals may influence the estimate of the SCs parameters. 
On this basis, we preferred to perform a visual isochrone fitting, using the PARSEC models \citep{Bressan2012}. This procedure allowed us to estimate the SCs ages through the magnitude of MS turn-off (TO) as well as to gauge their reddening and metal content through the position and inclination of the RGB, and red clump, RC.
However, it is difficult to disentangle the effect of reddening and metallicity when comparing the isochrones with observed RGB and RC stars. This occurrence causes an increase of the uncertainties of these two parameters, estimated by varying them until the isochrones no longer fit the RGB/RC stars and taking into account the above quoted degeneration. The resulting errors are $\Delta$E(B - V) = 0.04 mag and $\Delta$Z = 0.002.
To correct the isochrones for distance and extinction we used the relations: $g = g_{\rm iso} + (m - M)_{0} + R_{\rm g} \times E(B - V)$ and $E(g - i) = (R_{\rm g} - R_{\rm i}) \times E(B - V)$ with $R_{\rm g} = 3.303$ and $R_{\rm i} =  1.698$ \citep{Schlafly2011}.\\
\par To reduce the number of possible combinations of isochrones to fit the CMD of each SC, we chose to fix the distance modulus to $(m - M)_{\rm o} = 18.49 \pm 0.09$ mag $(49.90^{+2.10}_{-2.04}$ kpc) \citep{degrijs-wicker-bono-2014}; this value is very close to the recent very accurate measurement of the LMC distance obtained by \citet{Pietrzynski2019} from a sample of eclipsing binaries.
Given that the LMC has non-negligible depth along the line of sight ($3.14 \pm 1.16$ kpc measured by \citealt{Subramanian2009}, and about $\sim 7$ kpc, due to a recent measure made by \citealt{Choi2018}), taking into account also the three-dimensional structure of the LMC, the maximum error on the distance modulus is $\Delta(m - M) \sim 0.2$ mag.
This variation could seem significant, but our uncertainties on the age, estimated by observing the overall dispersion using the visual fitting procedure on the CMD, are typically $\sigma$log(t) $\sim$ 0.1 dex, a value that is equivalent to about 0.4 mag in the distance modulus, at the mean LMC distance. Therefore to assume a constant distance for the isochrone fitting in all the tiles analyzed is acceptable. 
It is worth to point out that $\sigma$log(t) $\sim$ 0.1 dex is an upper limit on the estimate errors, since younger SCs could have even a smaller error (i.e $\sigma$log(t) = 0.05 dex). To be more conservative, we preferred to rise all errors to this upper limit.

\begin{figure*}
    \centering
    \includegraphics[scale=0.5]{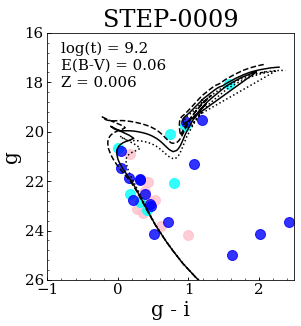}
    \includegraphics[scale=0.5]{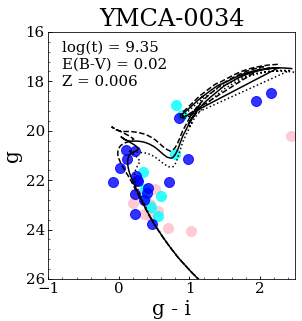}
    \includegraphics[scale=0.5]{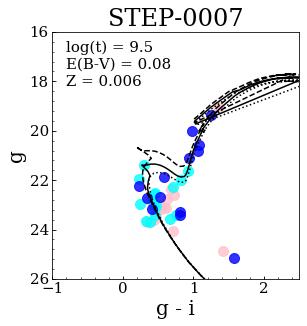}\\
    \includegraphics[scale=0.5]{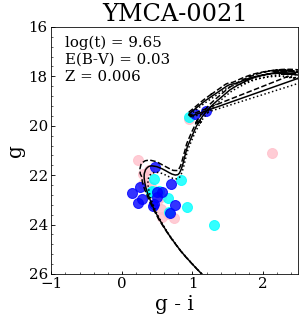}
    \includegraphics[scale=0.5]{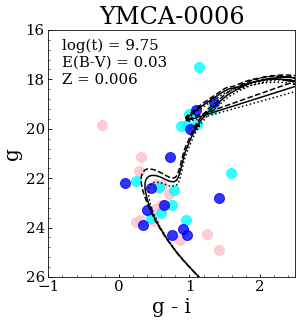}
    \includegraphics[scale=0.5]{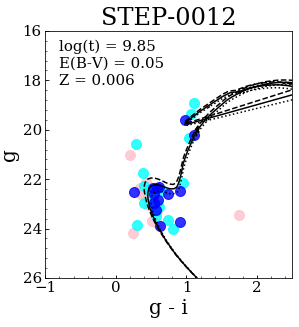}\\
    \caption{Isochrone fitting for six SCs. Solid lines represent the best fit isochrones obtained fixing the distance modulus (m - M = 18.49), while dashed and dotted lines are isochrones with log(t) = $\pm 0.1$ respect to the best one. Stars are color-coded by their membership probability, according the procedure described in Sect. \S\ref{sec:cleaning procedure}. The sky images of these SCs are in Fig. \ref{fig:cluster_images}}. 
    \label{fig:example_iso_fitting}
\end{figure*}

On this basis, we proceeded with the isochrone fitting, varying reddening, age and metal content to find the best match between isochrones and cluster stars, considering only objects with membership probability $P \geq 50 \%$. 
We note that some additional uncertainty on the color and magnitude in the MS could be due to the presence of binary stars that might shift MS stars to the red side of the isochrone. However, this effect is difficult to take into account with present data, hence we did not produce any correction due to binaries.
The SC parameters resulting from this analysis are listed in Table~\ref{tab:clusters}, while in  Fig.~\ref{fig:example_iso_fitting} it is shown an example of the fitting procedure for six (typical) SCs detected in this work. From the figure is also visible that some SCs, e.g.  YMCA-0021, might show differential reddening. In these cases we adopted a ``mean" reddening value as a study of differential reddening would be difficult with present dataset.
The CMDs for all the 85 candidate clusters are shown in Appendix~\ref{appendix2}. We discuss our findings in the next section.

\subsection{Absolute magnitudes}

We derived the absolute magnitudes in $g$-band of each candidate SC using the open source \emph{photutils} package \citep{bradley2019}. This software includes tools to perform aperture photometry on astronomical images and, being written in Python, can be efficiently customized for our purposes.\\
As a first step, for each SC we measured the total flux in a circular aperture centered on the SC and as large as its estimated radius. This flux includes a background that must be subtracted. We, therefore, measured the flux in eight circular apertures with a radius equal to 50 pixels, and placed around the SC, with centres located at 2 $\times$  $R_{\rm cl}$ from the center of the cluster. In each aperture we derived the median flux per pixel as the measure of the background and the mean background to be subtracted was obtained by taking the median of the eight estimates. 
Once the background is subtracted from the total flux, the instrumental magnitude in $g$-band of a given SC is $m_{\rm instr} = -2.5 * \log_{10} (flux)$.
These magnitudes were then calibrated into the APASS system, by evaluating, for each tile, the zero point between the instrumental and the APASS magnitudes.  
To this purpose, we calculated the instrumental magnitudes of all stars in each tile using the aperture photometry tool available in \emph{photutils}. We then calculated the difference between these values and the calibrated PSF photometry (see~\ref{sect:observations}). 
The zero point is then simply the mean of these differences, obtained using only sources in the magnitude interval $15 < g < 20$ mag to exclude bright saturated stars and faint, low S/N objects. 
We repeated this procedure for every tile, obtaining for each SC, a calibrated apparent magnitude in the $g$-band.
Finally, the absolute magnitude in $g$-band was derived as $M_{g} = m_{g} + DM_{\rm LMC} - A_{g}$, where $m_{g}$ is the apparent magnitude, $A_g$ is the extinction in the $g$-band, and $DM_{\rm LMC}$ is the LMC distance modulus adopted above. \\
The resulting $g$-band absolute magnitudes are listed in Table~\ref{tab:clusters} and briefly discussed in Sect.~\ref{sec: absolute mag}.

\subsection{Radial density profiles}
\label{app:density_profile}

\begin{figure}
    \centering
    \includegraphics[width=0.5\textwidth]{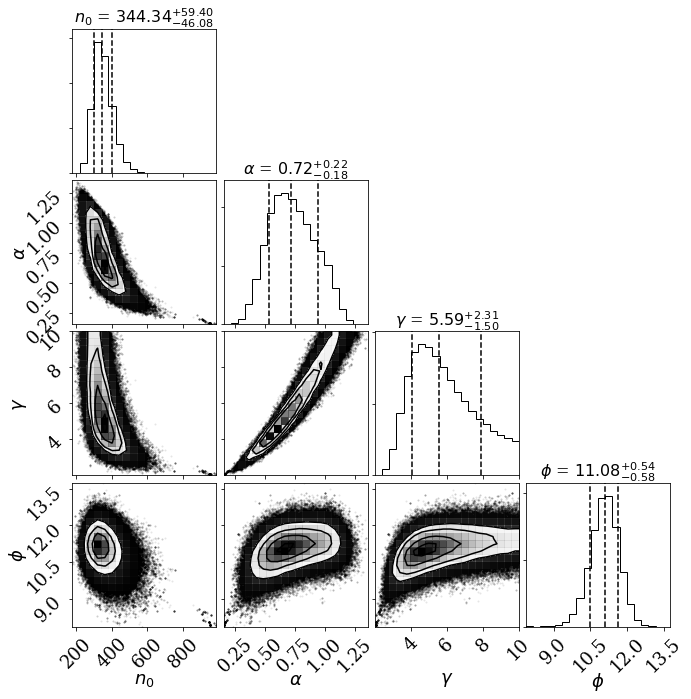}\\
    \includegraphics[width=0.5\textwidth]{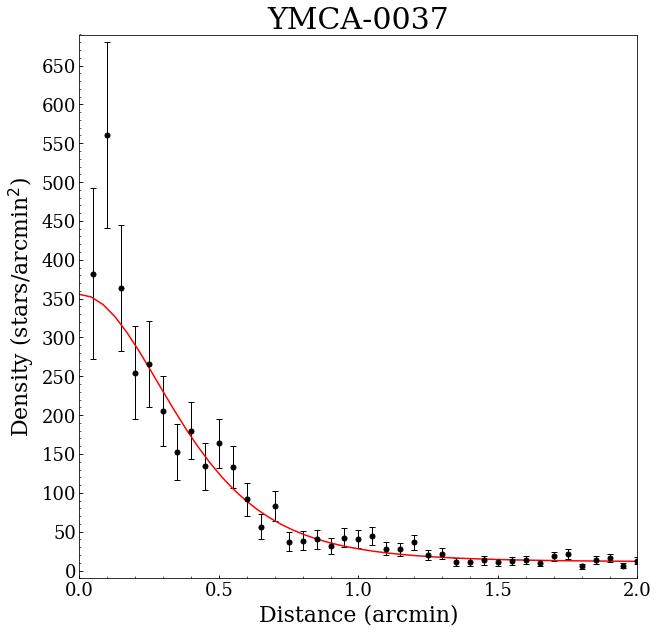}
    \caption{\emph{Top:} Contour plots and marginalized histogram of all the four parameters in the case of the SC YMCA-0037. The estimated values and their errors are also indicated. \emph{Bottom:} RDP with overlapped an EFF profile for the same SC.}
    \label{fig:mcmc}
\end{figure}

In this section we analyze SC radial density profiles (RDP) in order to better assess the physical reality of our newly detected objects.
All SCs follow analytical functions, which usually can be approximated with a flat core in the inner regions, and a power-law at higher cluster-centric distances.
King's family of curves \citep{King1962} and the Elson, Fall \& Freeman profile \citep[EFF,][]{Elson1987} are the two most employed analytical functions for the SCs.
The latter is identical to the former when the parameter indicating the slope of the curve $\gamma$ equals the value of 2 and the tidal radius goes to infinite.  
Even though these analytical formulations are useful to get insights into the SC dynamical evolution \citep[and LMC SCs have been used to this purpose, see e.g.][]{Elson1987,Mackey&Gilmore2003}, here we will only use these profiles as tools to substantiate the reality of our SCs, planning to carry out a dedicated study of their internal dynamics in a forthcoming paper.\\
As \citet{Elson1987} showed that most of the LMC SCs do not seem to have a tidally truncated radius, in the following we preferred to perform a fit by using the EFF profile, described by:
\begin{equation}
    \label{eq:EFF}
    n(r) = n_0 \times \{1 + (\frac{r}{\alpha})^2\}^{-\gamma/2} + \phi
\end{equation}{}
\noindent
where $n(r)$ is the number of stars per squared arcmin as a function of the distance from the cluster centre, $n_0$ is the central surface density, $\alpha$ is the core parameter, $\gamma$ is the slope parameter and $\phi$ is the background value, considered here as a free parameter of the fit.
To obtain RDPs we made use only of stars with P $\geq$75\% in the entire cleaned area (i.e. an area of radius $R_{ca}$ = $3R_{cl}$, see \S\ref{sec:cleaning procedure}).
This choice allow us to both: i) assess the cluster's existence, since if a central over-density persists after the cleaning procedure and the SC RDP is well reproduced by the EFF profile, then the SC reliability increases and ii) check the residuals of Piatti's method (i.e. stars with P $\geq$ 75\% beyond the cluster radius). Indeed, if significant residuals remain after the cleaning procedure we should obtain a nearly flat RDP.
SCs with a well fitted EFF profile on their RDPs derived with P $\geq$ 75\% membership stars will ensure a high reliability of these objects. This is the most conservative approach to take advantage of the SC RDP in order to determine its authenticity.
Anyway, the bulk of our candidates is very sparse and composed by very few stars, making it tricky to build reliable RDPs availing of only the most likely SC members. In addition, Piatti's cleaning procedure does not take into account the distance of the star to the cluster centre, affecting in some cases the shape of the RDP (see the detailed discussion below). 
Thus, we decided to make use also of all stars with P $\geq$ 50\% in the whole clean area to get RDPs for all SCs where a reliable fit was not achieved with the former procedure.
As mentioned before, we do not aim at investigating the internal structure of the SCs, but at statistically assessing our sample, and thus to provide other statistical parameters complementary to the parameter G introduced in \S\ref{sec:cleaning procedure}. In particular, we provide as further statistic parameter in tables \ref{tab:EFF_parameters} and \ref{tab:EFF_parameters_50}, the ratio between the estimated central surface density (background subtracted) and the estimated background (i.e. the ratio between the estimated $n_0$ and $\phi$ parameters of the EFF profile).\\ 
We derived the four parameters via a Markov Chain Monte Carlo technique (MCMC) through the {\it emcee} python package\footnote{https://emcee.readthedocs.io/en/stable/}. The MCMC approach is a widely used technique to sample probability distributions in high dimensions of the parameter space.
It is based on the idea that by using random sampling in a probabilistic space, after a number of steps (i.e. the length of the Markov chain) the chain will contain points that follow the target distribution.
Figure \ref{fig:mcmc} shows a contour plot of the parameters along with their marginalized histograms (top panel), and the RDP with overlapped the best fit obtained with the MCMC method (red line in the bottom panel) for the SC YMCA-0037.
We listed the derived parameters with their uncertainties for 67 SCs (79\% of the sample) with a RDP built employing only stars with P$\geq$75\% fitted by an EFF profile in table \ref{tab:EFF_parameters}, while their RDP are in the Fig.~\ref{fig:EFF_profiles}. 
{}
At first glance it is evident that more than half (precisely $\sim$ 65\%) of the SCs have the ratio between estimated central and background density higher than 5, and all of them, except one, above 3. The mean of this ratio is 11.50 while its median is 10.44.
Figure~\ref{fig:EFF_profiles_50} shows the RDP built by using all stars with P$\geq$50\%, meaning that for them a reliable fit by employing the most likely members has not been obtained.
Table~\ref{tab:EFF_parameters_50} lists the EFF estimated parameters and the $n_0$ and $\phi$ ratio for such SCs. In particular the ratio between the estimated central and background density has a mean of 3.95 and a median equals to 4.04.
Even though most of the SCs in our sample are very small and comprised with a handful of stars, almost all the objects looks like to follow a EFF profile very well, meaning that the majority of our candidate SCs should be genuine physical systems.
Anyway, because of the very low number of stars a few SCs have a profile that does not seem consistent with an EFF profile.
These cases suggest us that a deeper investigation is needed to confirm or reject them.
We want to point out that all 85 SCs are over-densities in the sky and each cleaned CMD is reasonably well fitted by a single isochrone. Both circumstances suggest that they are actual clusters.
Furthermore, as described in Sect.~\ref{sec:cleaning procedure}, Piatti's cleaning procedure removes stars considering only their position on the CMD, without using any information about the distance from the star to the cluster centre.
This approach should be preferred in the case of very small and poorly populated objects as our SCs \citep{Piatti2012}, but it has the effect that the cleaning procedure in some cases could be more severe in the inner regions with respect to the outer ones. If this happens the central density surface decreases, hence avoiding a reliable fit.\\
In the last column of Tab~\ref{tab:clusters} we reported a flag which indicates the statistical reliability of each SC. A SC has flag = 3 if it has $G$ $\geq$3 and $n_0/\phi \geq$5; flag = 2 and flag = 1 if only the latter or the former condition is satisfied, and flag = 0 if both conditions are false. 
We want to underline that the flag does not take into account CMDs, which also represent a  valuable tool to discern whether or not a group of stars is a real SC.
To conclude, 65\% of the SCs have a central to background ratio above 5 even considering only the most likely star members.

\section{Results}
\label{sec:results}

\begin{figure*}
    \centering
    \includegraphics[width=0.32\textwidth]{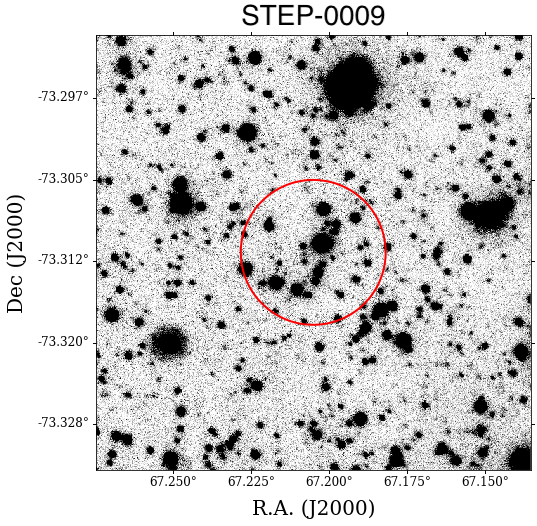}
    \includegraphics[width=0.32\textwidth]{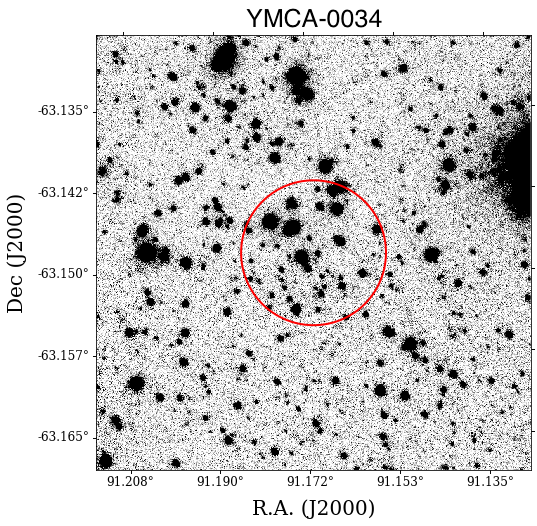}
    \includegraphics[width=0.32\textwidth]{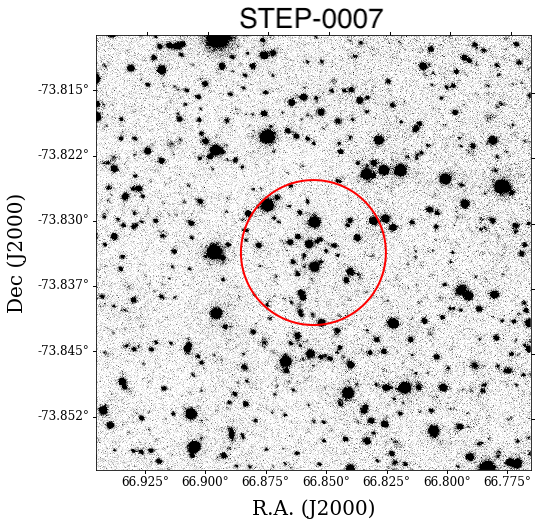}\\
    \includegraphics[width=0.32\textwidth]{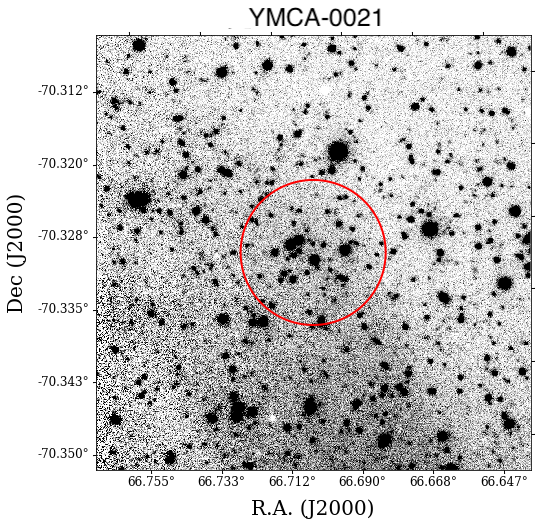}
    \includegraphics[width=0.34\textwidth]{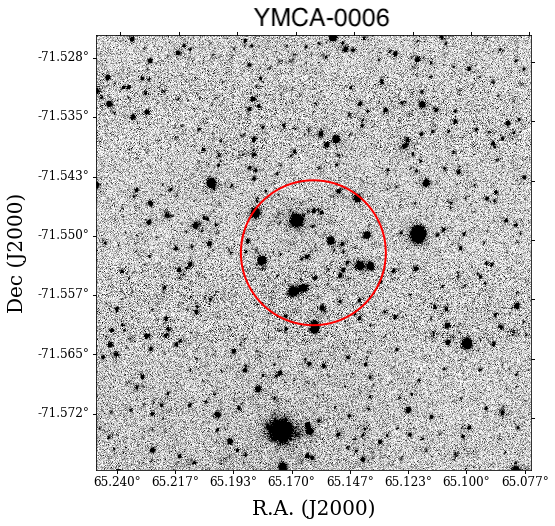}
    \includegraphics[width=0.32\textwidth]{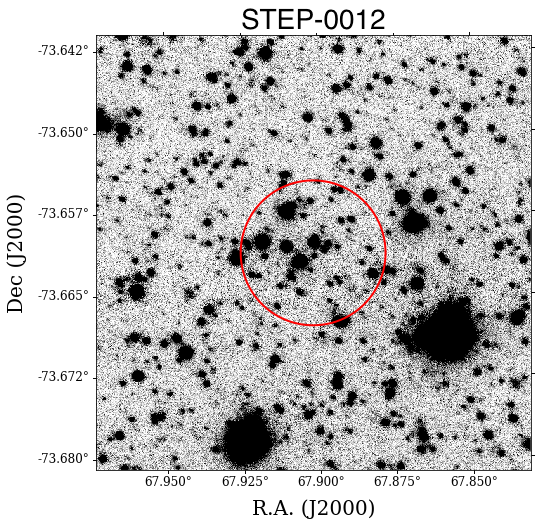}
    \caption{Sky images of the same SCs whose CMDs are displayed in Fig.~\ref{fig:example_iso_fitting}. Each panel has a size three times as large as the SC radius, which is represented by a red circle.}
    \label{fig:cluster_images}
\end{figure*}

With the procedures outlined in the previous sections we were able to detect 85 SCs, 78 of which are new candidates. In the investigated tiles there were eight already known clusters; although we recovered all of them as overdensities, we excluded one of them since its CMD did not match any isochrone (OGLE-LMC-CL-0757).
In Fig.\ref{fig:cluster_images} we show the sky images of the six SCs whose CMD has been presented in Fig. \ref{fig:example_iso_fitting}.
In the following section we discuss in some detail the characteristics of the 85 SCs. 

{}



\subsection{Comparison with literature}

Since the number of already known SCs in our observed fields is limited, it is hard to make a statistically significant comparison with the literature in order to test our methodology. 
Indeed, only two of the seven already discovered SCs have their main parameters estimated in the literature. For these two objects we verified that our estimates for age, reddening and metallicity are consistent with the literature. As for the ages we have logt 9.4 vs 9.35 for OGLE-LMC-CL-1133 and 9.3 vs 9.3 for SL842 (see also Tab~\ref{tab:clusters}). Concerning the 
reddening we have $E(B-V)$ = 0.08 mag vs 0.06 mag and 0.03 mag vs 0.03 mag for OGLE-LMC-CL-1133 and SL842, respectively. Finally, we found the literature metallicity estimate only for the SC SL842, finding that its value Z$\sim$0.005 is consistent within the uncertainties with our measure Z = 0.006.
Overall, our SCs have radius estimates smaller with respect to the literature ones. We argued that this difference is due to the different methods employed to derive such a measure. We discuss this point more in detail in \S\ref{sec:discussion}.

\subsection{Spatial distribution}

\begin{figure}
    \centering
    \includegraphics[width=0.5\textwidth]{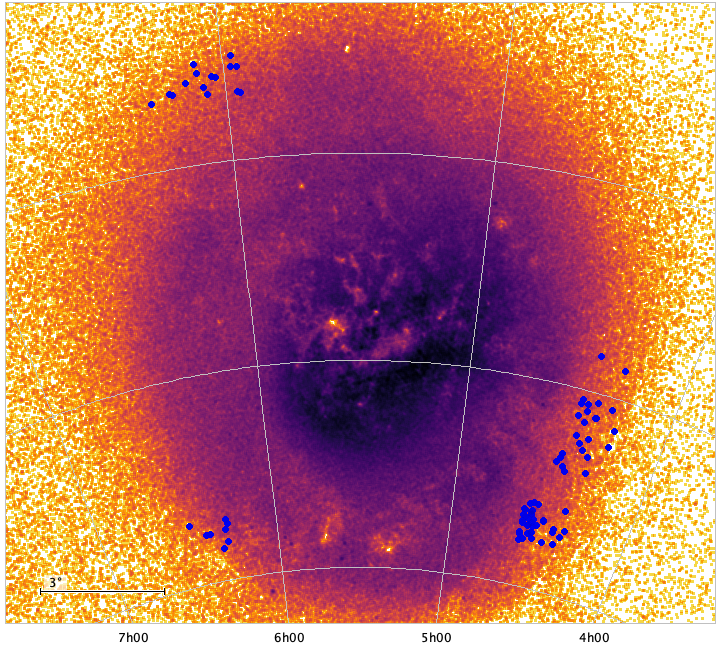}
    \caption{Position of the 85 clusters in the sky showed as blue points. The LMC galaxy is depicted using Red Clump (RC) stars taken from {\it Gaia} Data Release 2 \citep{Brown2018}.}
    \label{fig:cluster positions}
\end{figure}

Figure~\ref{fig:cluster positions} shows the position of all 85 SCs detected in this work. It can be seen that the majority (62 objects) is placed in the West-South-West region (37 and 25 objects in STEP and YMCA tiles, respectively). Among the remaining SCs, 15 and 8 were found in the North-East and at South-East of the LMC, respectively.
The high number of SCs found in the STEP 3\_21 tile (N. SCs = 35) with respect to the nearby tiles at similar R.A. (9 SCs in the tile YMCA 3\_21 and 14 SCs in the tile YMCA 4\_22) is remarkable. A natural explanation is that the STEP 3\_21 tile includes a region of higher star density ($\sim$ 30\% and 70\% more stars than YMCA 3\_21 and YMCA 4\_22, respectively) with respect to the other two tiles. However, such discrepancy alone does not justify a number of SCs 4/2.5 times larger than YMCA 3\_21/YMCA 4\_22.
The STEP 3\_21 tile lays at the end of the Bridge connecting the LMC with the SMC and some substructures have been found in this region likely due to the repeated interaction between the MCs \citep[e.g.][]{mackey2018,belokurov2019}. Hence we can speculate that the increased number of SCs in the quoted tile can be due to these interactions. A definitive interpretation of this SCs over-density will probably be possible only when all the other tiles in YMCA will be analysed.\\

To measure the angular distance of the SCs from the LMC centre we used Eq. 1 from \citet[][]{Claria2005}:

\begin{equation}
    d = d_0\{1 + [sin(p - p_n)^2][tan(i)]^2\}^{0.5}
\end{equation}{}

\noindent
where \emph{d} is the angular de-projected distance from the LMC optical centre, \emph{$d_0$} is the angular distance on the plane of the sky, \emph{p} is the position angle of the cluster, \emph{$p_n$} is the position angle of the line of nodes and \emph{i} is the inclination of the LMC disk with respect to the plane of the sky. We assumed as LMC centre coordinates the optical centre $(\alpha, \rho) = (79.91, -69.45)$ taken from \citet{devaucoulerus1972}, whereas to compute the deprojected distance we used $p_n = 149.23$ and $i = 25.86$ estimated by \citet{Choi2018}.

{}

Even though our tiles cover regions up to $\sim 12$ degrees, we did not find any SC beyond $\sim$ 9 degrees from the LMC centre, although the disk of the LMC extends up to about 15 kpc \citep{Saha2010, Balbinot2015}. This result is in agreement with the works by \citet{Piatti2017b} and \citet{Pieres2016}.

Since it is known that both MCs contain a numerous population of binary clusters \citep[e.g.][]{Pietrzynski2000b}, we performed an internal research to look for those candidate SCs that are closer than one arcminute. We found three pairs that satisfied this condition. The first couple of candidate SCs, namely STEP-0008 and STEP-0009, has a high probability to be a binary SC, indeed, their centres are separated, in projection, by $\sim$ 0.5\arcmin, and  their estimated ages are very similar (log(t) = 9.2 and log(t) = 9.25, respectively). The other two couples have a lower chance to be binary SCs, and we report them here for completeness: the centres of STEP-0018 and STEP-0020 are separated by 0.94\arcmin  and have a slightly different estimated age (log(t) = 9.2 and log(t) = 9.3, respectively); STEP-0022 and STEP-0024 are separated by 0.86\arcmin and have estimated ages of log(t) = 9.45 and log(t) = 9.6, respectively.

The detection of numerous new SCs in all three regions analyzed in this work means that most likely there are many SCs still undetected in the still unexplored outskirts of the LMC, at least within 9 degrees.
In particular, excluding the STEP tiles\footnote{The over-density of SCs found in the tile STEP 3\_21 would influence the following considerations.}, we detected 42 new SCs in the six innermost YMCA tiles (at distances in the range  4.8\degr-9.0\degr from the LMC centre). Therefore, the expected number of SCs in this range of distances is of the order of $\sim$ 7 SCs per sq. degrees. This implies that, in a rough approximation, $\sim$ 70 new SCs in the LMC periphery are still awaiting to be detected in the remaining YMCA tiles.

\subsection{Age distribution: first evidence of clusters in the "age gap"}

\begin{figure}
    \centering
    \includegraphics[width=0.48\textwidth]{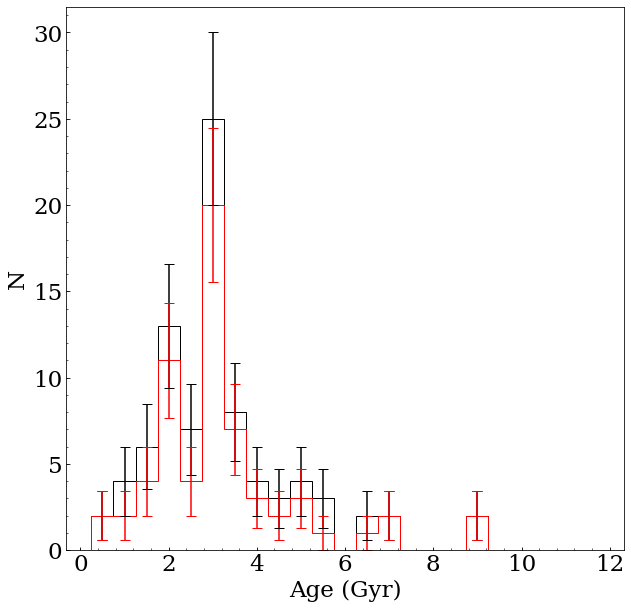}
    \caption{Age distribution of the whole cluster sample (black histogram) and that one of the 64 SCs with a G $\geq$ 3 (red histogram). Errors are poissonian.}
    \label{fig:cluster ages}
\end{figure}

Figure~\ref{fig:cluster ages} displays the distribution in age of all SCs detected in this work (black contours). Since we are aware that the visual inspection of the cleaned CMD suffers from an unavoidable amount of subjectivity, we plot in the same figure also the age distribution of the 64 SCs having a G $\geq$ 3 (red contours), and hence statistically more reliable.
The ages range from $8.80$ to $10.05$ log(t), but with the exception of three objects, all SCs are older than 1 Gyr. This result suggests that there was no SC formation activity in the last Gyr in the LMC periphery, at least in the analysed regions. A result which, on the one side, is consistent with the existing literature on the SFH and age-metallicity relationship in the LMC outer disk, which demonstrates that these regions are mainly composed by an old (and metal-poor) population \citep[e.g.][]{Saha2010,Piatti2013}.
On the other side, however, this occurrence is at odds with what is known for the inner part of the LMC, where the SCs distribution shows an enhancement of formation about $\sim 100 - 300$ Myr ago \citep{Pietrzynski2000a, Glatt2010,Nayak2016}. 
These authors suggested that such increase in the SCs formation rate was due to a recent close passage between the MCs which led to the formation of the connecting bridge, as confirmed by Montecarlo simulations \citep{Besla2012,Diaz&Bekki2012,Kallivayalil2013,Zivick2018}.\\

{}

An inspection of Fig.~\ref{fig:cluster ages} reveals that the SC age distribution has a peak around 3 Gyr, very close to the secondary peak found by \citet[][]{Pieres2016} at 2.7 Gyr in their investigation of the Northern regions of the LMC.
A possible explanation for this enhancement implies a previous close encounter between the LMC and SMC. Indeed, since the MCs should be on their first infall onto the MW \citep{Besla2007,Kallivayalil2013}, it is difficult to invoke a tidal interaction with the Galaxy to explain this peak.
On the other hand, current simulations of the Magellanic System, taking into account the recent accurate proper motion measurements of the MCs \citep{Kallivayalil2013,Zivick2018}, agree that the MCs became an interacting pair just a few Gyr ago \citep[see][]{Besla2012,Diaz&Bekki2012,Pardy2018,Tepper-Garcia2019}.
Such models, depending on the assumed initial conditions on the parameters support the occurrence of a first close encounter 2 - 3 Gyr ago: consistent with the peak found at $\sim 3$ Gyr.
It is worth noticing that the LMC field stars also show an enhancement of star formation at an age of 2-3 Gyr \citep{Tosi2004,Harris2009,Rubele2011,Weisz2013}. 
A LMC-SMC close passage is expected to increase also the SC and stellar field formation of the SMC. Unfortunately, most of the literature works focused on SC in the SMC do not have sufficiently deep photometry to estimate the age of cluster older than 1 Gyr. However, focusing on studies about SFH of the field, several literature works reported peaks in the SFH, some are consistent with our main peak \citep[e.g.][who revealed a peak at 2-3 Gyr and 3.5 Gyr, respectively]{Harris2004,Weisz2013}, others not \citep[e.g.][who found a peak at 1.5 Gyr]{Rubele2015,Rubele2018}. This interesting point deserves further investigations that will be possible when a complete and homogeneous characterization of the SCs belonging to both MCs will be available.
It is also visible a likely secondary peak at 2 Gyr. Since an uncertainty of $\sigma$log(t) = 0.1 at 2 Gyr corresponds to $\sim$ 500 Myr, there is a probability that such a peak might represents the tail of the main peak. A definitive answer is only possible when all YMCA tiles will be explored.\\
It is worth to point out that \citet[][]{Pieres2016} found a main peak at $\sim$1.2 Gyr. Even considering SC age uncertainties, our secondary and their main peak are very unlikely to be associated. This difference is very interesting and might reveal a different SC formation history in different LMC regions due to a past interaction between the MCs. Indeed, fields analyzed by \citet[][]{Pieres2016} lie very far away from the majority of our tiles (with the exception of the very few tiles in the North-East side). Anyway, a definitive answer could be achieved only with the analysis of the remaining YMCA tiles all around the LMC.

An additional interesting aspect in Fig.~\ref{fig:cluster ages}, is the presence of a number of SCs in the well known LMC SCs age gap (see the Introduction). Indeed, we find 16 SCs (corresponding to 19\% of our sample) in the age interval 9.6 $<$ log(t) $\leq 10.0$ ($\sim 4 \leq$ t (Gyr) $\leq 10$), even if only four objects have been found between 7 and 10 Gyr. 
Until few years ago, only one age gap SC was known in the literature \citep[e.g.][ESO 121-SC03]{Rich2001}. Recently, \citet{Pieres2016} reported a few SCs with estimated ages falling in the range interval between 4 and 10 Gyr, even if they do not addressed it. This finding is even more significant by considering that they performed a visual research of undiscovered SCs, thus sparse and less populous SCs might still be missing.
The absence of SCs in this range of ages has been a debated question for more than 25 years, suggesting different formation and evolution paths for the SCs and the stellar field, since in the latter a similar gap is not observed \citep{Tosi2004,Carrera2011,Piatti2013}.
Furthermore, also the SMC does not have an age gap, despite in other periods their SC share a similar evolution, like the enhanced periods of SC formation \citep[e.g.][]{Glatt2010,Nayak2018}\\

This gap might have been an observational bias. As already mentioned, most surveys focused on the inner LMC, leaving the periphery almost unexplored. 
Since central areas might have stronger destruction effects due to a shorter evaporation time at smaller galactocentric radius \citep{Baumgardt2003}, a lower number of old SCs is expected in the central part of LMC with respect to young ones. 
Indeed, in these regions a very high number of young SCs is present \citep{Pietrzynski2000a,Glatt2010,Piatti2013,Nayak2016,Piatti2015a,Piatti2018}. 
Furthermore, the SC luminosity function decreases with age \citep[e.g.][]{Degrijs2003,Hunter2003}, making it hard to detect old SCs (including SCs in the age gap), hence only a deep photometry can reveal them.
It is worth mentioning again that most surveys had not enough depth to reveal such clusters \citep{Pietrzynski2000a, Glatt2010, Nayak2016}.
The presence of a subsample of SCs belonging to the age gap represents an important opportunity to disclose the past LMC evolution, hence such objects may be the follow-ups target for a deeper photometry in order to better constrain their physical properties.
As the age distribution of the SCs with G $\geq$ 3 looks similar to that of the entire sample (it is just shifted down), all the above considerations are still valid even if some candidate SCs turn out not to be physical systems. In particular, the SCs falling in the age gap with G $\geq$ 3 are 11.\\

\begin{figure}
    \centering
    \includegraphics[width=0.48\textwidth]{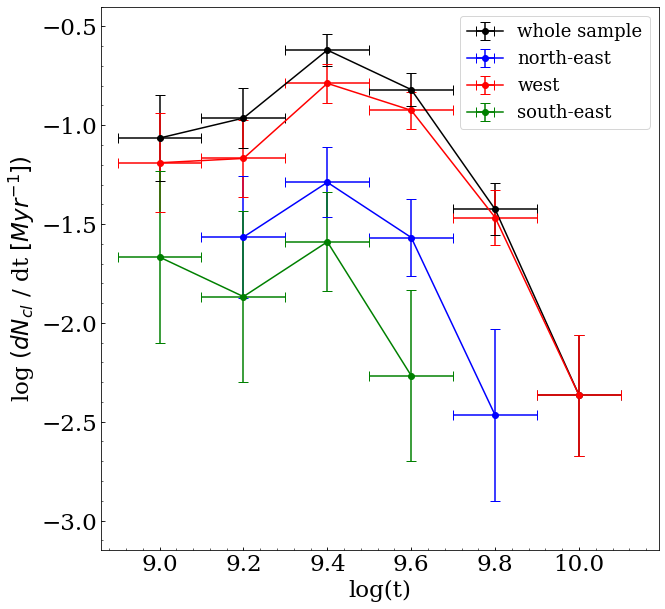}
    \caption{Number of SCs per bin of age as a function of age. The black solid line corresponds to the whole cluster sample, while coloured solid lines represent clusters in different regions of the LMC.}
    \label{fig:formation rate}
\end{figure}

To be more confident with our results, we tested the reality of our detection of SC in the ``age gap" against the possibility that we underestimated the uncertainties on the age for the SCs with log(t) $\geq$ 9.5\footnote{It is very unlikely that for SCs younger that 3 Gyr our errors are underestimated (see also discussion at \S\ref{sec:isochrone fitting}).}.
Indeed, if such SCs had a larger age uncertainty, i.e. $\sigma$log(t) = 0.2 dex  with respect to our estimate $\sigma$log(t) = 0.1 dex, it could happen that some SCs falling in the age gap could actually be younger.
We basically aim at verifying whether it is statically possible to observe a consistent number of SC in the age gap starting with a parent distribution with no SCs at all in this age interval, if age uncertainties are greater with respect our evaluation.
To verify this possibility, we random generated three distributions: the first one is a Gaussian peaked at $\sim$ 3 Gyr, with standard deviation equal to 0.1 dex in log(t) and with no SCs in the age gap, called $D0$ hereafter.
The other two distributions ($D1$ and $D2$) differ from the former in having 8 and 16 SCs in the age gap, i.e. half and the total number of SCs detected in the age gap in this work, respectively.
We added Gaussian errors to the objects in the three distributions with $\sigma$=0.1 dex or 0.2 dex according to whether their ages are below or above the threshold of log(t)=9.5 dex, respectively.
We performed a Kolmogorov-Smirnov test to select the distribution that better approximate our data.
We repeated the whole procedure 1000 times and as result, in more than 88\% of the cases the distributions that best approximate the observed one were those with a consistent number of age gap SCs (445 and 440 cases for $D1$ and $D2$, respectively). On these basis, we can rule out the hyphotesis that our subsample of SCs in the age gap is generated by an underestimation of their age uncertainties with a $\sim$90\% confidence.
Furthermore, the distribution $D0$ yielded a mean number of SCs in the age gap (after errors have been added) of $8.7 \pm 2.9$, i.e. at $\sim$ 2.5 $\sigma$ from the actual value.
Hence $D0$ could explain about half of the SCs detected in the age gap, but never all of them, demonstrating that it is very unlikely that $D0$ is the parent distribution of our data.
For completeness, $D1$ and $D2$ had a mean number of age gap SCs of $13.0 \pm 3.0$ and $13.1 \pm 3.1$, well consistent within the errors with the actual number of age gap SCs.
Concluding, this test further strengthen our results,  supporting the reality of the presence of a consistent number of SCs in the age gap. \\
In Fig.~\ref{fig:formation rate} we display the number of SCs per Myr (SC frequency) as a function of the age expressed as log(t). Note that due to the very low number of SCs younger than 1 Gyr, we put log(t) = 9.0 as the minimum value on the x-axis. The figure shows the SC frequency for the entire sample (black solid line) as well as that for the three analysed regions, namely North-East, West and South-East of the LMC (blue, red and green lines, respectively).
The peak of SC formation is evident in all three regions at $\sim$2-3 Gyr\footnote{Considering our uncertainties of $\sigma$log(t) = 0.1 on the estimated cluster ages and the bin interval of log(t) = 0.2, the peak could be at slightly lower or higher values.}. 
In the age gap (9.65$\leq$ log(t) $\leq $ 10.0), looking at the whole sample (black solid line), the SC frequency seems to be consistent with the one measured at more recent epochs (9.0$\leq$ log(t) $\leq $ 9.3). Instead, beyond log(t)$\sim$9.8 the SC frequency has a sharp decrease, as expected from the fact that only a few SCs are present in the age range 7-10 Gyr.
However, it is important to emphasize that the SC frequency also depends on disruption effects, such as two-body relaxation or disc and bar shocking \citep{Spitzer1987,Zhang1999,Baumgardt2002,Baumgardt2003}. Even if the LMC periphery is a low-density environment, such effects could be present \citep[e.g.][]{Lamers2005}, leading to the dissolution of low-mass SCs. 
On this basis, the data presented in this work suggest that the SC frequency in the age gap does not reflect an epoch of quenched SC formation as it is commonly assumed, but more likely is the result of observational bias (at least in the outskirt of the LMC). 

{}
\begin{figure}
    \centering
    \includegraphics[width=0.5\textwidth]{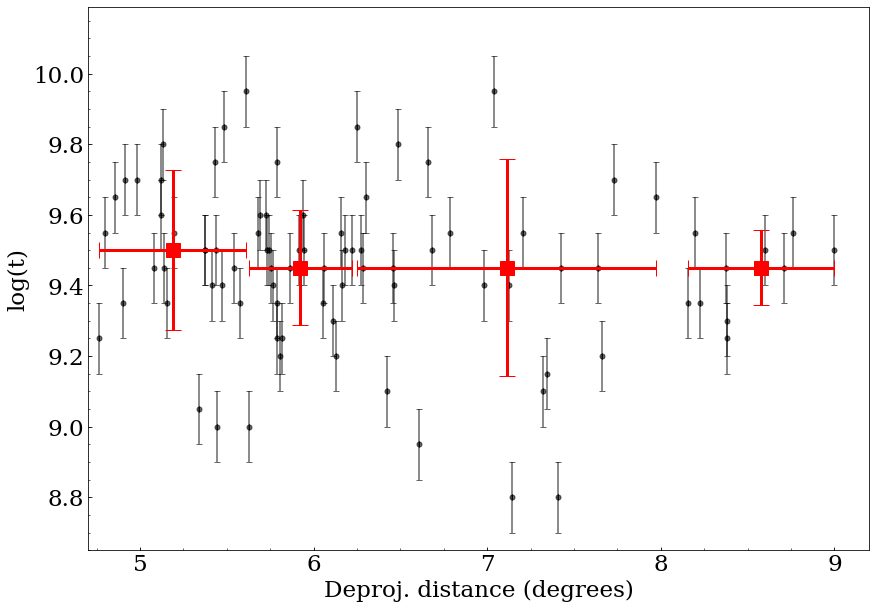}
    \caption{Estimated ages of SCs as a function of distance from the LMC center (black points). Red squares indicate the median of the log(t) per distance bin.}
    \label{fig:age vs distance}
\end{figure}{}

Finally, we do not find any correlation between SC ages and distance from the LMC centre, as shown in Fig.~\ref{fig:age vs distance}, where we display the estimated SC age vs the galactocentric deprojected distance, along with the median of the age in each distance interval, properly calculated to have 25 SCs in each interval. This outcome confirms a similar result obtained by \citet[][]{Pieres2016} in the Northern side of the LMC.

\subsection{Cluster metallicity}

\begin{figure*}
    \centering
    \includegraphics[scale=0.5]{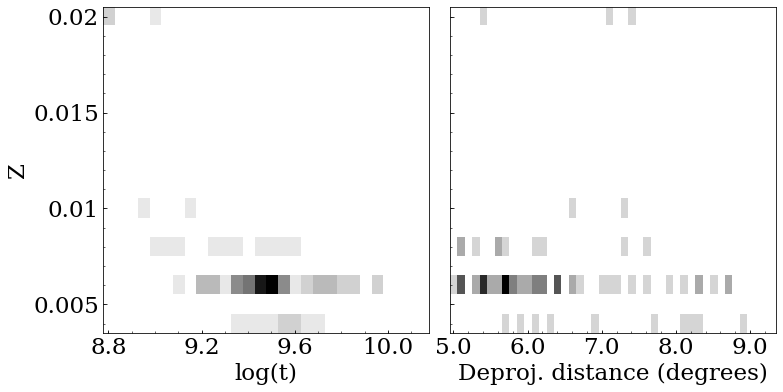}
    \caption{{\it Left:} Density plot of the metal content as a function of the age. The map is 28$\times$17 pixels. {\it Right:} Density plot of the metal content as function of the deprojected distance from the LMC centre. The map is 44$\times$17 pixels.}
    \label{fig:metallicities}
\end{figure*}

The metallicity estimated for the SCs from isochrone fitting ranges from Z = 0.004 up to 0.02. The youngest SCs show a solar metallicity, whereas the metallicity of the majority of newly discovered SCs is Z = 0.006, corresponding to the mean LMC metallicity value for the last 2-3 Gyr \citep{Piatti2013}.
As it is shown in Fig.~\ref{fig:metallicities} (right panel), there is apparently no relationship between the SC metal content and the corresponding deprojected distance from the center of the LMC, even if there is no cluster more metal rich than Z = 0.006 beyond 8\degr. 

{}

Concerning the age-metallicity relation displayed in Fig.~\ref{fig:metallicities} (left panel), we do not note any clear correlation for ages older than $\sim$1.5 Gyr (log(t) $\sim$ 9.2). There are a few SCs younger than log(t) = 9.2 that have Z $\geq$ 0.006, but the statistic is too poor to devise any possible age-metallicity relationship.\\

\subsection{Absolute magnitudes}
\label{sec: absolute mag}

\begin{figure}
    \centering
    \includegraphics[width = 0.5\textwidth]{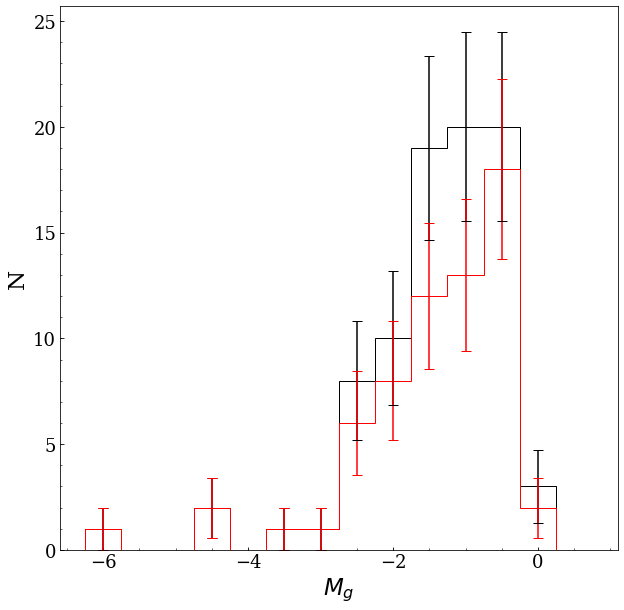}
    \caption{Distribution of the absolute magnitudes of the entire SC sample (black histogram) along with SCs having G $\geq$ 3 (red histogram). Errors are poissonians.}
    \label{fig:mag}
\end{figure}

Estimating SC absolute magnitudes is usually the starting point to estimate the corresponding masses, and, in turn, to study the SC mass function.
Information about the masses of the complete sample of LMC SCs would allow to probe fading or evaporation effects, as well as to derive how their destruction time-scale depends on the SC mass.
This investigation will be possible only when YMCA is completed, and is beyond the scopes of the present paper, hence in this section we briefly discuss the absolute magnitude distribution of the 85 detected SCs displayed in Fig.~\ref{fig:mag}, along with that of the SCs with G $\geq$ 3.
Since the two distributions have similar shapes, the following considerations will hold for both cases.\\
We note that most of the SCs have a $M_g$ in the range between -3 and 0 mag.
This interval of magnitudes is well below the magnitude limit of other works present in the literature \citep[e.g.][their fig. 4]{Hunter2003}, suggesting that we are sampling the low end of the SC luminosity function.
Indeed, the distribution has a peak around -1 mag, then it sharply decreases. This value might reflects our observational limit or could have a physical origin, i.e. can be regarded as the lower limit absolute magnitude (mass) for bound objects, and/or as the limiting mass needed by a SC to survive destruction for at least 1 Gyr. 
Exploring such intervals of low masses might help to constrain the formation and the dynamical evolution of the SCs, and it will be investigated in another paper.

{}

\section{Discussion and conclusion}
\label{sec:discussion}

\begin{figure*}
    \centering
    \includegraphics[scale=0.6]{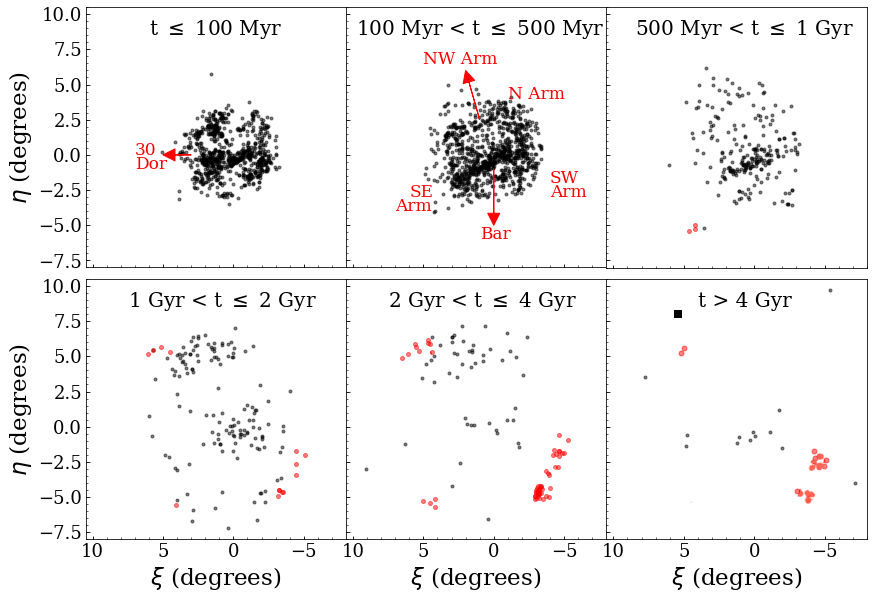}
    \caption{SCs relative position, with respect to the LMC centre, detected in this work (red points) and present in the literature (grey points), divided per age bins. Black square in the bottom right panel marks the position of ESO121-SC03, the first known SC of the age gap. Besides ESO121-SC03, the only other previously known clusters in the > 4 Gyr age range are the old globulars.}
    \label{fig:spatial_position_allclusters}
\end{figure*}

In order to get some insight into the global evolution of the LMC it is necessary to combine our SC sample with the others available in the literature.
To this aim and to produce a rather homogeneous sample, we selected only SCs whose ages were estimated through isochrone fitting. In particular,  we used the catalogues by \citet{bonatto2010}, \citet[][]{Glatt2010}, \citet[][]{popescu2012}, \citet{Baumgardt2013}, \citet[][]{Piatti2015a,Piatti2017b,Piatti2018} \citet[][]{Nayak2016}, \citet[][]{Pieres2016}.
In order to avoid duplicates, we identified all SC pairs whose centres were closer than 20\arcsec to each other, retaining only the cluster from the most recent work. At the end of this procedure, we were left with 2610 clusters, including those newly discovered by us, whose spatial distribution is displayed in Fig.~\ref{fig:spatial_position_allclusters}. In this figure, each panel shows the position with respect to the LMC centre of a sample of SC in a given age interval: black points represent SCs taken from the literature while red points are our new SC candidates. It can be seen that the youngest SCs displayed in the top three panels of Fig.~\ref{fig:spatial_position_allclusters}, are mainly located along the bar of the LMC, SCs with age in the interval $500 \leq t \leq 1000$ Myr are particularly concentrated, contrarily to SCs older than 1 Gyr, that are more evenly distributed.
{}

\begin{figure}
    \centering
    \includegraphics[scale=0.38]{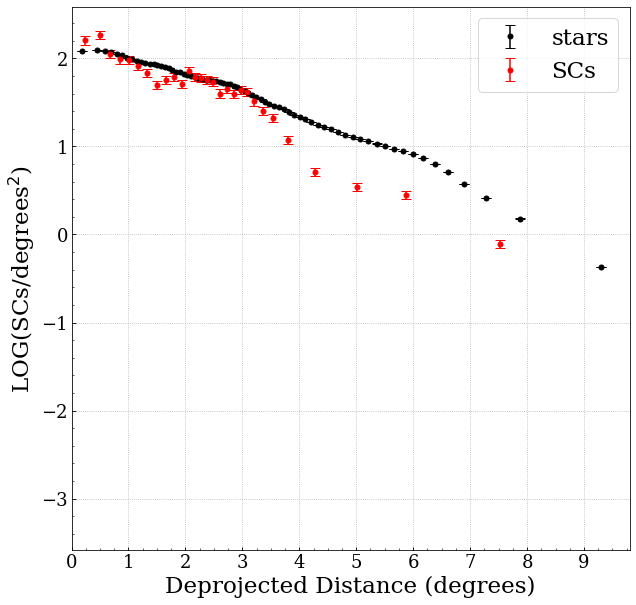}
    \caption{Number density profile of SCs (red points) and field stars (black points) taken from the Gaia Data Release 2.}
    \label{fig:cluster radial distr.}
\end{figure}

Interestingly, such SCs are completely absent in some regions of the LMC, despite we doubled the number of known SCs with t $\geq$ 2 Gyr (see bottom central and bottom right panels of the figure). This is a clear observational bias, as most of the surveys are not deep enough to detect SCs older than $\sim$ 1-1.5 Gyr \citep{Pietrzynski2000a,Glatt2010,Nayak2016}, unless they are very rich. Furthermore, the majority of the surveys explored the inner regions of the LMC, as it is evident from the figure (bottom left panel). The only exception is the work by \citet{Pieres2016} who used the deep DES photometry to explore the Northern part of the LMC (see the bottom left and bottom central panels of the figure). 
Therefore, we can conclude that the census of SCs in the LMC is still quite incomplete, not only in the outskirts, but also in more central regions of the galaxy, were SCs surveys were shallower.\\
Figure~\ref{fig:cluster radial distr.} shows the number density profile of all 2610 SCs (red points) along with that of LMC field stars (black points) in order to infer analogies and/or differences between these two population. LMC stars were taken from the Gaia Data Release 2 \citep[DR2,][]{Gaia2016,Gaia2018}, and have been selected according to their proper motions (PMs) to remove likely MW stars, in particular, since we were more interested in having a clean sample rather than a complete one, we estimated the peak of LMC's stars PMs and retained the stars in a rectangular box with size ($\mu_{\alpha^*}$, $\mu_{\delta}$) = (1.8, 1.2) mas yr$^{-1}$ centred around the above peak.\\
The star number density profile has been scaled in order to overlap with that of the SCs, thus the values on the y-axis refer to the SCs. An inspection of the figure reveals that in the LMC innermost 3\degr, the slope of the SC and star distributions agree very well, with two exceptions: i) there is a central peak in the SC number density profile, likely due to the very high number of young SCs located in the LMC bar, while the star distribution is flat, likely a consequence of lower completeness of Gaia measures at the LMC centre; ii) the SC number density drops at about 1.5\degr, indicating perhaps the low density regions between the LMC bar and the spiral arms, where the SC surface density quickly increases. However, since previous literature works aiming at SC detection were mainly devoted to the regions of the bar and spiral arms, the SC low density at $\sim$1.5\degr could be an observational bias. \\
An interesting feature in Fig.~\ref{fig:cluster radial distr.} is a slope change in both SC and star distributions (although the SC profile becomes steeper with respect to the stellar counterpart), at $\sim$3\degr from the LMC centre, indicating the distance from the LMC centre where the spiral arms are no longer visible. All these similarities could be a signature of a common evolutionary path between SCs and stellar field population.
Even though the two distributions look different beyond $\sim$3-4\degr, we cannot compare them anymore, due to the already mentioned high SC incompleteness in the LMC periphery (clearly visible also in Fig.\ref{fig:spatial_position_allclusters}).
Interestingly enough, Fig.~\ref{fig:cluster radial distr.} shows a flattening of the SC number density profile beyond 5 degrees, an obvious result of the newly SCs discovered in this work and by \citet{Pieres2016}, suggesting again that many SCs are still missing.

{}

\begin{figure*}
    \centering
    \includegraphics[scale=0.5]{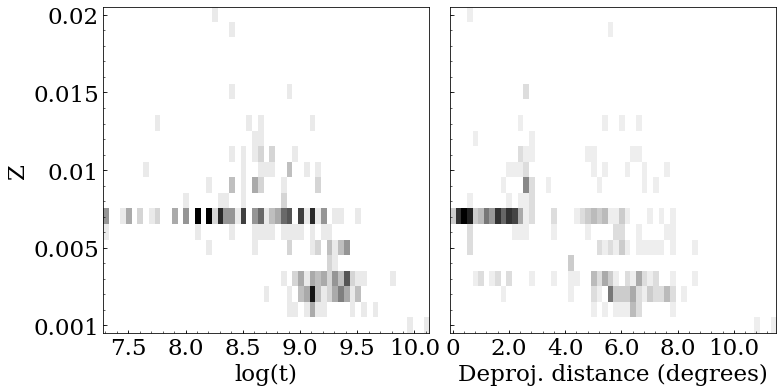}
    \caption{{\it Left:} Density plot of the metal content as a function of the age. The map is 57$\times$21 pixels. {\it Right:} Density plot of the metal content as function of the deprojected distance from the LMC centre. The map is 58$\times$21 pixels.}
    \label{fig:Z_catalogue}
\end{figure*}

\begin{figure}
    \centering
    \includegraphics[scale=0.35]{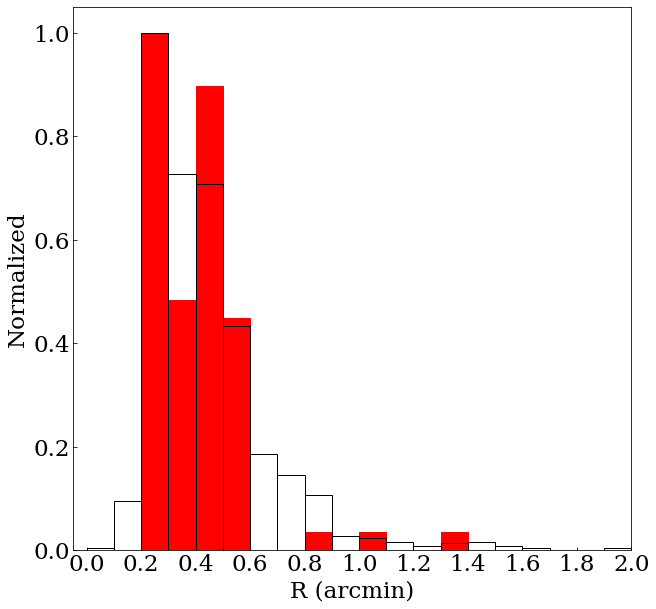}
    \caption{Radius distribution of all SCs from literature (black histogram) along with our sample (red filled histogram), normalized at maximum.}
    \label{fig:radius_hist}
\end{figure}

In Fig.~\ref{fig:Z_catalogue} we investigate the correlation between metallicity-age and metallicity-LMC distance using SCs studied in the literature. To obtain a catalog as homogeneous as possible, we selected SCs from works that published metallicities for a large number of objects. Therefore, we collected data from \citet[][]{Piatti2014}, \citet{Palma2016}, \citet{Pieres2016}, obtaining a sample of 328 SCs, spanning a range of ages from $\sim$10 Myr to $>$ 10 Gyr and covering all distances from the LMC centre up to $\sim$11\degr.
All metallicities extracted from the above cited works have been estimated through isochrone fitting of the SC CMD, and results were given in [Fe/H], and hence we converted them using the relation $Z = 10^{[Fe/H]} * Z_\odot$ where $Z_\odot = 0.02$.
An inspection of Fig.~\ref{fig:Z_catalogue} reveals a trend between metallicities, age, and distances, with the more metal-poor SCs being also the oldest ones and located at larger distances with respect to the metal-rich counterpart. This is more evident in the plot metallicities vs. age (left panel), as almost all SCs younger than 1 Gyr have Z $\geq$ 0.006 and all those older than 1 Gyr have Z lower than that. The correlation between metal content and galactocentric distance is less evident (right panel), but again most of the SCs with Z $\geq$ 0.006 are concentrated in the inner 5 degrees, whereas those more metal poor are beyond that radius. This outcome is also confirmed by the age-metallicity relationship (AMR) in the stellar field population. In particular, \citet{Carrera2011} found that metal-poor stars have mostly been formed in the outer disk, while the more metal-rich ones preferentially formed in the inner disk. This scenario have been confirmed by \citet{Piatti2013} using 5.5 million stars belonging to LMC main body.
The figure also shows many SCs having Z = 0.007, i.e. a value very close to the LMC metallicity value for the last 2-3 Gyr \citep[Z = 0.006:][]{Piatti2013}.
A comparison with the SCs detected in this work is not immediate since our sample is mainly composed by old SCs that are all located farther than 5\degr~from the LMC centre. Anyway, the majority of the 85 SCs studied here have a metallicity of Z = 0.006, consistent with many SCs distributed throughout the LMC, 
although at a first glance our SCs appear more metal-rich compared to the literature at the same age i.e. Z $\sim$ 0.006 vs Z $\sim$ 0.003.
However, considering the uncertainties on metallicities in our and in the cited works, the above quoted difference is consistent within the errors. Indeed, we estimated an error of $\Delta$Z = 0.002, while literature SCs have a mean uncertainty of $\Delta$Z $\sim$ 0.001 - 0.002.
Furthermore, we cannot rule out the possibility that the reddening-metallicity degeneracy could enhance this disagreement. For example, if we underestimated the reddening while previous authors overestimated it, the difference in metallicity naturally arise. 
\\
Finally, it is instructive to compare the SC radii calculated in this work with those in the literature. 
However, defining the edge of a SC (the distance from the SC centre beyond which no star belongs to the SC anymore) is not a trivial task. In fact, there are many different definitions of a SC radius, depending on how it is estimated. Hence, comparing radii measured by different authors means to deal with possible diverse definitions.
Bearing this in mind, we tried to build a sample of literature SCs as large as possible, but avoiding to mix catalogs built adopting very different ways to measure the radii. To this aim, we checked the average difference between the estimated radii of SCs in common in each pairs of investigations. 
At the end of the procedure, we remained with 2315 SC radii, including the works by \citet[][]{Glatt2010}, \citet{bonatto2010}, \citet{Nayak2016}, whose radii are consistent or homogeneizable within less than 0.1\arcmin. Instead, we excluded the catalogs by  \citet[e.g.][]{Palma2016,Pieres2016,Sitek2016}, because the spread around the mean of the radii differences was considerably high ($\sim$ 0.3$\arcmin$). 
Figure~\ref{fig:radius_hist} displays the radius distribution (up to $2\arcmin$) of our 85 SCs (red filled histogram) along with the literature SCs sample (black histogram). Both histograms are normalized to their maximum to facilitate the comparison.  
An inspection of the figure reveals that most of the SCs in the LMC have $0.2\arcmin \leq$ R $\leq1.0\arcmin$, being the mode of the distribution placed at $\sim 0.2-0.3\arcmin$. Our SCs have radii in the same range of those taken from literature, as the bulk of them have a size comprised between 0.2\arcmin and 0.6\arcmin, with a peak at $\sim$0.2-0.3\arcmin, in agreement with the literature.

{}

{}

In this work we explored 23 sq. deg. in the outskirt of the LMC, using data from the YMCA and STEP surveys, to detect new SCs and thus constrain the LMC evolutionary history.
To this purpose, we developed a procedure that we can basically resume in two steps. The first step is to search for over-densities in the sky (pixels with a S/N ratio above a given threshold), through a density map built with a KDE, and then derive their centers and radii.
We exploited Montecarlo simulations to both define a threshold depending on the field densities and simulate artificial SCs to quantify the efficiency of the adopted method. As a result, a detection percentage around 90\% was obtained even in the worst case, e.g. high stellar field density and sparse SCs.
The second step consisted in using the cleaning CMD procedure developed by \citet[][]{Piatti2012}, in order to recognize real physical systems from false positives.  In the end, we were left with 85 candidate SCs, among which 78 were not catalogued in the literature.
We estimated age, reddening, metallicity of each cluster through an isochrone fitting technique, keeping the distance modulus constant to 18.49 mag. \citep{degrijs-wicker-bono-2014}. We also measured the integrated absolute magnitudes for each SCs.
Finally, we fitted their RDP built by using only stars with P$\geq$75\% with an EFF profile to adduce further robustness to their physical reality. About 70\% of the SC RDPs are well  fitted with an EFF profile. The remaining SCs have been fitted by employing stars with P$\geq$50\%.

From the SC parameters we derived the following results:
\begin{itemize}
    \item 
The age distribution has a sharp peak at $\sim$ 3 Gyr, likely due to a close encounter between the MCs, that might have enhanced the SC formation activity. Such interaction is expected from simulations \citep{Besla2012, Diaz&Bekki2012, Pardy2018, Tepper-Garcia2019}, which take into account the recent proper motion measurements of the MCs.
Furthermore, an increase in the star formation rate $\sim$ 2-3 Gyr ago has been observed also in the stellar field \citep{Harris2009,Rubele2011,Weisz2013}.\\

\item
For the first time, we detected a consistent number of candidate SCs in the 'age gap', a period ranging from 4 to 10 Gyr lacking of SCs \citep{DaCosta1991}. Given the high number of SCs in the age gap (19\% of our sample) and from the analysis of the SC frequency, the natural outcome is that the age gap is not an interval of minimal SCs formation as it has been believed so far. 
On the contrary, the age gap is likely the product of an observational bias, due to surveys using too shallow photometry and unable to detect clusters older than 1-1.5 Gyr \citep{Pietrzynski2000a,Glatt2010,Nayak2016}. Moreover most observations so far were focused on the LMC centre/bar where the extremely high stellar field density makes it hard to detect old and faint SCs.
A more accurate analyses of these SCs through follow-ups will provide a relevant opportunity to shed light on this evolutionary period of the LMC, and even on the whole MC system.\\

\item
There is no trend either between the age and the distance from the LMC centre, or between the galactocentric distance and the cluster metallicity.
Indeed, even though young clusters have a higher metal content with respect to old ones, the very few young clusters detected do not allow us to establish any correlation between cluster ages and their metallicities.\\
\end{itemize}

This work is the first of a series aiming at completing the census of SCs around MCs. As demonstrated here, many SCs are still undetected but their census and the estimation of their parameters are vital to get insights into the recent and past evolution of the MCs. As the YMCA survey is complete, we will be able, along with data from STEP, to explore the surroundings of the LMC, the SMC and their connecting bridge and thus to trace the evolutionary history of the entire Magellanic system.

\section*{Acknowledgments}

We are grateful to the referee for the useful suggestions that helped clarify our paper.
We thank L. Limatola and A. Grado for their help with data reduction.\\
M.G. and V.R. acknowledge support from the INAF fund "Funzionamento VST" (1.05.03.02.04).\\ 
This work has been partially supported by the following INAF projects; i) ``Main Stream SSH program" (1.05.01.86.28); progetto premiale ``MITiC: MIning The Cosmos Big Data and Innovative Italian Technology for Frontier Astrophysics and Cosmology" (1.05.06.10).\\
M.R.C. acknowledges support from the European Research Council (ERC) under the European Union's Horizon 2020 research and innovation programme (grant agreement No 682115).\\
G.Longo acknowledges support from the European
Union's Horizon 2020 research and innovation programme under Marie
Sk{\l}odowska-Curie grant agreement No 721463 to the SUNDIAL ITN
network.
This research has made use of the APASS database, located at the AAVSO web site. Funding for APASS has been provided by the Robert Martin Ayers Sciences Fund.\\
This work has made use of data from the European Space Agency (ESA) mission
{\it Gaia} (\url{https://www.cosmos.esa.int/gaia}), processed by the {\it Gaia}
Data Processing and Analysis Consortium (DPAC,
\url{https://www.cosmos.esa.int/web/gaia/dpac/consortium}). Funding for the DPAC
has been provided by national institutions, in particular the institutions
participating in the {\it Gaia} Multilateral Agreement.\\

\section*{Data availability statement}

The data underlying this article are available in the article and in its online supplementary material.



\bibliographystyle{mnras}
\bibliography{mybibliography_ymca} 



\appendix

\section{Definition of a threshold through Montecarlo simulations}
\label{appendix:appendix_a}

\begin{figure*}
    \centering
    \includegraphics[width=0.49\textwidth]{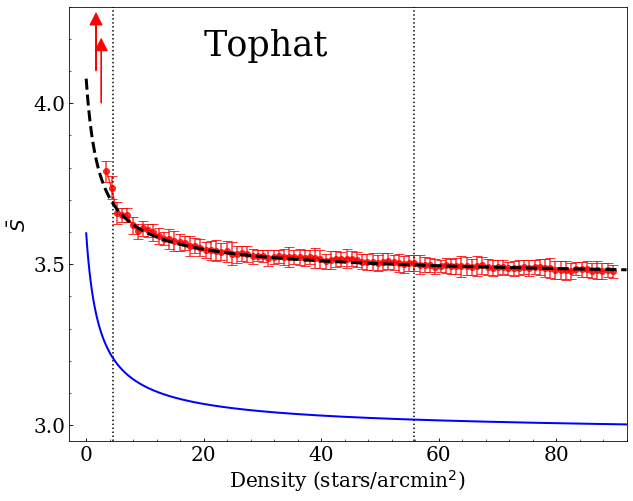}
    \includegraphics[width=0.49\textwidth]{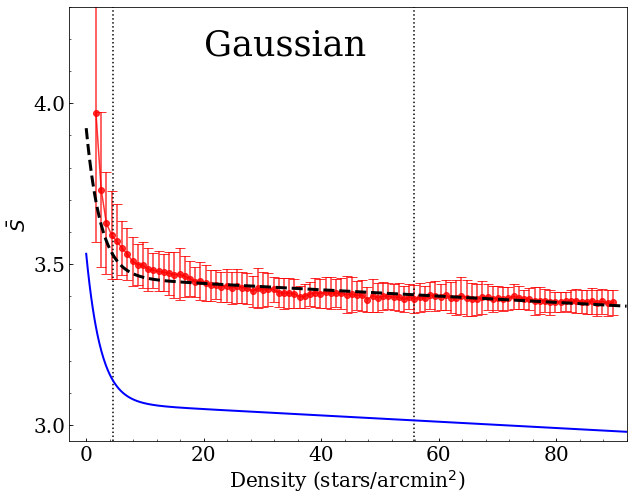}\\
    \caption{Mean of the significance (red points) with their standard deviation (red lines) of the false positives as function of the stellar field density. Dashed lines represent the fit of these curves for the tophat kernel (\emph{left figure}) and the gaussian one (\emph{right figure})). In both panels, the  blue lines represent a rescaling to a threshold=3 of the best-fitting lines. Dotted vertical lines define the lowest and highest density value of our observed fields.}
    \label{fig:significance}
\end{figure*}

\begin{figure*}
    \centering
    \includegraphics[width=0.49\textwidth]{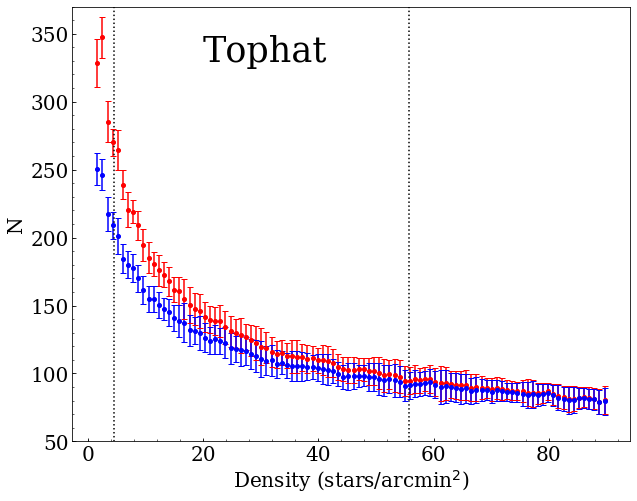}
    \includegraphics[width=0.49\textwidth]{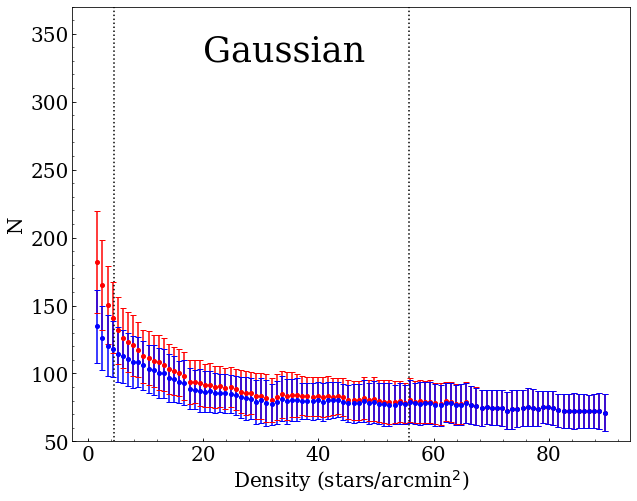}\\
    \caption{Total number of spurious over-densities, averaged in each density bin, as function of the stellar density field (see text).}
    \label{fig:count_struct}
\end{figure*}

The efficiency and reliability of the algorithm for the SC detection rely heavily on the threshold value of the significance ($S_{\rm th}$). It is therefore important to set properly this value. In fact, a low value of $S_{\rm th}$ is useful to reveal sparse and faint structures, but at the same time, it might increase the number of false detections, especially in low density fields. On the contrary, a high cut-off value would raise the purity, but at sample completeness' expenses.
Moreover, in several tiles, the density is not constant and increases towards the centre of the LMC. This means that we cannot use a fixed threshold along the whole tile, thus complicating the choice of the correct ($S_{\rm th}$). To face this problem, we decided to test the detection algorithm against a suite of Montecarlo simulations built with different stellar densities.
More in detail, we generated 2000 mock stellar fields with size 30\arcmin$\times$30\arcmin, distributing randomly the stars (uniform distribution). To simulate different field densities, we added 100 different amounts of stars, ranging from 1500 ($\sim$ 1.7 stars/arcmin$^2$) up to 80700 stars ($\sim 89.7$ stars/arcmin$^2$). These values encompass the lowest and highest values measured in the investigated tiles (vertical dotted lines in Fig.~\ref{fig:significance}). 
Subsequently, we carried out the over-density search method on these simulated fields exactly as it was done for the actual images (see Sect.\ref{ch:algorithm}), setting an initial threshold value of $S_{\rm th} = 3$. This value usually represents a reasonable compromise in maximizing the discovery of faint and sparse SCs while minimizing at the same time the spurious detection.
We recall that no artificial SC has been added in these simulated images, and therefore, every positive detection must be regarded as spurious. 
Figure~\ref{fig:significance} shows the results of these tests. Both panels display the significance (S/N) of the spurious over-densities (averaged over each density bin) as a function of the star field densities. The left and right panels show the results for the two KDE kernels used in this work, tophat and gaussian, respectively. 
In both panels, we can notice that the significance of the spurious over-densities is constant up to stellar field densities as low as 20 stars/arcmin$^2$. Below this value, this quantity starts to rise exponentially, especially for the tophat kernel, revealing that we cannot use the same $S_{\rm th}$ at low stellar density values. 
We can model this trend with a simple curve (black dashed line in Fig.~\ref{fig:significance}) to derive the correct $S_{\rm th}$ for each measured stellar field density. Note that the final $S_{\rm th}$ was obtained by re-scaling the black dashed curves in Fig.~\ref{fig:significance}) in order to have a threshold equal to three at high densities (solid blue line in the figure). In this way, we are confident that at low densities a higher $S_{\rm th}$ is able to remove most false detections. However, fixing $S_{\rm th} = 3$ at higher density values provides a great level of completeness, as demonstrated in  Sect. \ref{ch:efficiency_algorithm} by means of the tests on the recovery of artificial clusters.\\
To further support the usefulness of using a threshold variable with the stellar field density, we display in Fig.~\ref{fig:count_struct} the average number of spurious detections at varying densities for the two kernel functions (left and right panels, tophat and gaussian kernels, respectively) using fixed and variable thresholds (red and blue dot, respectively). 
An inspection of the figure reveals that the Gaussian kernel (\emph{right figure}) is more stable against density variations. Moreover, the variable threshold removes a maximum of $\sim$ 15-20\% of false over-densities.
On the contrary, when using the Tophat kernel (\emph{left figure}), there is an exponential increase of spurious detections found and low field densities. The use of a variable threshold, in this case, allows decreasing false positives by 30\%.
As a final consideration, the number of false positives expected in each tile goes from $\sim$ 300 up to $\sim$ 800. However, it is worth noticing that our definitive spurious removal relies on the efficient cleaning procedure described in Sect.~\ref{sec:cleaning procedure}, hence we are confident that even if the false over-densities will represent a significant amount of the total, our methodology allows us to remove the large majority of them.

\section{Candidate cluster properties}
\label{appendix2}

Table~\ref{tab:clusters} lists all 85 candidate SCs identified in this work with their estimated parameters: centre coordinates, radius, age, reddening, metallicity, apparent and absolute magnitudes.\\
Figure \ref{fig:all_clusters_cmd} displays the CMDs of the 85 candidate SCs along with three isochrones: the best fitting one (solid line) and those with log(t) = $\pm$ 0.1 (dashed and dotted lines, respectively). 

\begin{table*}
    \centering
    \caption{Estimated parameters of all the 85 SCs identified in this work. The columns in the table indicate:
    1) name of the cluster; 2-3) R.A. and Dec; 4) estimated cluster radius; 5) age; 6) reddening; 7) metallicity; 8) apparent magnitude in $g$-band 9) absolute magnitude in $g$-band; 10) G as defined in \S\ref{sec:center and radius estimation}; 11) number of stars within the SC radius; 12) the tile name where the cluster lie; 13) flag to assess the statistical reliability of a SC: 0 is minimum, 3 is maximum (see \S\ref{app:density_profile} for details)}
    \label{tab:clusters}
    \begin{tabular}{l|c|c|c|c|c|c|c|c|c|c|c|c}
    \hline
    ID & R.A. & Dec & R & log(t) & E(B-V) & Z & $m_g$ & $M_g$ & G & N. stars & Tile & Flag\\
     & (J2000) & (J2000) & (\arcmin) & & (mag) & & (mag) & (mag) & & & & \\
    \hline

ymca-0001 & 63.8302 & -70.7697 & 0.35 & 9.75 & 0.04 & 0.006 & 17.79 & -0.83 & 2.72 & 14 & 4\_22 & 2\\
ymca-0002 & 63.8712 & -71.1712 & 0.25 & 9.65 & 0.18 & 0.004 & 18.78 & -0.31 & 5.77 & 14 & 4\_22 & 3\\
ymca-0003 & 64.3757 & -69.3018 & 0.25 & 9.5 & 0.03 & 0.006 & 18.27 & -0.32 & 4.11 & 10 & 5\_23 & 3\\
ymca-0004 & 64.4336 & -70.2989 & 0.25 & 9.6 & 0.03 & 0.006 & 18.81 & 0.22 & 4.36 & 13 & 4\_22 & 3\\
ymca-0005 & 64.991 & -71.9376 & 0.4 & 9.25 & 0.06 & 0.008 & 17.69 & -1.00 & 3.84 & 30 & 3\_21 & 3\\
ymca-0006 & 65.1621 & -71.5519 & 0.55 & 9.75 & 0.03 & 0.006 & 16.24 & -2.35 & 2.84 & 42 & 3\_21 & 0\\
step-0001 & 65.2362 & -73.4328 & 0.3 & 9.75 & 0.04 & 0.006 & 18.21 & -0.41 & 4.82 & 22 & 3\_20 & 3\\
ymca-0007 & 65.3561 & -70.5944 & 0.5 & 9.7 & 0.02 & 0.006 & 17.00 & -1.56 & 3.46 & 42 & 4\_22 & 3\\
ymca-0008 & 65.4038 & -70.5862 & 0.25 & 9.6 & 0.03 & 0.008 & 18.10 & -0.49 & 4.07 & 16 & 4\_22 & 3\\
ymca-0009 & 65.4748 & -71.1374 & 0.35 & 9.05 & 0.15 & 0.008 & 17.84 & -1.15 & 5.12 & 29 & 4\_22 & 3\\
step-0002 & 65.4784 & -73.5847 & 0.3 & 9.55 & 0.03 & 0.006 & 18.29 & -0.30 & 3.56 & 18 & 3\_20 & 3\\
ymca-0010 & 65.5203 & -70.232 & 0.5 & 9.4 & 0.03 & 0.006 & 17.38 & -1.21 & 2.94 & 37 & 4\_22 & 2\\
step-0003 & 65.6166 & -72.9424 & 0.55 & 9.5 & 0.05 & 0.008 & 17.33 & -1.33 & 6.36 & 76 & 3\_21 & 1\\
ymca-0011 & 65.6829 & -71.4134 & 0.3 & 9.5 & 0.03 & 0.006 & 18.05 & -0.54 & 4.44 & 26 & 3\_21 & 1\\
step-0004 & 65.885 & -73.798 & 0.35 & 9.95 & 0.06 & 0.006 & 18.14 & -0.54 & 3.37 & 23 & 3\_21 & 3\\
ymca-0012 & 66.0329 & -71.2766 & 0.2 & 9.95 & 0.03 & 0.006 & 18.58 & -0.01 & 4.58 & 16 & 3\_21 & 3\\
ymca-0013 & 66.1188 & -70.7725 & 0.25 & 9.8 & 0.07 & 0.006 & 18.15 & -0.57 & 3.03 & 16 & 4\_22 & 3\\
ymca-0014 & 66.1419 & -70.4832 & 0.45 & 9.35 & 0.06 & 0.008 & 15.70 & -2.99 & 3.13 & 38 & 4\_22 & 3\\
step-0005 & 66.1542 & -73.4368 & 0.4 & 9.8 & 0.03 & 0.006 & 17.93 & -0.65 & 2.90 & 31 & 3\_21 & 0\\
ymca-0015 & 66.1664 & -70.3033 & 0.5 & 9.25 & 0.06 & 0.006 & 16.73 & -1.96 & 4.21 & 52 & 4\_22 & 3\\
step-0006 & 66.2021 & -73.5264 & 0.3 & 9.55 & 0.05 & 0.004 & 18.17 & -0.48 & 4.09 & 23 & 3\_21 & 3\\
ymca-0016 & 66.2049 & -69.1095 & 0.4 & 9.55 & 0.02 & 0.006 & 17.90 & -0.66 & 3.39 & 23 & 5\_23 & 1\\
ymca-0017 & 66.4461 & -71.1112 & 0.4 & 9.7 & 0.02 & 0.006 & 16.18 & -2.38 & 3.37 & 34 & 4\_22 & 1\\
ymca-0018 & 66.6028 & -70.218 & 0.4 & 9.55 & 0.01 & 0.006 & 16.96 & -1.56 & 4.13 & 38 & 4\_22 & 3\\
ymca-0019 & 66.6152 & -72.0306 & 0.5 & 9.5 & 0.03 & 0.006 & 17.33 & -1.25 & 3.18 & 55 & 3\_21 & 3\\
ymca-0020 & 66.6736 & -70.6311 & 0.45 & 9.45 & 0.06 & 0.006 & 17.07 & -1.62 & 3.14 & 42 & 4\_22 & 3\\
ymca-0021 & 66.7019 & -70.3302 & 0.45 & 9.65 & 0.03 & 0.006 & 17.00 & -1.58 & 2.85 & 39 & 4\_22 & 2\\
ymca-0022 & 66.8356 & -71.9218 & 0.45 & 9.45 & 0.1 & 0.006 & 17.11 & -1.71 & 2.54 & 47 & 3\_21 & 0\\
step-0007 & 66.8558 & -73.8335 & 0.5 & 9.5 & 0.08 & 0.006 & 17.50 & -1.25 & 2.50 & 45 & 3\_21 & 0\\
ymca-0023 & 67.0673 & -71.62 & 0.45 & 9.7 & 0.05 & 0.006 & 16.39 & -2.26 & 2.21 & 46 & 3\_21 & 0\\
ymca-0024 & 67.1417 & -71.7341 & 0.35 & 9.0 & 0.1 & 0.02 & 17.76 & -1.06 & 1.32 & 27 & 3\_21 & 2\\
step-0008 & 67.1827 & -73.3176 & 0.25 & 9.25 & 0.06 & 0.006 & 17.90 & -0.79 & 1.94 & 16 & 3\_21 & 0\\
step-0009 & 67.2051 & -73.3117 & 0.4 & 9.2 & 0.06 & 0.006 & 16.95 & -1.73 & 1.87 & 35 & 3\_21 & 0\\
ymca-0025$^a$ & 67.395 & -71.8408 & 0.8 & 9.35 & 0.05 & 0.006 & 14.11 & -4.55 & 18.98 & 302 & 3\_21 & 3\\
step-0010 & 67.6849 & -73.4533 & 0.45 & 9.45 & 0.08 & 0.006 & 16.93 & -1.82 & 3.98 & 56 & 3\_21 & 1\\
step-0011 & 67.8165 & -73.8014 & 0.3 & 9.5 & 0.06 & 0.006 & 18.11 & -0.58 & 4.29 & 29 & 3\_21 & 1\\
step-0012 & 67.9018 & -73.6612 & 0.4 & 9.85 & 0.05 & 0.006 & 17.21 & -1.45 & 3.50 & 45 & 3\_21 & 3\\
step-0013 & 67.9237 & -72.9487 & 0.3 & 9.5 & 0.03 & 0.006 & 18.06 & -0.53 & 3.59 & 36 & 3\_21 & 3\\
step-0014 & 67.9965 & -73.5009 & 0.4 & 9.4 & 0.05 & 0.006 & 17.87 & -0.79 & 2.94 & 45 & 3\_21 & 0\\
step-0015$^b$ & 68.0319 & -73.6709 & 0.4 & 9.1 & 0.02 & 0.006 & 17.41 & -1.15 & 2.98 & 43 & 3\_21 & 2\\
step-0016 & 68.0349 & -73.345 & 0.25 & 9.4 & 0.05 & 0.006 & 18.24 & -0.41 & 3.19 & 23 & 3\_21 & 1\\
step-0017 & 68.153 & -73.43 & 0.2 & 9.5 & 0.07 & 0.006 & 18.12 & -0.60 & 4.34 & 20 & 3\_21 & 3\\
step-0018$^c$ & 68.2506 & -73.2275 & 0.3 & 9.2 & 0.04 & 0.006 & 16.67 & -1.95 & 7.56 & 54 & 3\_21 & 3\\
step-0019 & 68.2628 & -73.7053 & 0.35 & 9.4 & 0.02 & 0.006 & 17.58 & -0.97 & 4.12 & 39 & 3\_21 & 1\\
step-0020 & 68.2673 & -73.2126 & 0.55 & 9.3 & 0.08 & 0.008 & 16.81 & -1.94 & 3.02 & 88 & 3\_21 & 1\\
step-0021 & 68.2948 & -72.9248 & 0.45 & 9.45 & 0.02 & 0.006 & 16.16 & -2.40 & 3.69 & 74 & 3\_21 & 1\\
step-0022 & 68.3022 & -73.1334 & 0.25 & 9.45 & 0.05 & 0.006 & 17.71 & -0.94 & 4.02 & 29 & 3\_21 & 1\\
step-0023 & 68.3077 & -73.4732 & 0.45 & 9.45 & 0.08 & 0.006 & 17.14& -1.61 & 4.65 & 74 & 3\_21 & 3\\
step-0024 & 68.319 & -73.1468 & 0.4 & 9.6 & 0.08 & 0.004 & 17.34 & -1.42 & 2.27 & 49 & 3\_21 & 2\\
step-0025 & 68.386 & -73.5211 & 0.3 & 9.35 & 0.07 & 0.006 & 17.55 & -1.17 & 4.13 & 36 & 3\_21 & 3\\
step-0026 & 68.4947 & -73.3416 & 0.4 & 9.35 & 0.02 & 0.006 & 17.40 & -1.15 & 3.13 & 55 & 3\_21 & 1\\
step-0027 & 68.5198 & -73.8413 & 0.35 & 9.5 & 0.06 & 0.006 & 18.35 & -0.34 & 3.37 & 36 & 3\_21 & 1\\
step-0028 & 68.6028 & -72.98 & 0.25 & 9.5 & 0.03 & 0.006 & 17.22 & -1.37 & 2.85 & 27 & 3\_21 & 0\\
step-0029 & 68.7477 & -73.3113 & 0.25 & 9.85 & 0.02 & 0.006 & 18.23 & -0.33 & 3.20 & 28 & 3\_21 & 1\\
step-0030 & 68.845 & -73.1543 & 0.35 & 9.55 & 0.05 & 0.008 & 16.77 & -1.88 & 4.17 & 49 & 3\_21 & 3\\
step-0031 & 68.855 & -73.9019 & 0.3 & 9.45 & 0.03 & 0.006 & 18.06 & -0.52 & 3.14 & 26 & 3\_21 & 3\\
step-0032 & 68.8971 & -73.4728 & 0.45 & 9.5 & 0.03 & 0.006 & 16.75 & -1.84 & 2.64 & 64 & 3\_21 & 2\\
step-0033$^d$ & 68.8974 & -73.4169 & 0.35 & 9.45 & 0.02 & 0.006 & 17.32 & -1.24 & 4.80 & 54 & 3\_21 & 3\\
step-0034$^e$ & 68.9113 & -73.7331 & 1.05 & 9.35 & 0.06 & 0.006 & 14.43 & -4.26 & 18.02 & 450 & 3\_21 & 3\\

\hline
\end{tabular}
\end{table*}
\begin{table*}{}
\centering
    \contcaption{}
    
    \begin{tabular}{l|c|c|c|c|c|c|c|c|c|c|c|c}
    \hline
    ID & R.A. & Dec & R & log(t) & E(B-V) & Z & $m_g$ & $M_g$ & G & N. stars & Tile & Flag\\
     & (J2000) & (J2000) & (\arcmin) & & (mag) & & (mag) & (mag) & & & & \\
    \hline
    
step-0035 & 68.9257 & -73.6976 & 0.35 & 9.6 & 0.06 & 0.004 & 17.11 & -1.58 & 2.04 & 32 & 3\_21 & 0\\
step-0036 & 68.9323 & -73.1327 & 0.35 & 9.0 & 0.12 & 0.008 & 18.10 & -0.78 & 3.78 & 47 & 3\_21 & 3\\
step-0037 & 68.9358 & -73.2805 & 0.5 & 9.4 & 0.04 & 0.006 & 16.18 & -2.44 & 6.70 & 104 & 3\_21 & 1\\
ymca-0026 & 89.2224 & -63.3843 & 0.2 & 9.4 & 0.02 & 0.004 & 18.48 & -0.07 & 1.89 & 15 & 11\_41 & 2\\
ymca-0027 & 89.3118 & -62.7561 & 0.25 & 9.35 & 0.02 & 0.004 & 17.85 & -0.70 & 2.74 & 19 & 11\_41 & 0\\
ymca-0028 & 89.3897 & -63.3484 & 0.3 & 9.1 & 0.08 & 0.008 & 17.58 & -1.17 & 3.19 & 37 & 11\_41 & 3\\
ymca-0029 & 89.5528 & -62.4595 & 0.4 & 9.45 & 0.02 & 0.004 & 17.17 & -1.38 & 3.97 & 42 & 11\_41 & 3\\
ymca-0030 & 89.5887 & -62.7257 & 0.35 & 9.55 & 0.04 & 0.004 & 16.42 & -2.20 & 5.31 & 41 & 11\_41 & 3\\
ymca-0031 & 90.4314 & -62.957 & 0.35 & 9.65 & 0.03 & 0.006 & 17.27 & -1.32 & 4.07 & 35 & 11\_41 & 3\\
ymca-0032 & 90.6697 & -62.9244 & 0.3 & 9.2 & 0.03 & 0.006 & 17.31 & -1.28 & 4.07 & 27 & 11\_41 & 1\\
ymca-0033 & 90.9858 & -63.3337 & 0.5 & 9.7 & 0.03 & 0.004 & 15.18 & -3.41 & 3.68 & 60 & 11\_41 & 1\\
ymca-0034 & 91.1703 & -63.1483 & 0.4 & 9.35 & 0.02 & 0.006 & 16.75 & -1.80 & 3.09 & 40 & 11\_41 & 3\\
ymca-0035 & 91.4224 & -62.7954 & 0.2 & 9.55 & 0.02 & 0.006 & 18.10 & -0.46 & 4.55 & 13 & 11\_42 & 3\\
ymca-0036 & 91.5201 & -62.569 & 0.5 & 9.5 & 0.02 & 0.006 & 16.76 & -1.79 & 3.36 & 35 & 11\_42 & 3\\
ymca-0037$^f$ & 92.0658 & -62.9875 & 1.4 & 9.3 & 0.03 & 0.006 & 12.70 & -5.89 & 48.75 & 847 & 11\_42 & 3\\
ymca-0038 & 92.8553 & -63.2278 & 0.45 & 9.25 & 0.03 & 0.006 & 17.46 & -1.13 & 3.42 & 34 & 11\_42 & 3\\
ymca-0039 & 93.0118 & -63.2107 & 0.3 & 9.45 & 0.06 & 0.006 & 16.14 & -2.55 & 4.04 & 20 & 11\_42 & 3\\
ymca-0040 & 94.0711 & -63.3558 & 0.45 & 9.5 & 0.02 & 0.004 & 17.19 & -1.36 & 3.06 & 32 & 11\_43 & 3\\
ymca-0041 & 94.2216 & -73.6311 & 0.5 & 9.55 & 0.02 & 0.006 & 16.70 & -1.86 & 2.93 & 75 & 1\_27 & 2\\
ymca-0042 & 94.297 & -73.5258 & 0.4 & 8.95 & 0.1 & 0.01 & 17.09 & -1.73 & 2.13 & 46 & 1\_27 & 0\\
ymca-0043 & 94.3731 & -74.0865 & 0.35 & 9.15 & 0.07 & 0.01 & 17.69 & -1.03 & 2.23 & 31 & 1\_27 & 2\\
ymca-0044 & 94.4404 & -73.773 & 0.2 & 8.8 & 0.2 & 0.02 & 16.81 & -2.34 & 3.48 & 18 & 1\_27 & 3\\
ymca-0045 & 94.8468 & -74.214 & 0.45 & 9.45 & 0.08 & 0.008 & 17.77 & -0.98 & 4.02 & 44 & 1\_27 & 1\\
ymca-0046$^g$ & 95.8541 & -73.8284 & 0.5 & 9.4 & 0.05 & 0.006 & 16.02 & -2.63 & 12.00 & 106 & 1\_27 & 3\\
ymca-0047 & 96.1662 & -73.8367 & 0.3 & 8.8 & 0.2 & 0.02 & 18.23 & -0.92 & 4.01 & 24 & 1\_27 & 3\\
ymca-0048 & 97.3995 & -73.5321 & 0.35 & 9.45 & 0.02 & 0.006 & 17.67 & -0.89 & 2.93 & 22 & 1\_27 & 2\\       
\hline

\multicolumn{12}{l}{
  \begin{minipage}{15cm}
    \tiny \textbf{Reference names.\\}a: NGC1629,SL3,LW3,KMHK4\\
b: OGLE-LMC-CL-0824\\
c: OGLE-LMC-CL-0827\\
d: OGLE-LMC-CL-0826\\
e: OGLE-LMC-CL-1133,SL5,LW8,KMHK14\\
f: SL842,LW399,KMHK1652,ESO86SC61\\
g: OGLE-LMC-CL-0849
  \end{minipage}}

\end{tabular}  
\end{table*}

\begin{figure*}
\hspace{-1.2cm}
    \includegraphics[width=0.33\textwidth]{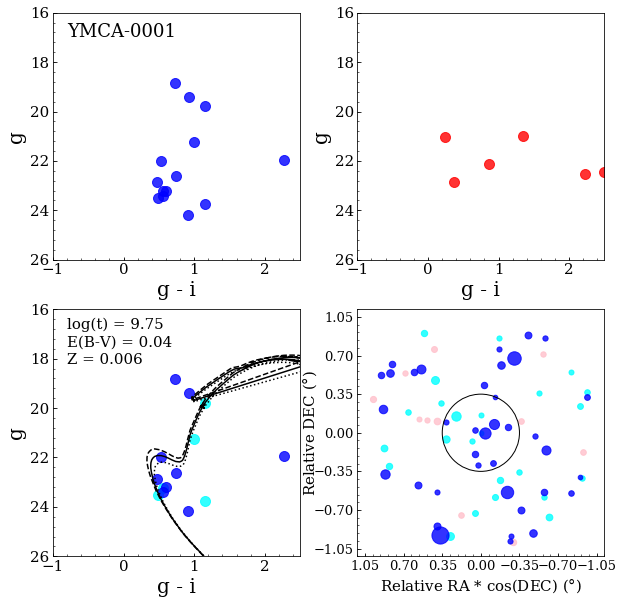}
    \includegraphics[width=0.33\textwidth]{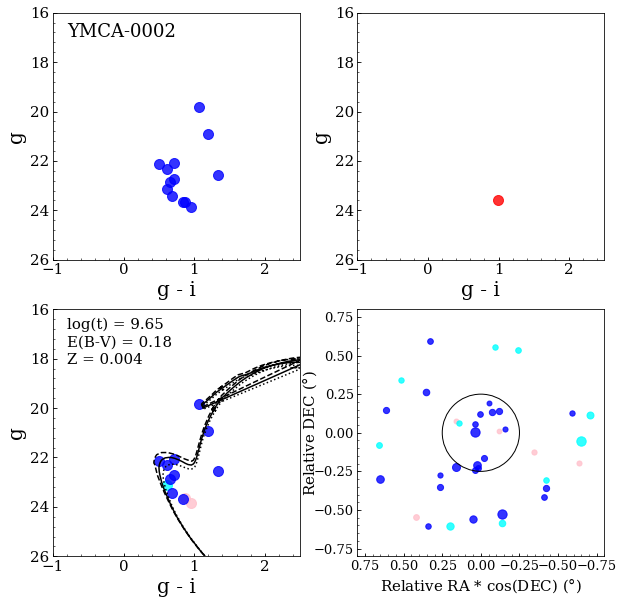}
    \includegraphics[width=0.33\textwidth]{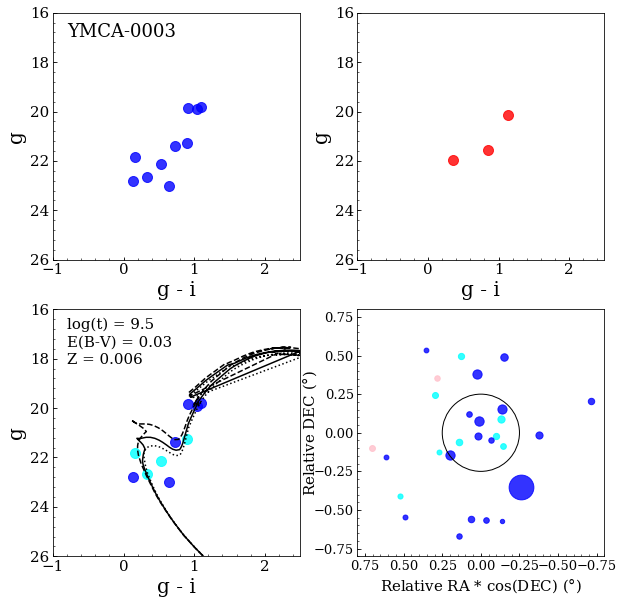}\\
\hspace{-1.2cm}
    \includegraphics[width=0.33\textwidth]{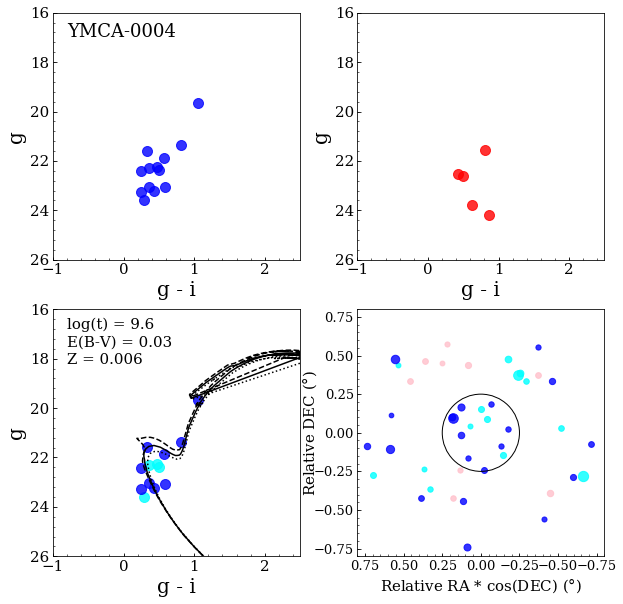}
    \includegraphics[width=0.33\textwidth]{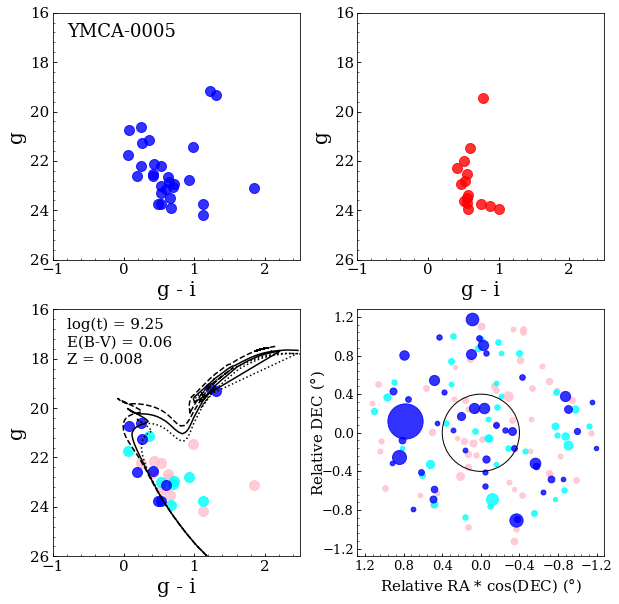}
    \includegraphics[width=0.33\textwidth]{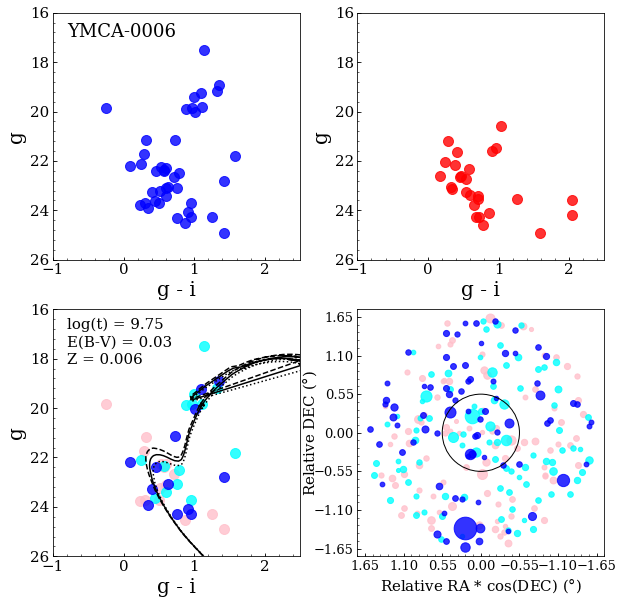}\\
\hspace{-1.2cm}
    \includegraphics[width=0.33\textwidth]{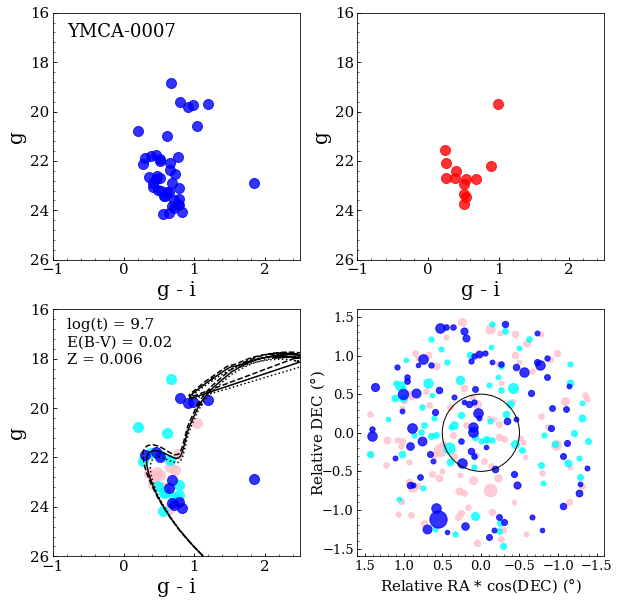}
    \includegraphics[width=0.33\textwidth]{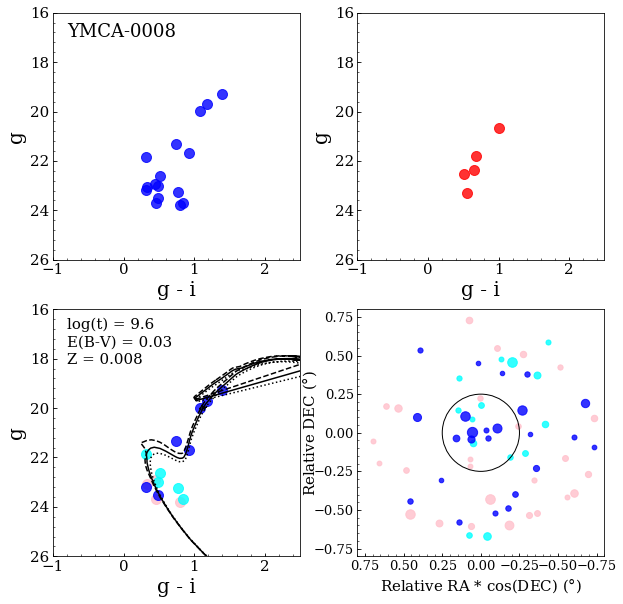}
    \includegraphics[width=0.33\textwidth]{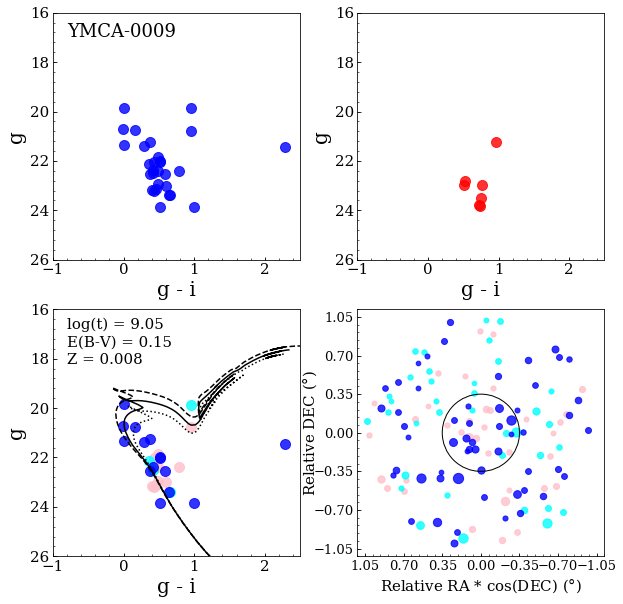}\\
\hspace{-1.2cm}
    \includegraphics[width=0.33\textwidth]{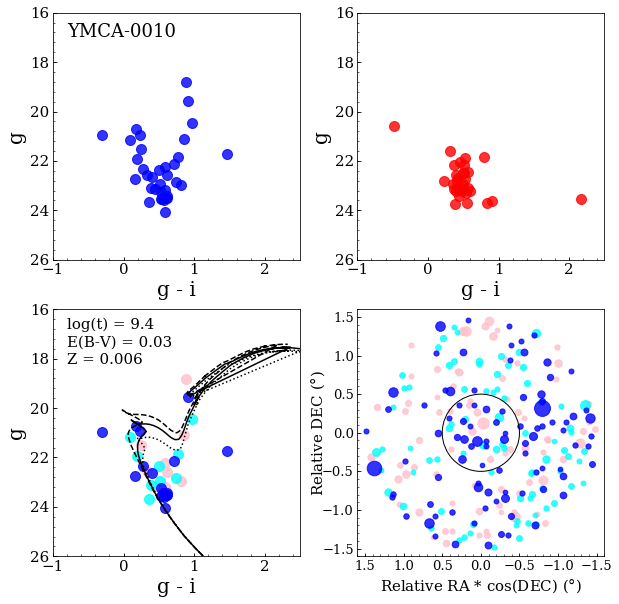}
    \includegraphics[width=0.33\textwidth]{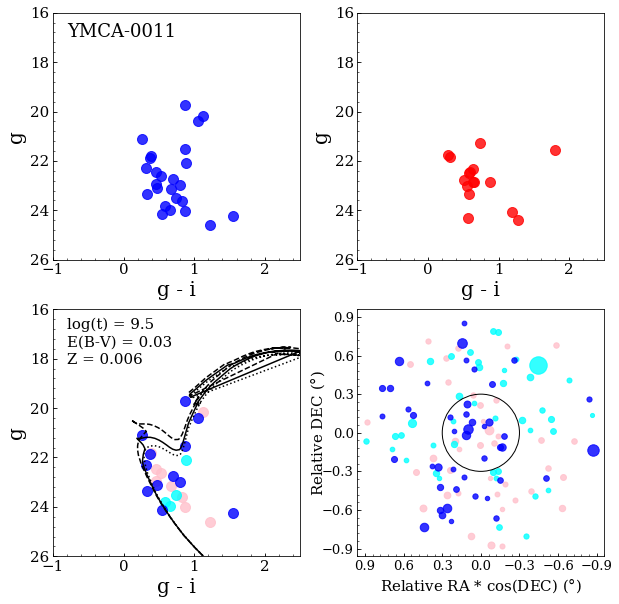}
    \includegraphics[width=0.33\textwidth]{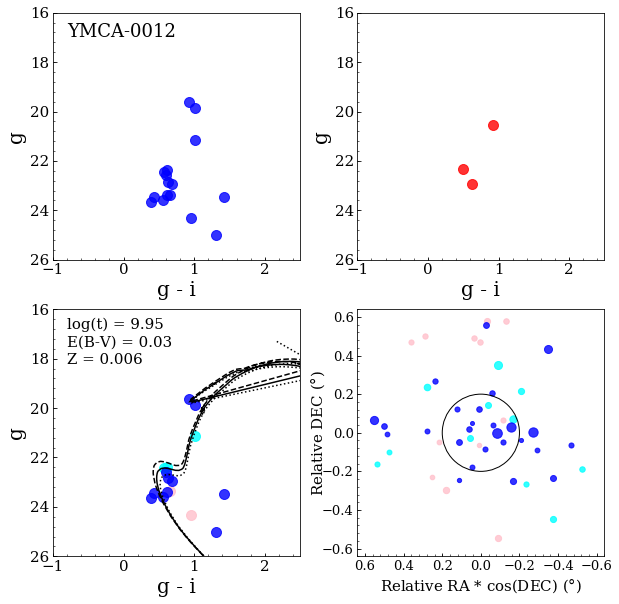}\\
\vspace{-0.35cm}
    \caption{Cleaning procedure for the 85 candidate SCs detected in this work (see Fig. \ref{fig:cleaned_cmd} to the explanation of each figure). The stars are colored on the basis of their probability to belong to the cluster. Blue points are stars with $P \geq$75\%, cyan points have P = 50\% while pink points indicate stars with $P \leq$ 25\%. The solid line shows the best fitting isochrone, whereas dashed and dotted lines are the isochrones shifted by log(t) = $\pm$0.1, respectively.} 
    \label{fig:all_clusters_cmd}
\end{figure*}

\begin{figure*}
\hspace{-1.2cm}
    \includegraphics[width=0.35\textwidth]{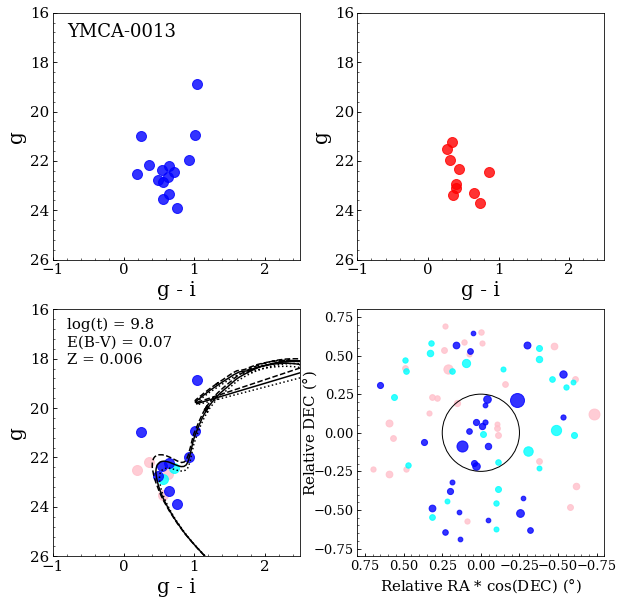}
    \includegraphics[width=0.35\textwidth]{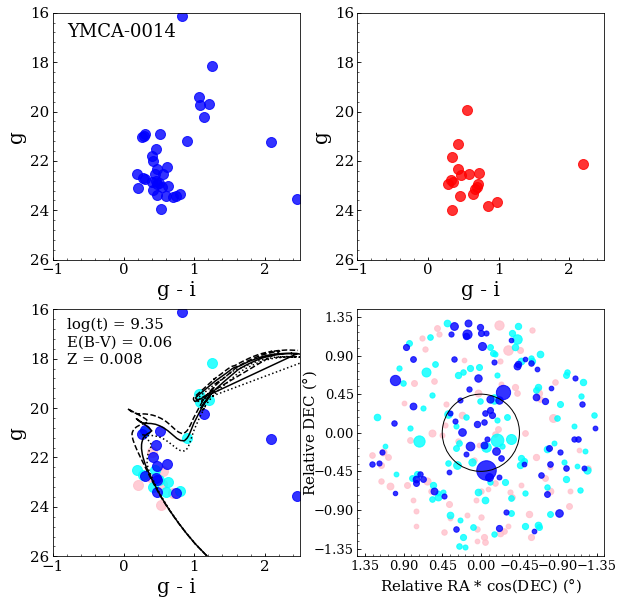}
    \includegraphics[width=0.35\textwidth]{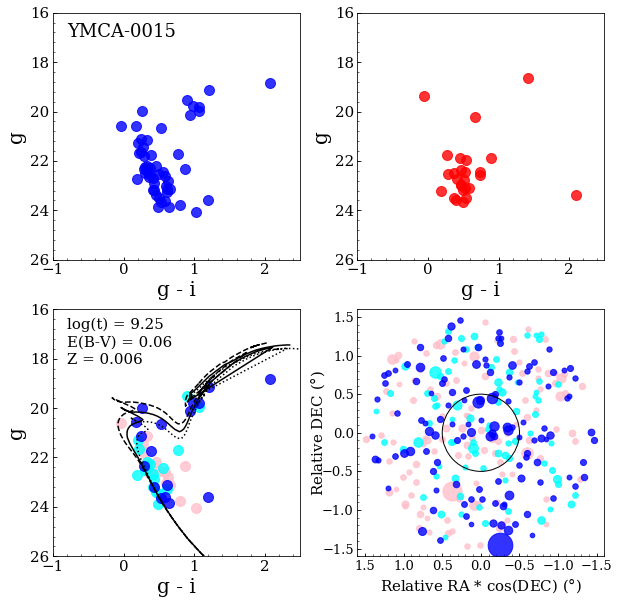}\\
\hspace{-1.2cm}
    \includegraphics[width=0.35\textwidth]{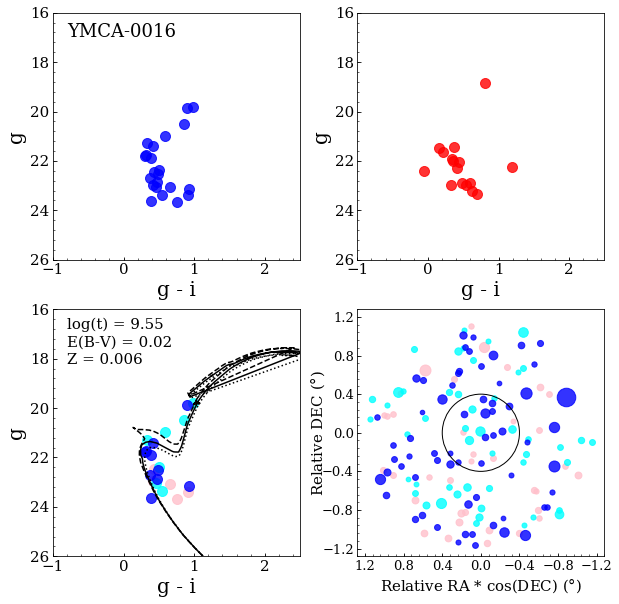}
    \includegraphics[width=0.35\textwidth]{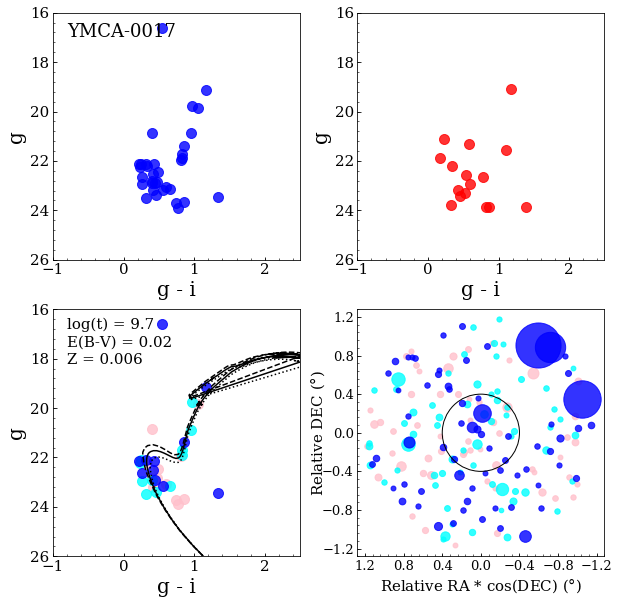}
    \includegraphics[width=0.35\textwidth]{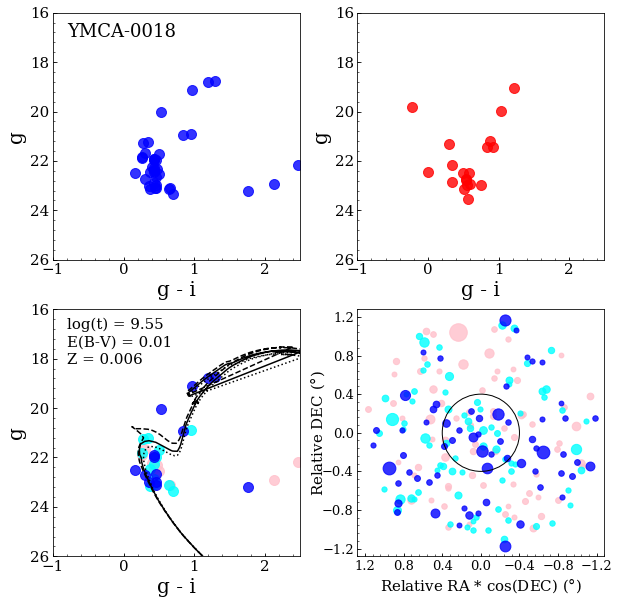}\\
\hspace{-1.2cm}
    \includegraphics[width=0.35\textwidth]{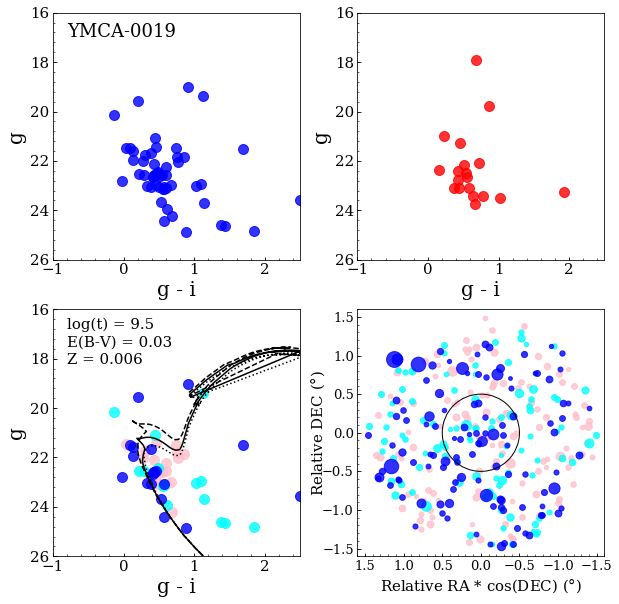}
    \includegraphics[width=0.35\textwidth]{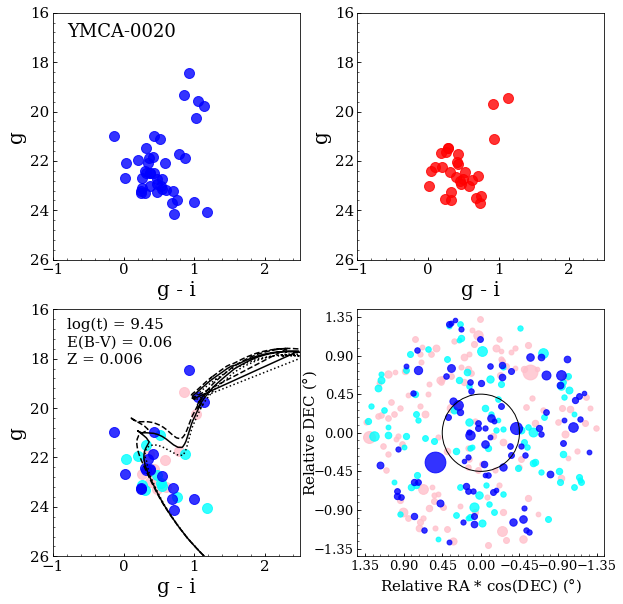}
    \includegraphics[width=0.35\textwidth]{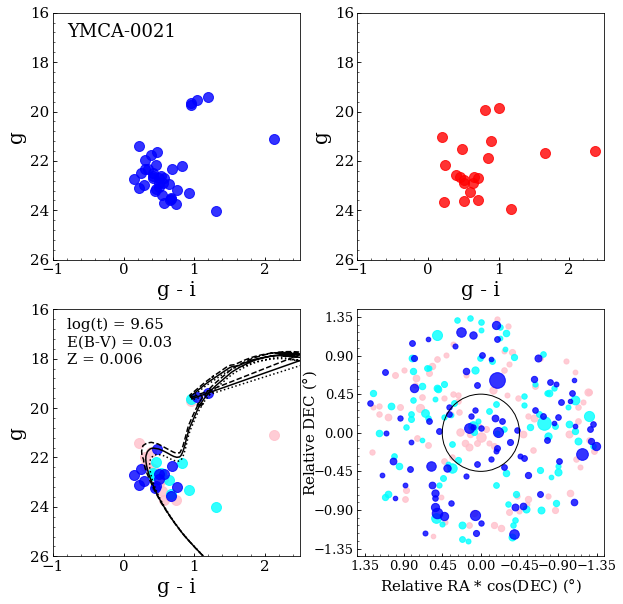}\\
\hspace{-1.2cm}
    \includegraphics[width=0.35\textwidth]{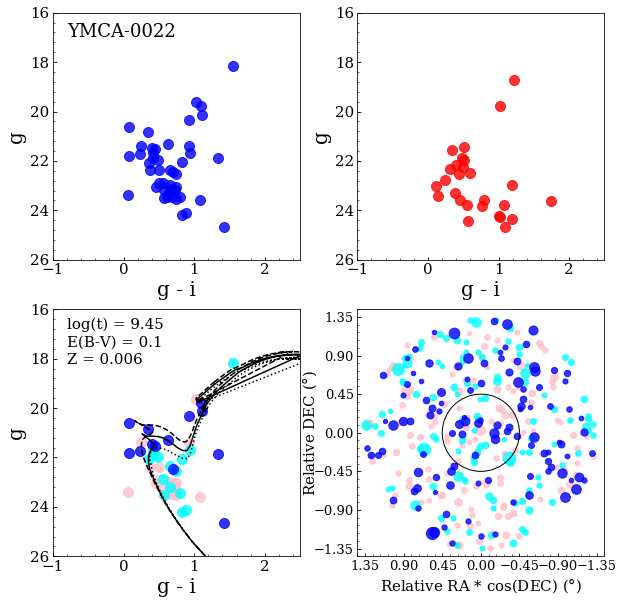}
    \includegraphics[width=0.35\textwidth]{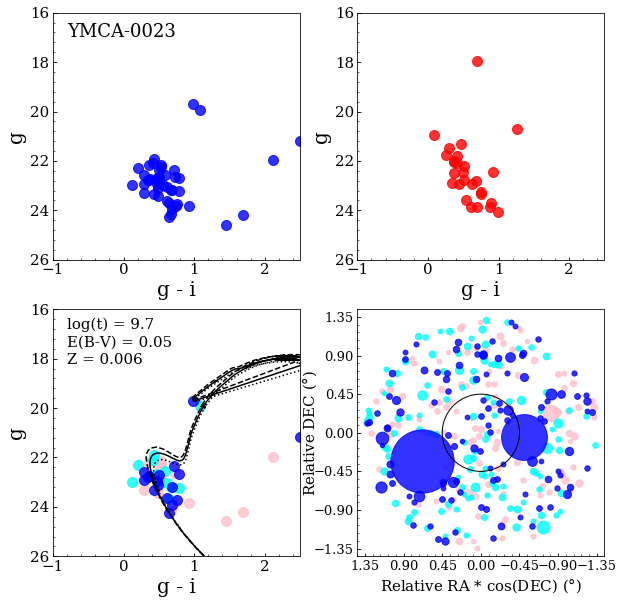}
    \includegraphics[width=0.35\textwidth]{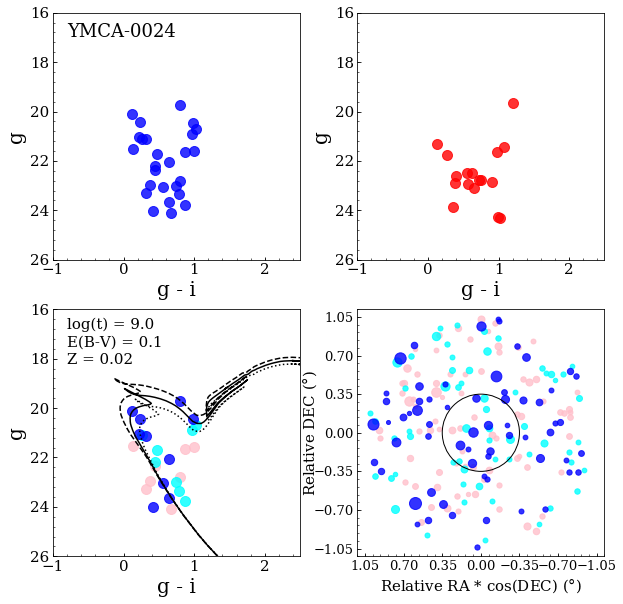}\\
    \contcaption{}
\end{figure*}

\begin{figure*}
\hspace{-1.2cm}
    \includegraphics[width=0.35\textwidth]{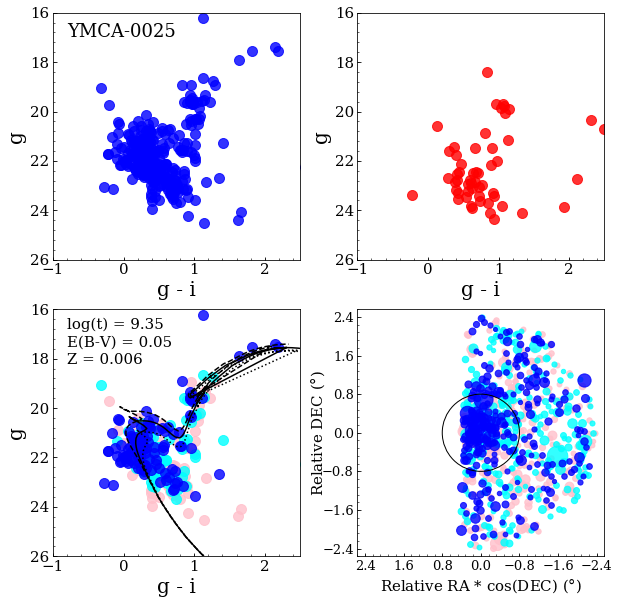}
    \includegraphics[width=0.35\textwidth]{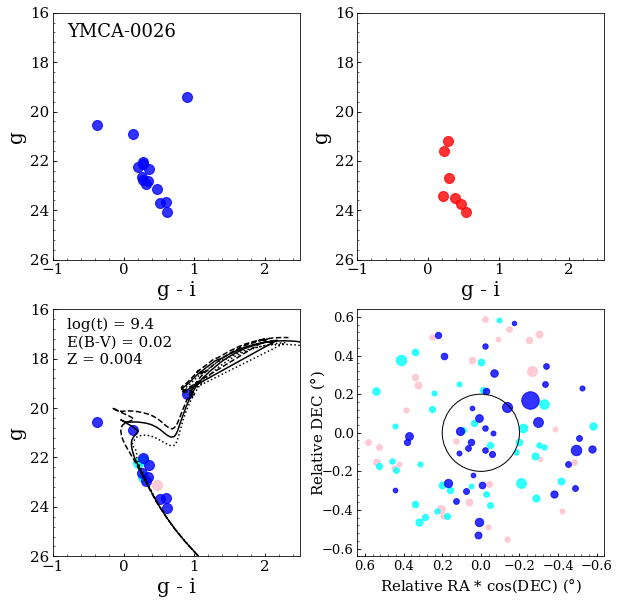}
    \includegraphics[width=0.35\textwidth]{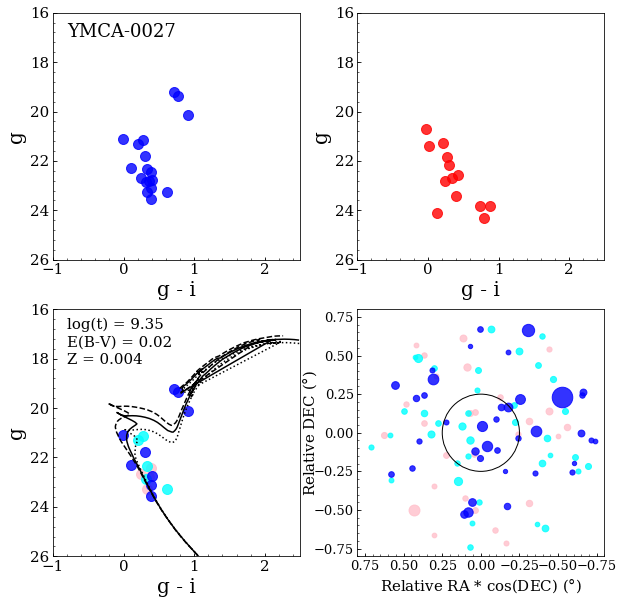}\\
\hspace{-1.2cm}
    \includegraphics[width=0.35\textwidth]{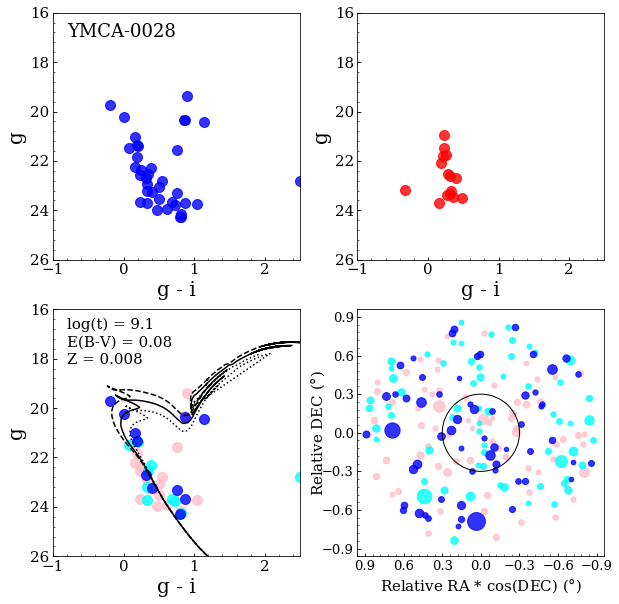}
    \includegraphics[width=0.35\textwidth]{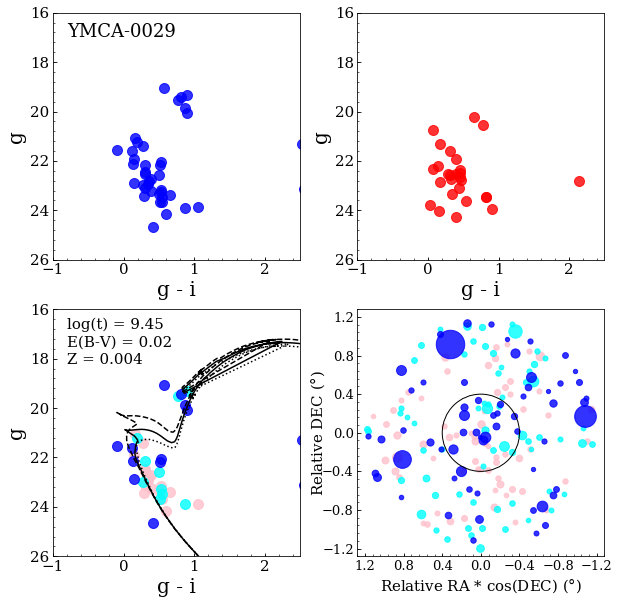}
    \includegraphics[width=0.35\textwidth]{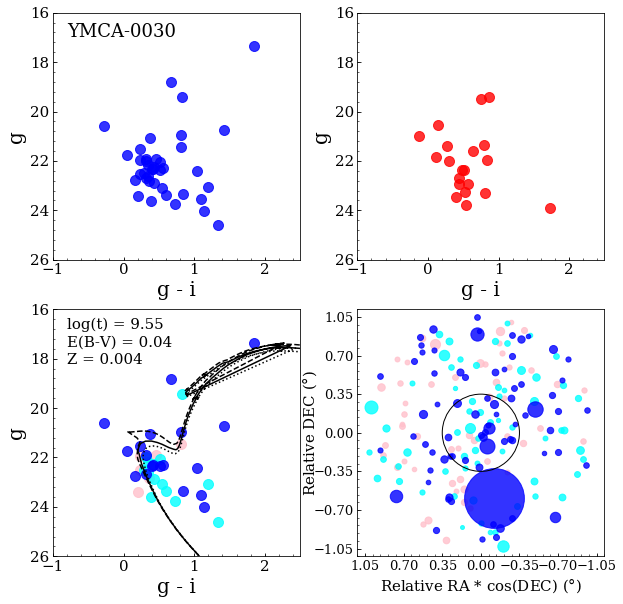}\\
\hspace{-1.2cm}
    \includegraphics[width=0.35\textwidth]{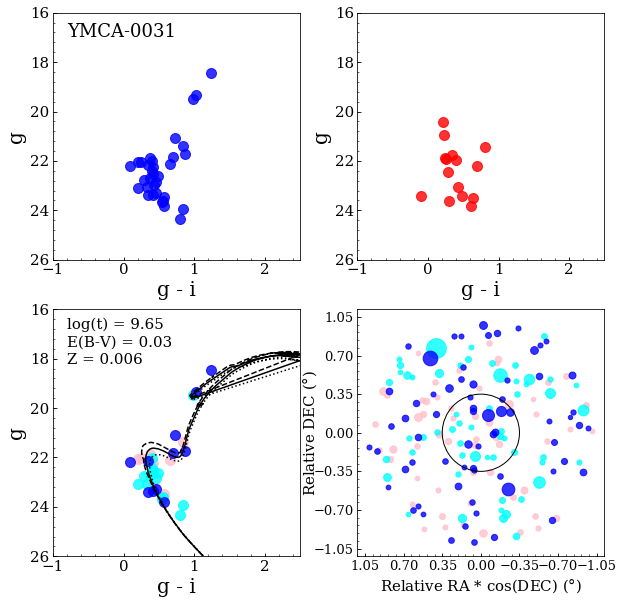}
    \includegraphics[width=0.35\textwidth]{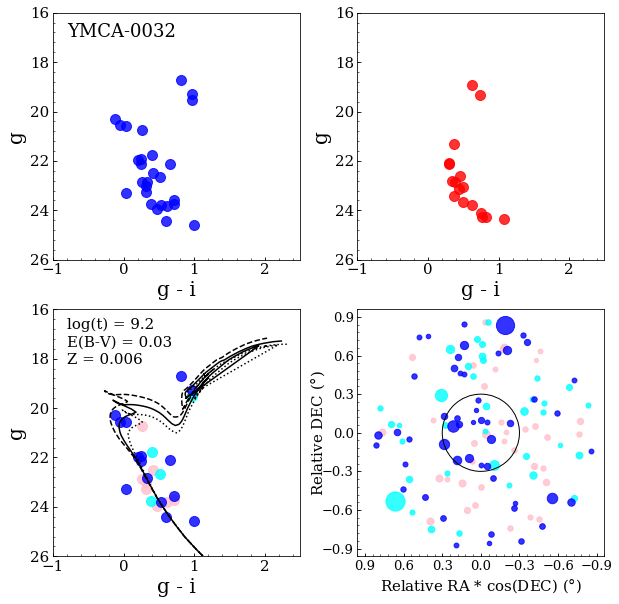}
    \includegraphics[width=0.35\textwidth]{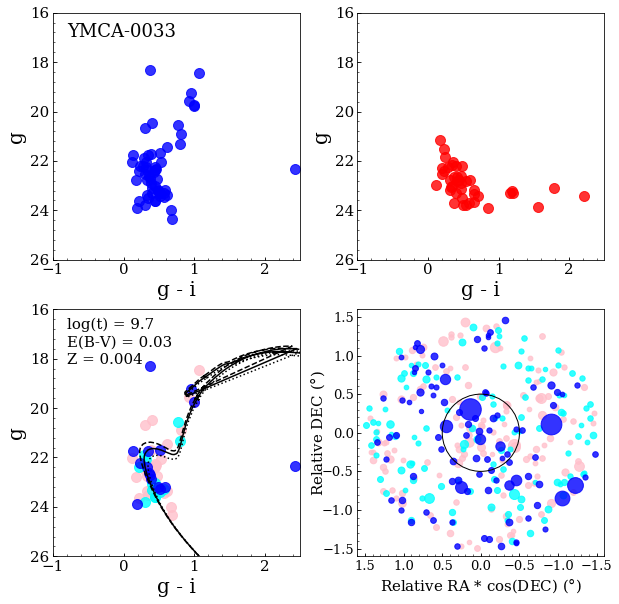}\\
\hspace{-1.2cm}
    \includegraphics[width=0.35\textwidth]{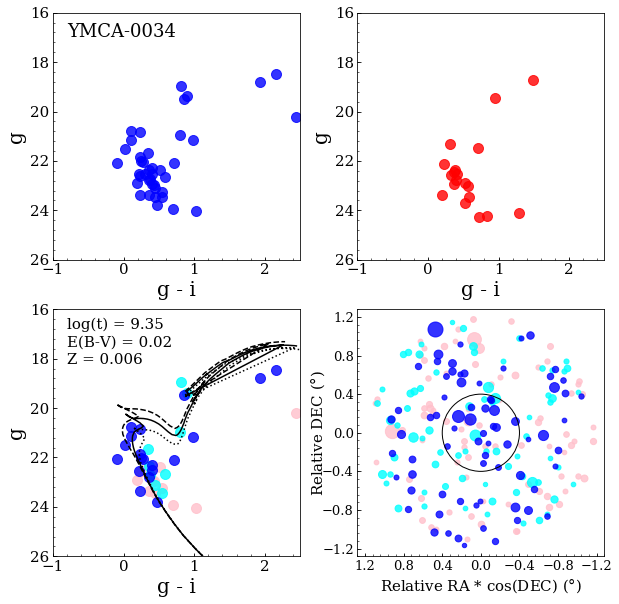}
    \includegraphics[width=0.35\textwidth]{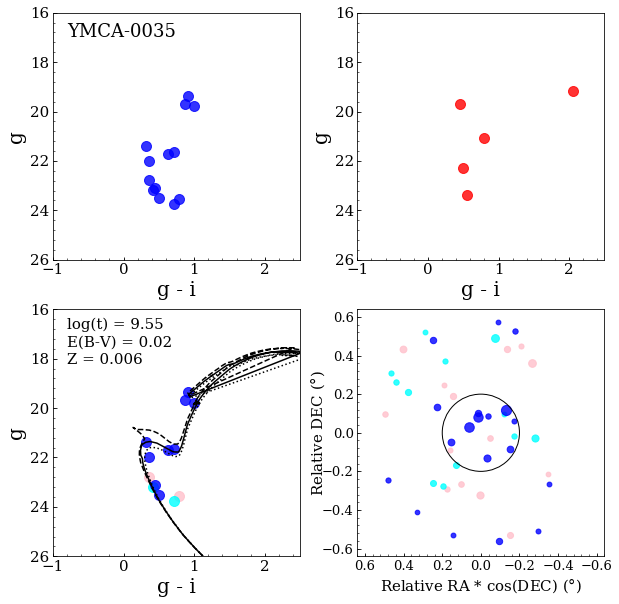}
    \includegraphics[width=0.35\textwidth]{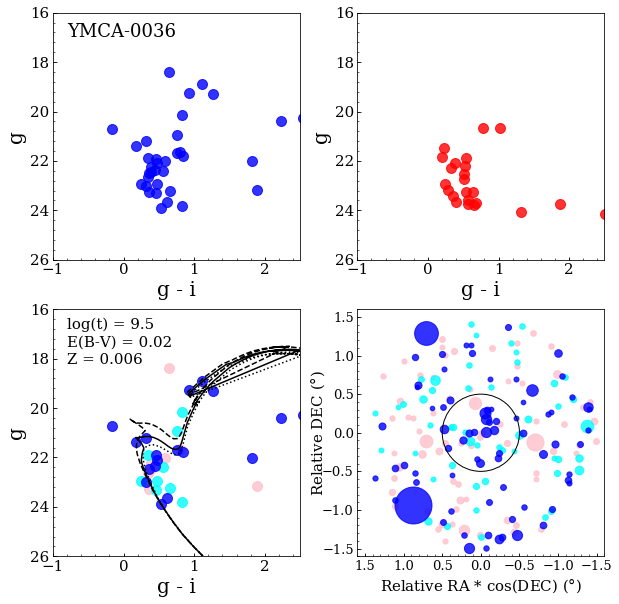}\\
    \contcaption{}
\end{figure*}

\begin{figure*}
\hspace{-1.2cm}
    \includegraphics[width=0.35\textwidth]{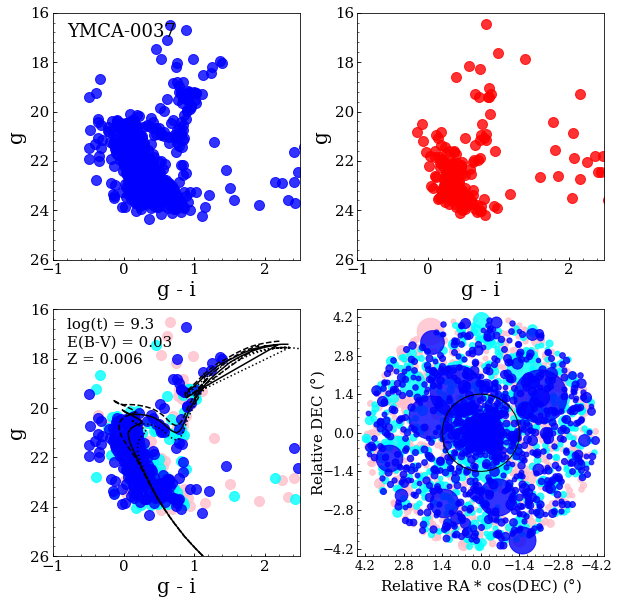}
    \includegraphics[width=0.35\textwidth]{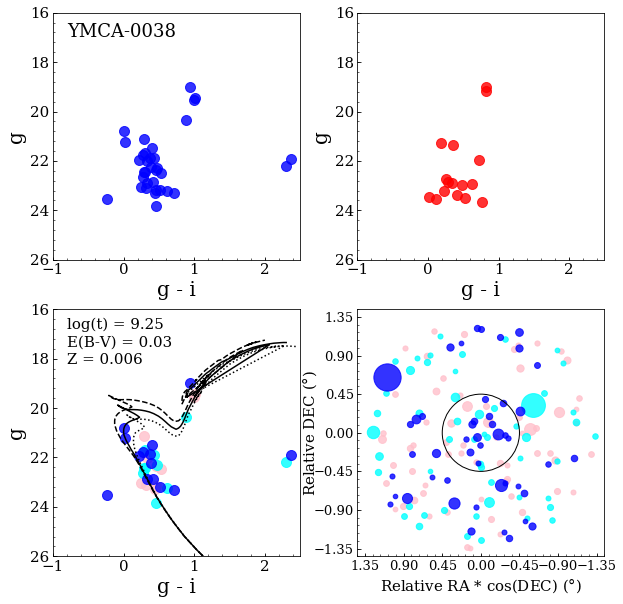}
    \includegraphics[width=0.35\textwidth]{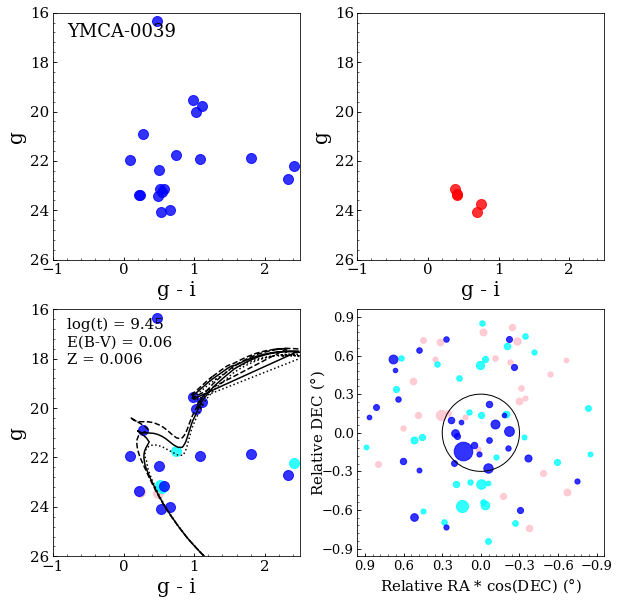}\\
\hspace{-1.2cm}
    \includegraphics[width=0.35\textwidth]{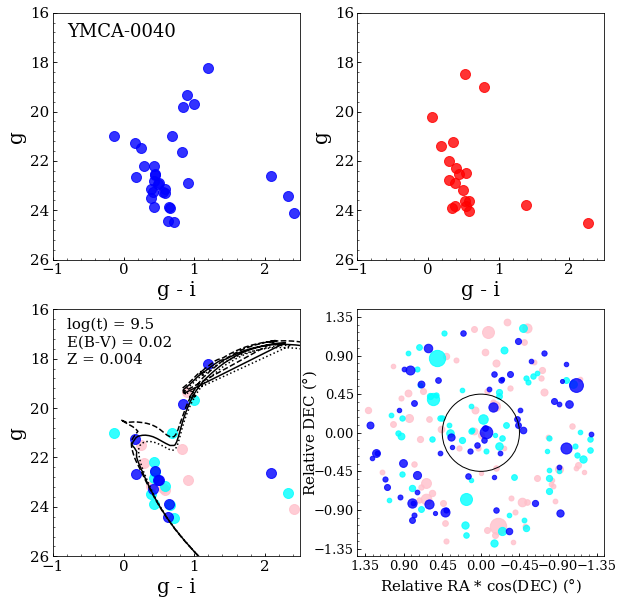}
    \includegraphics[width=0.35\textwidth]{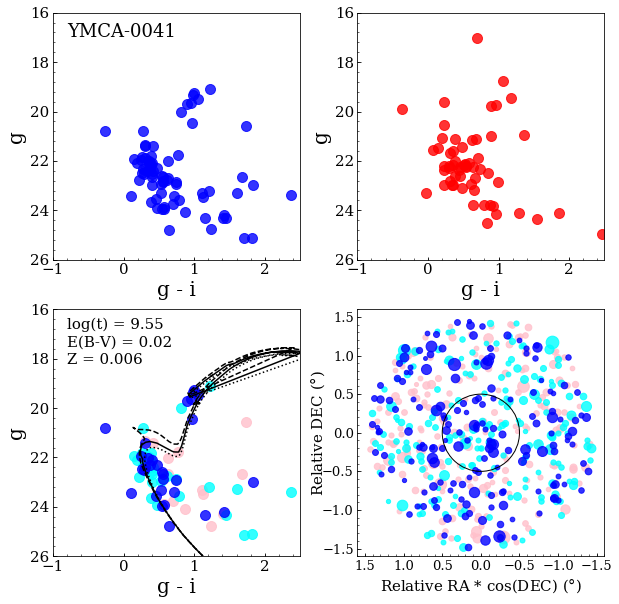}
    \includegraphics[width=0.35\textwidth]{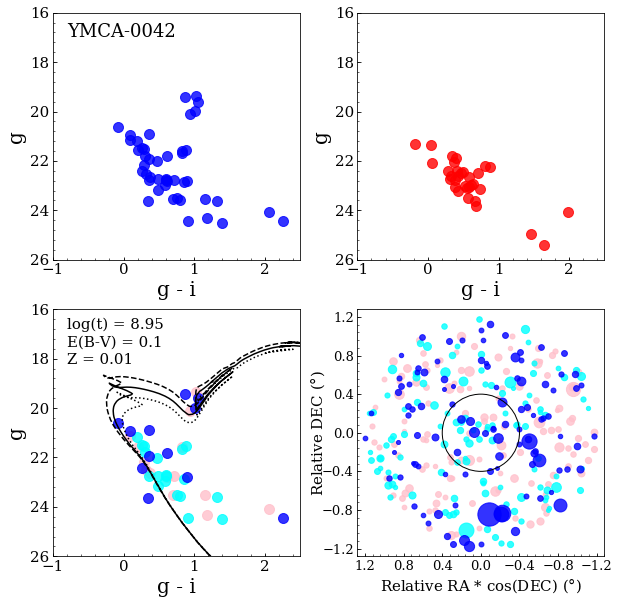}\\
\hspace{-1.2cm}
    \includegraphics[width=0.35\textwidth]{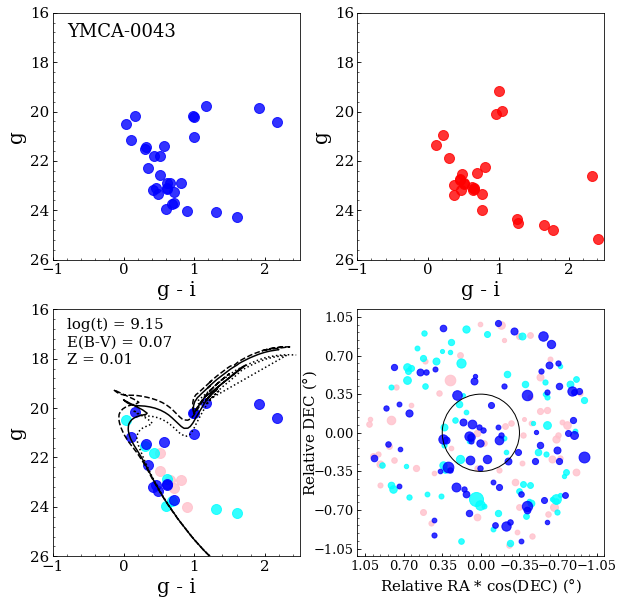}
    \includegraphics[width=0.35\textwidth]{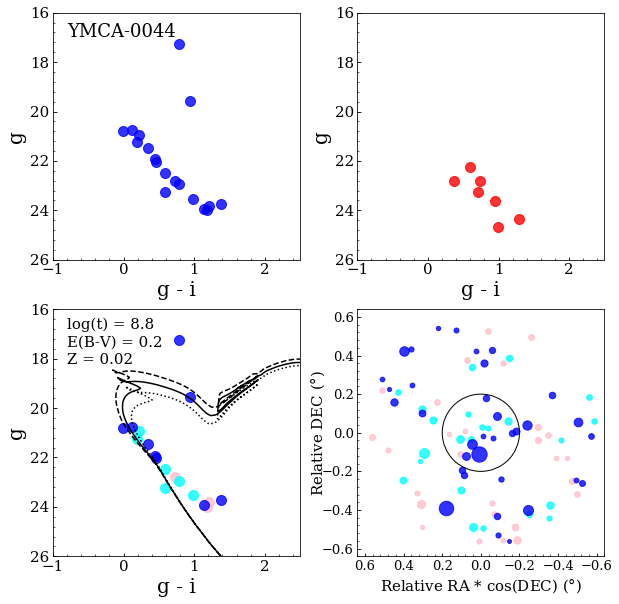}
    \includegraphics[width=0.35\textwidth]{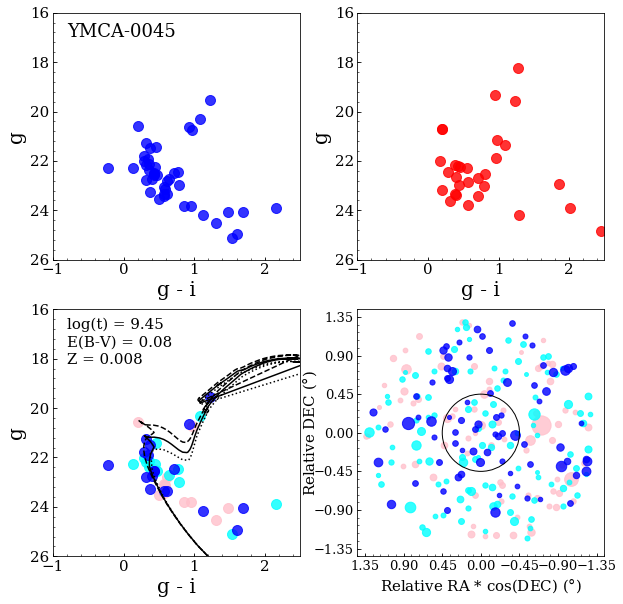}\\
\hspace{-1.2cm}
    \includegraphics[width=0.35\textwidth]{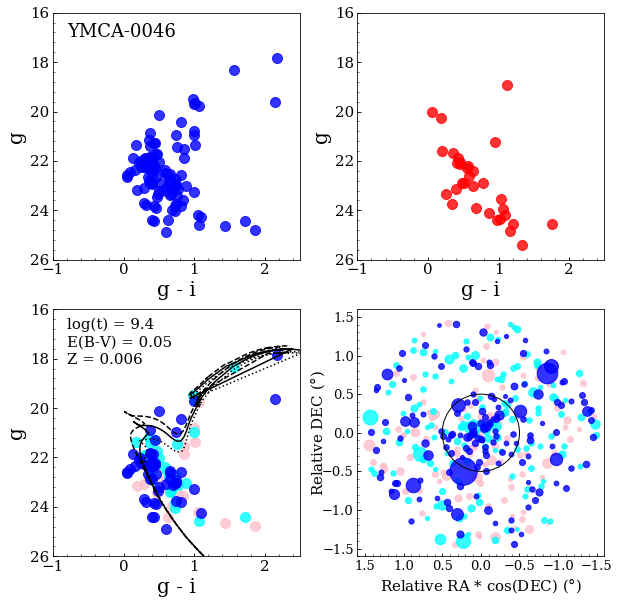}
    \includegraphics[width=0.35\textwidth]{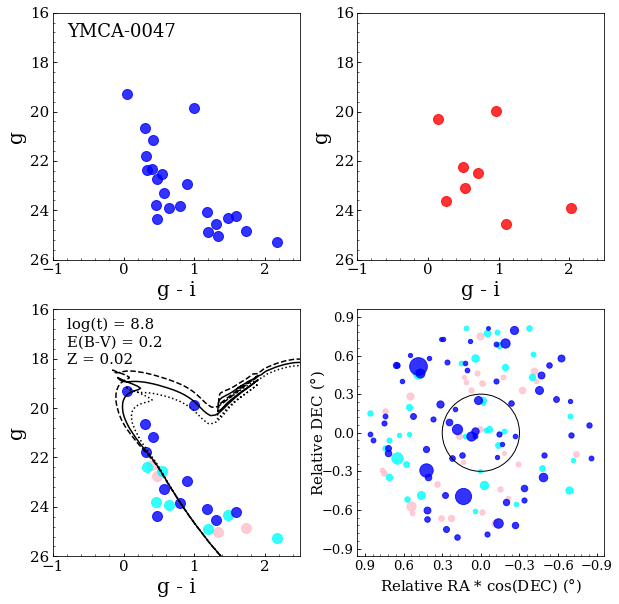}
    \includegraphics[width=0.35\textwidth]{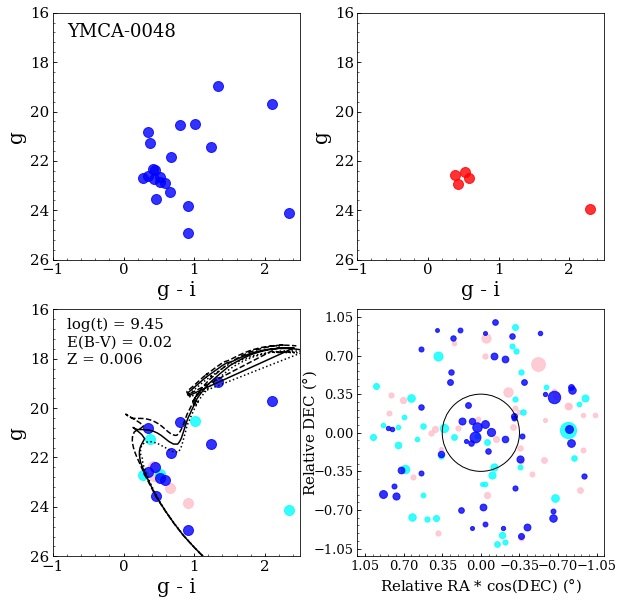}\\
    \contcaption{}
\end{figure*}

\begin{figure*}
\hspace{-1.2cm}
    \includegraphics[width=0.35\textwidth]{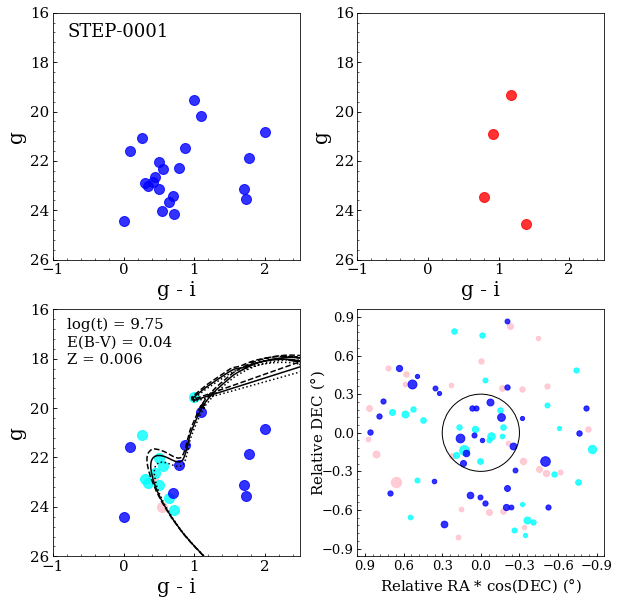}
    \includegraphics[width=0.35\textwidth]{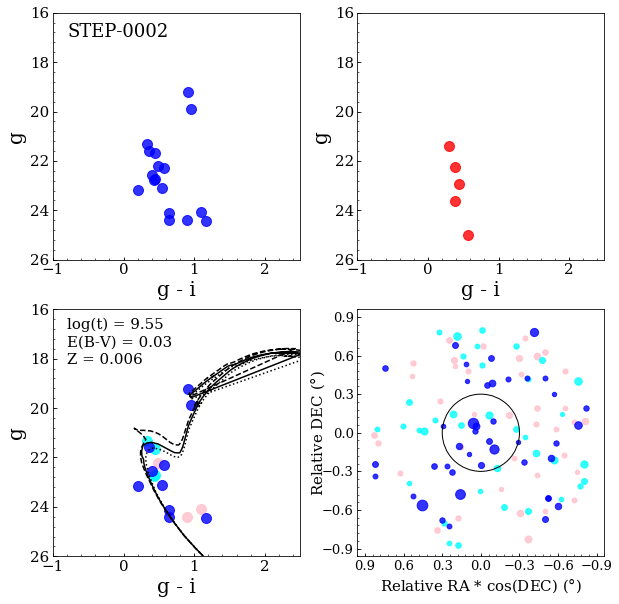}
    \includegraphics[width=0.35\textwidth]{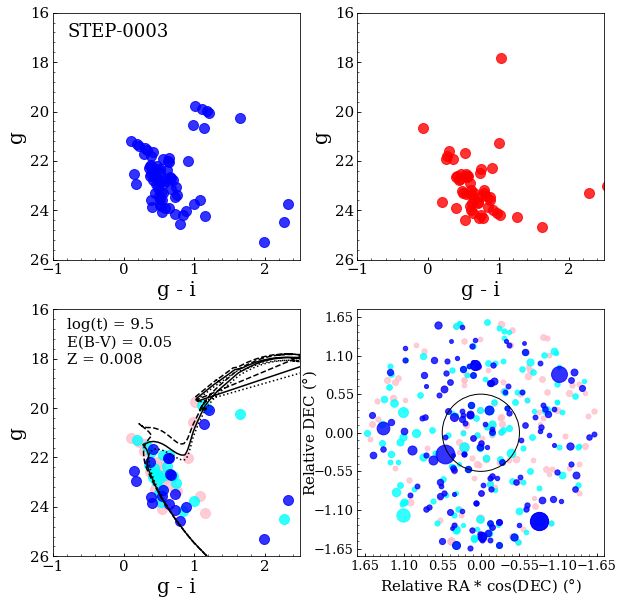}\\
\hspace{-1.2cm}
    \includegraphics[width=0.35\textwidth]{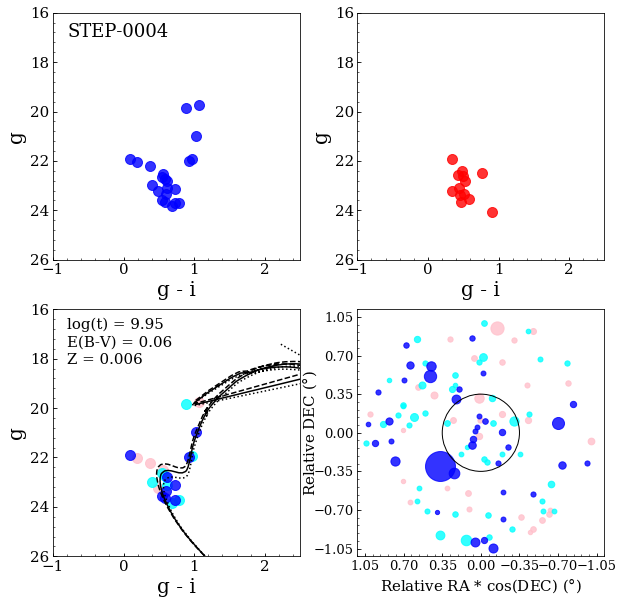}
    \includegraphics[width=0.35\textwidth]{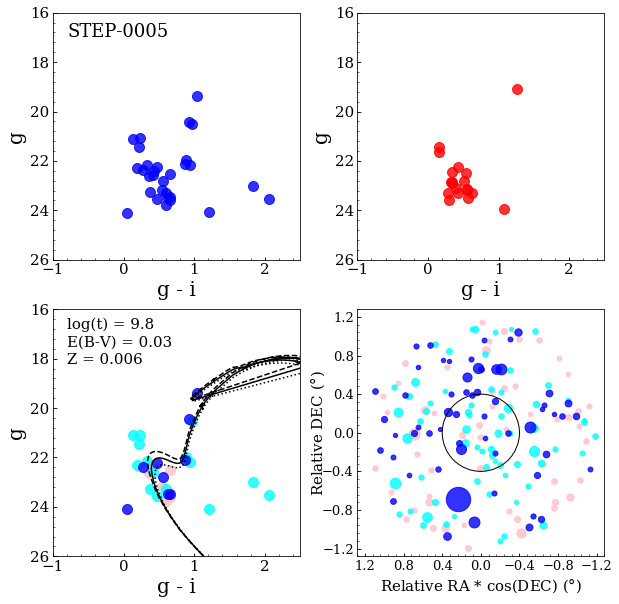}
    \includegraphics[width=0.35\textwidth]{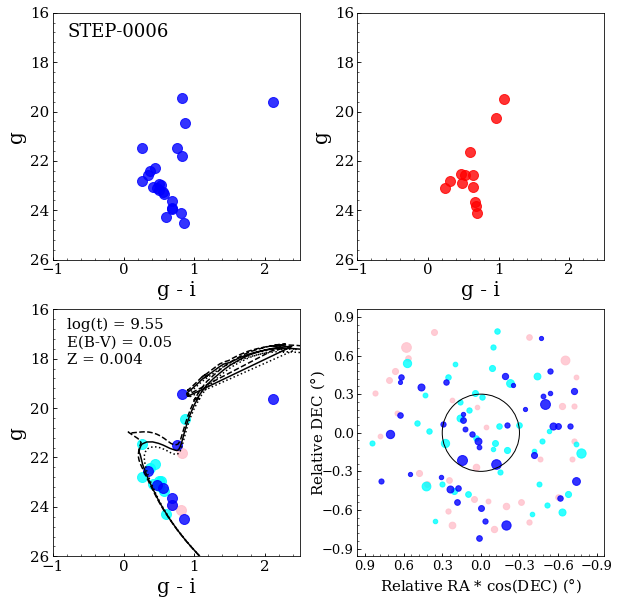}\\
\hspace{-1.2cm}
    \includegraphics[width=0.35\textwidth]{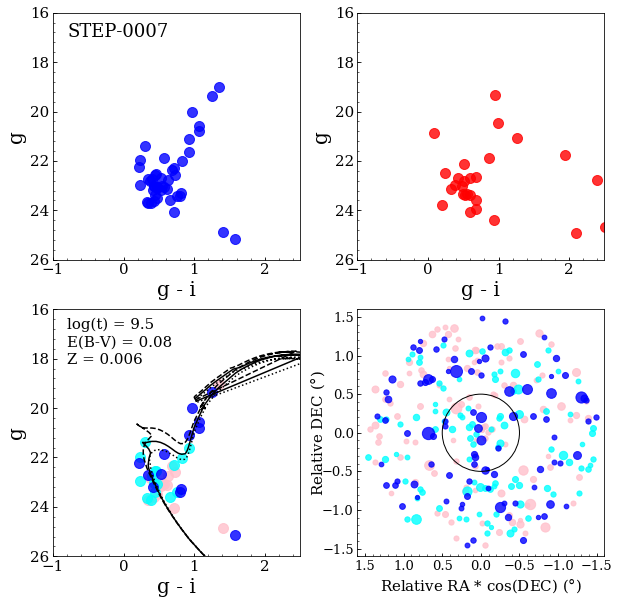}
    \includegraphics[width=0.35\textwidth]{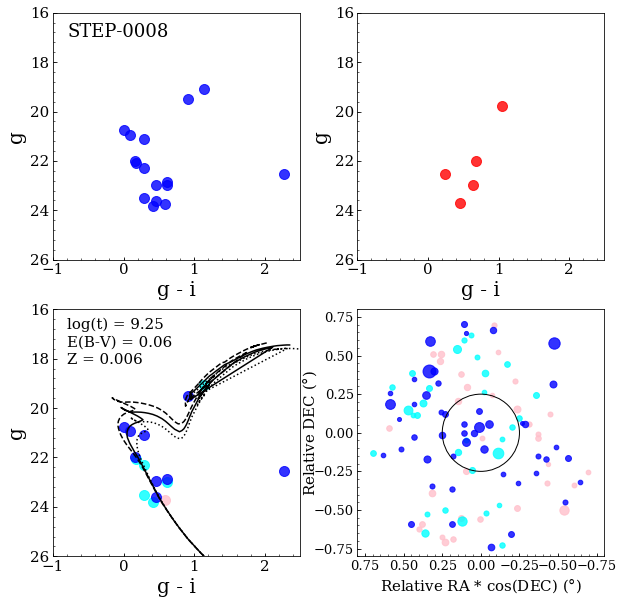}
    \includegraphics[width=0.35\textwidth]{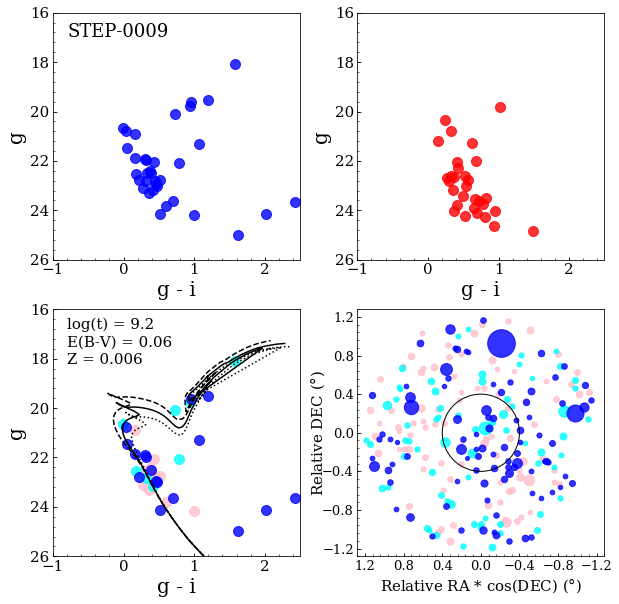}\\
\hspace{-1.2cm}
    \includegraphics[width=0.35\textwidth]{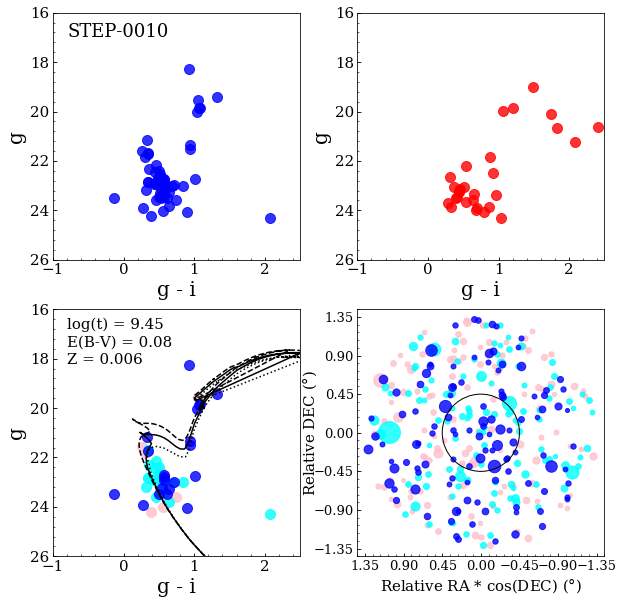}
    \includegraphics[width=0.35\textwidth]{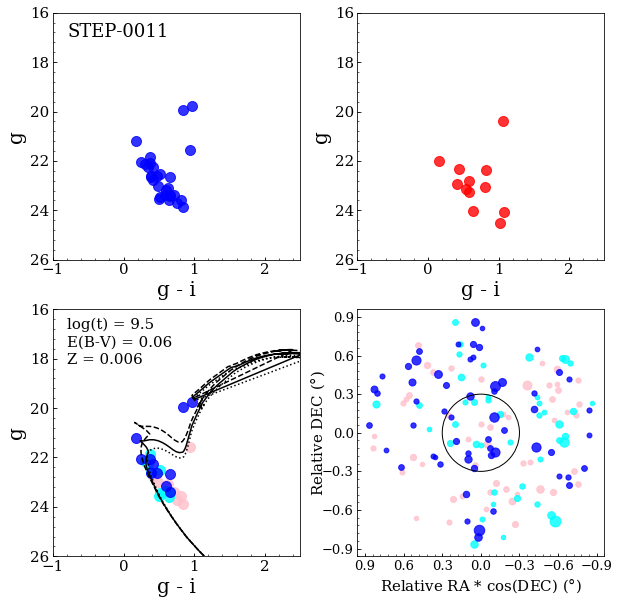}
    \includegraphics[width=0.35\textwidth]{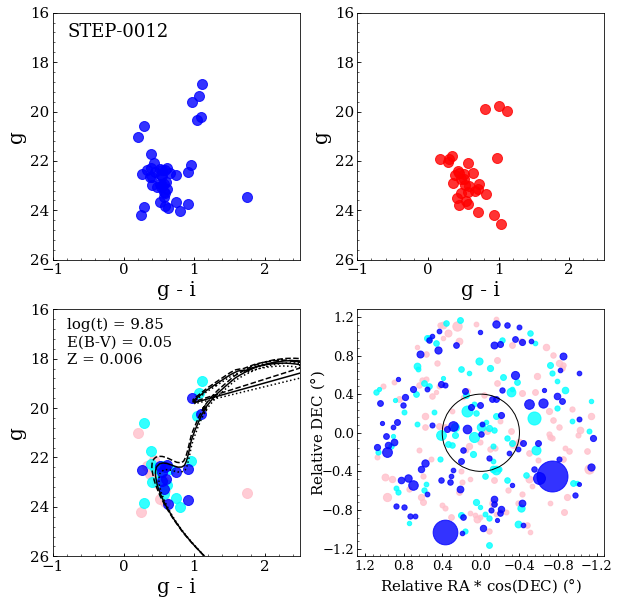}\\
    \contcaption{}
\end{figure*}

\begin{figure*}
\hspace{-1.2cm}
    \includegraphics[width=0.35\textwidth]{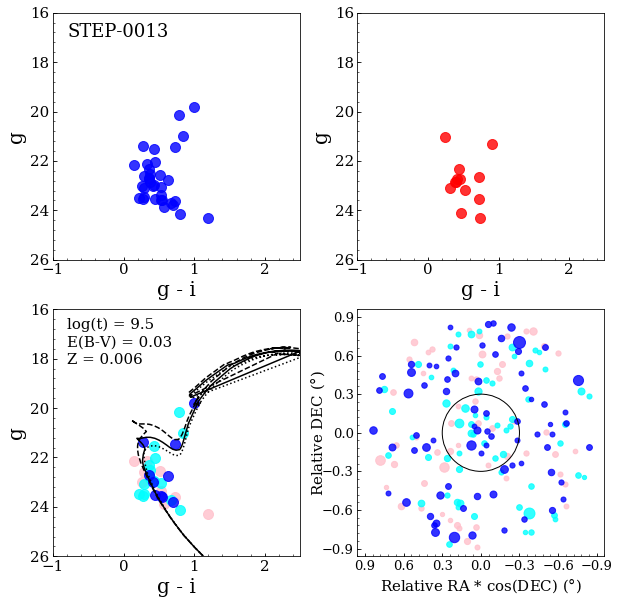}
    \includegraphics[width=0.35\textwidth]{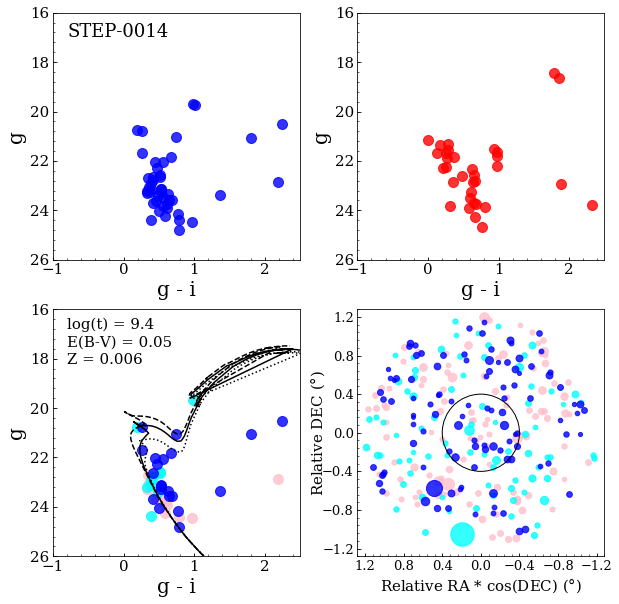}
    \includegraphics[width=0.35\textwidth]{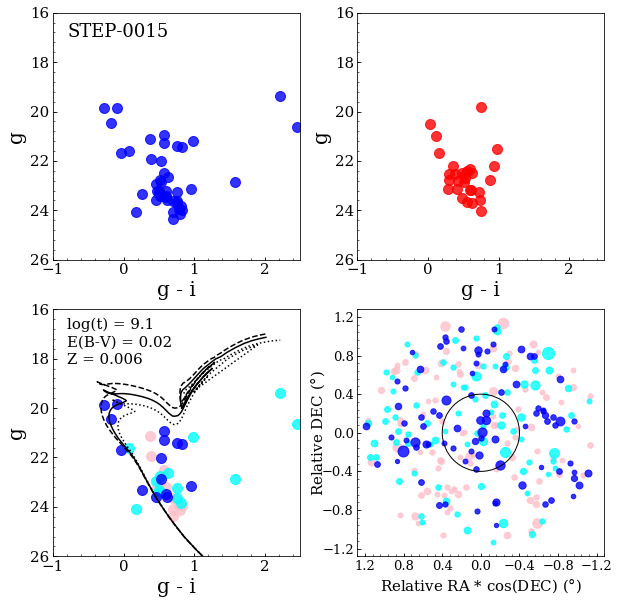}\\
\hspace{-1.2cm}
    \includegraphics[width=0.35\textwidth]{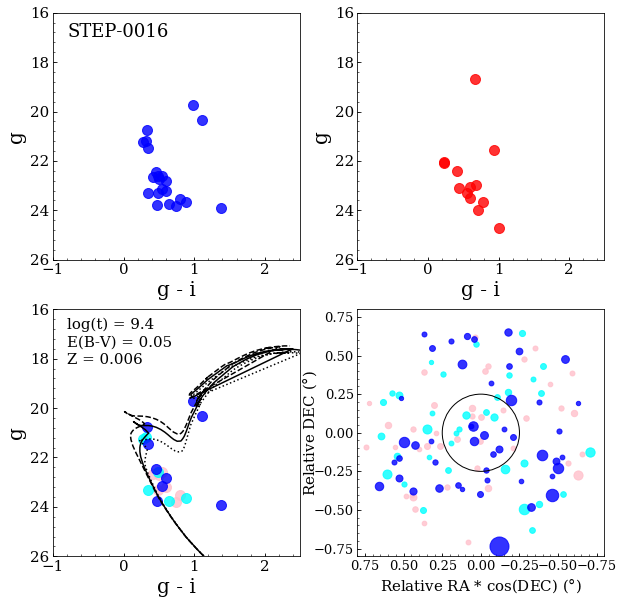}
    \includegraphics[width=0.35\textwidth]{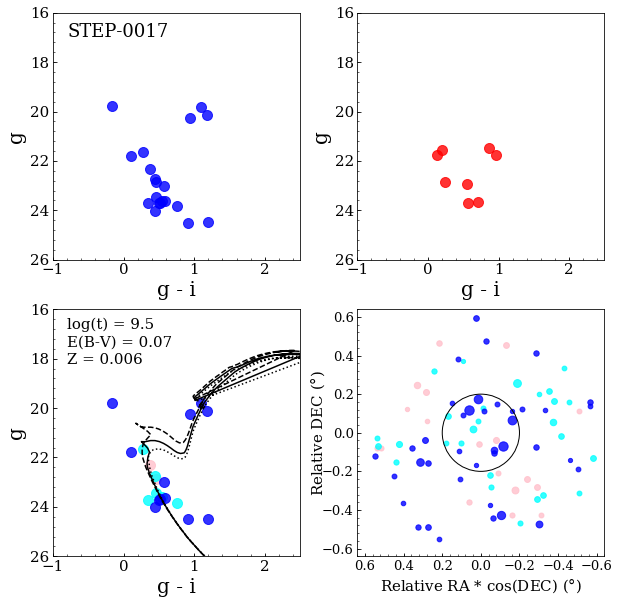}
    \includegraphics[width=0.35\textwidth]{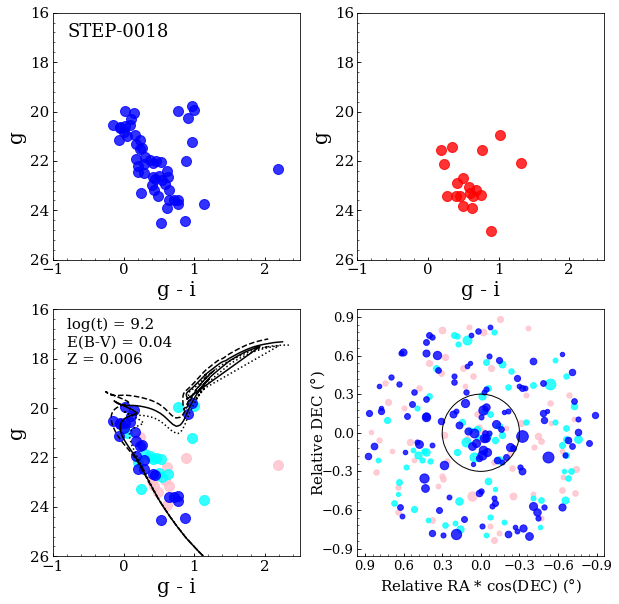}\\
\hspace{-1.2cm}
    \includegraphics[width=0.35\textwidth]{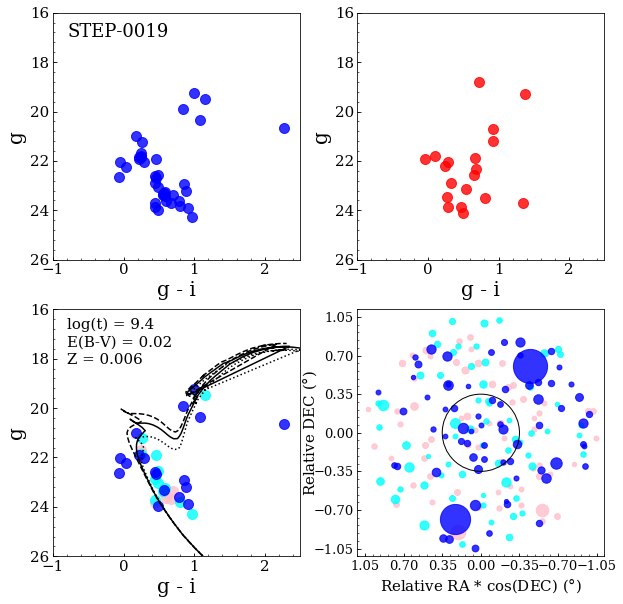}
    \includegraphics[width=0.35\textwidth]{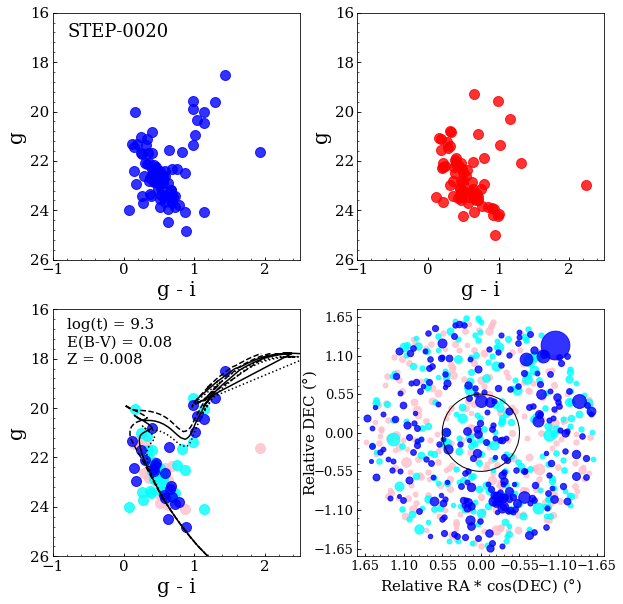}
    \includegraphics[width=0.35\textwidth]{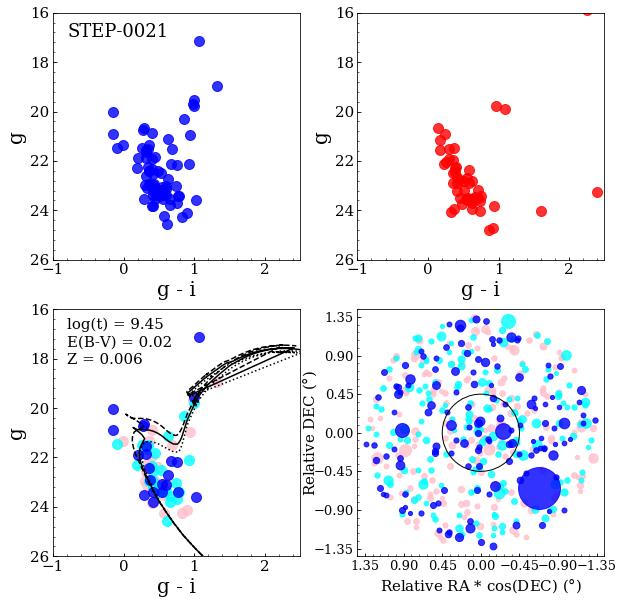}\\
\hspace{-1.2cm}
    \includegraphics[width=0.35\textwidth]{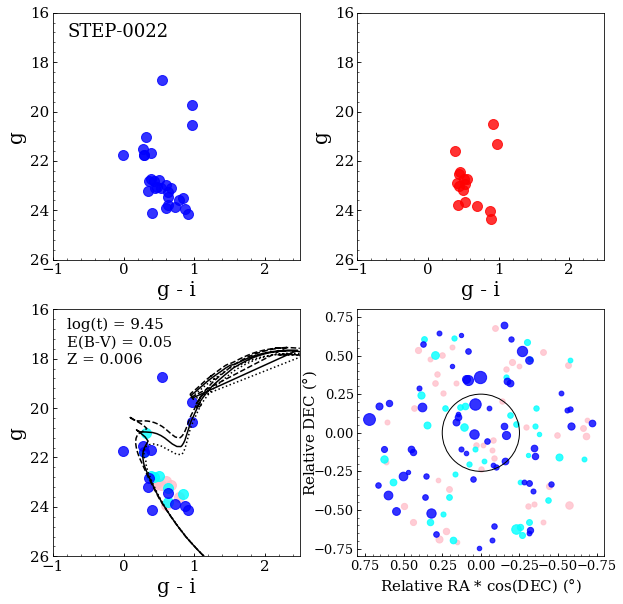}
    \includegraphics[width=0.35\textwidth]{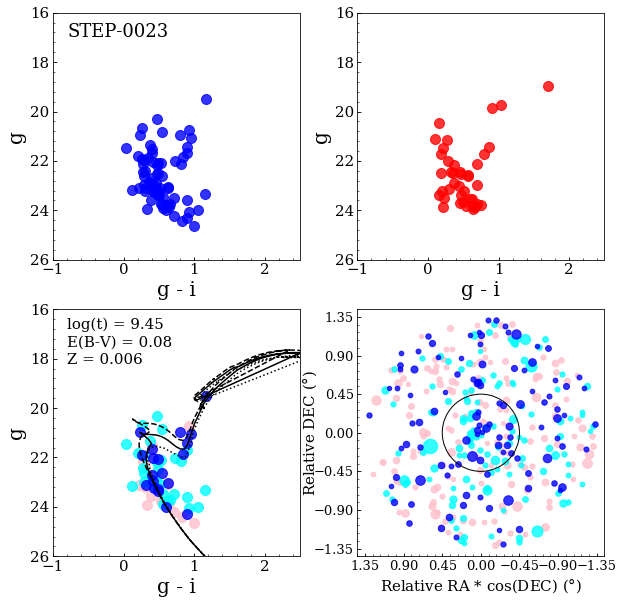}
    \includegraphics[width=0.35\textwidth]{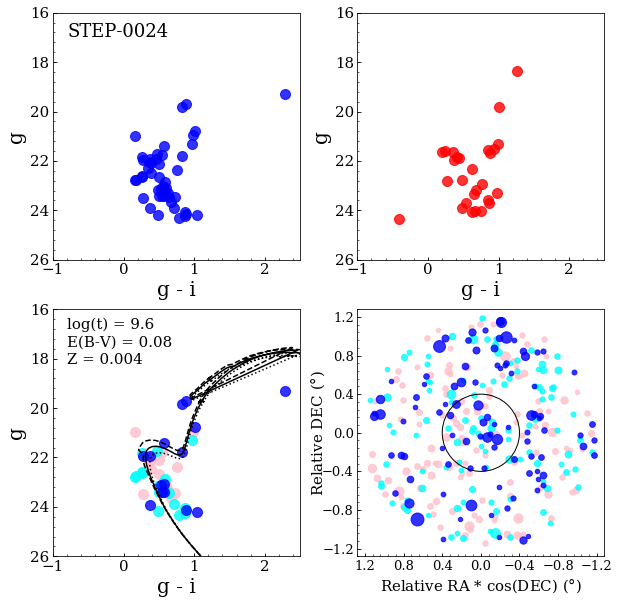}\\
    \contcaption{}
\end{figure*}

\begin{figure*}
\hspace{-1.2cm}
    \includegraphics[width=0.35\textwidth]{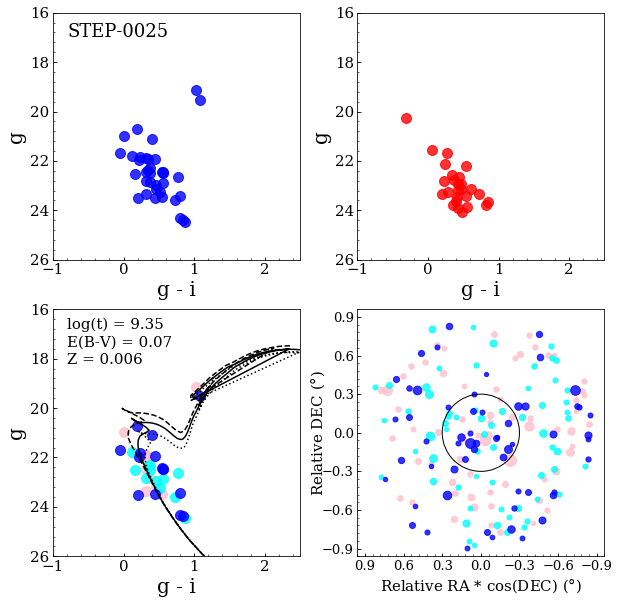}
    \includegraphics[width=0.35\textwidth]{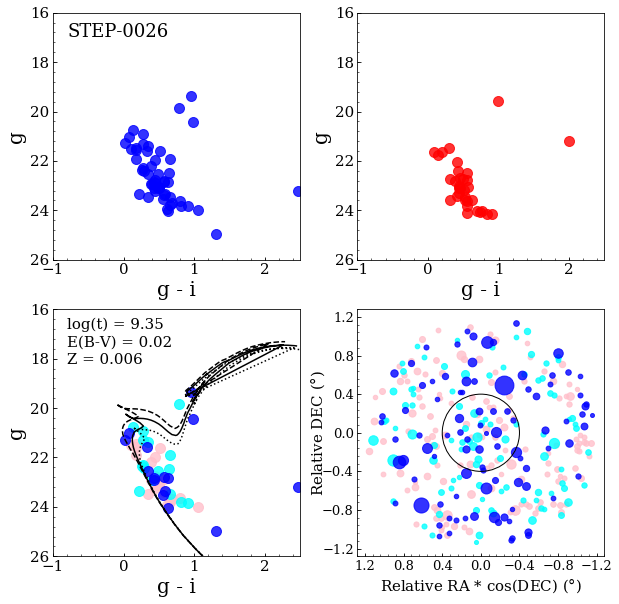}
    \includegraphics[width=0.35\textwidth]{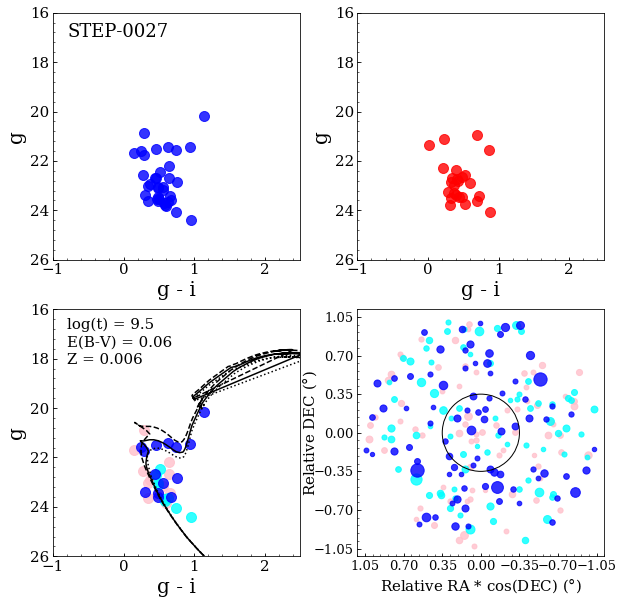}\\
\hspace{-1.2cm}
    \includegraphics[width=0.35\textwidth]{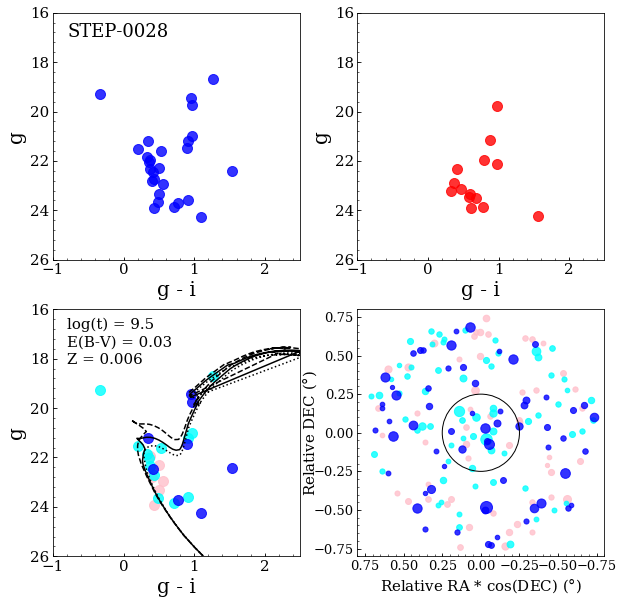}
    \includegraphics[width=0.35\textwidth]{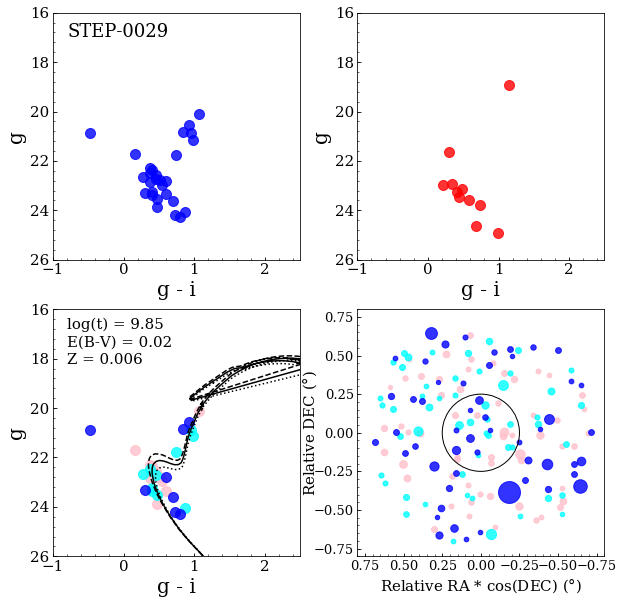}
    \includegraphics[width=0.35\textwidth]{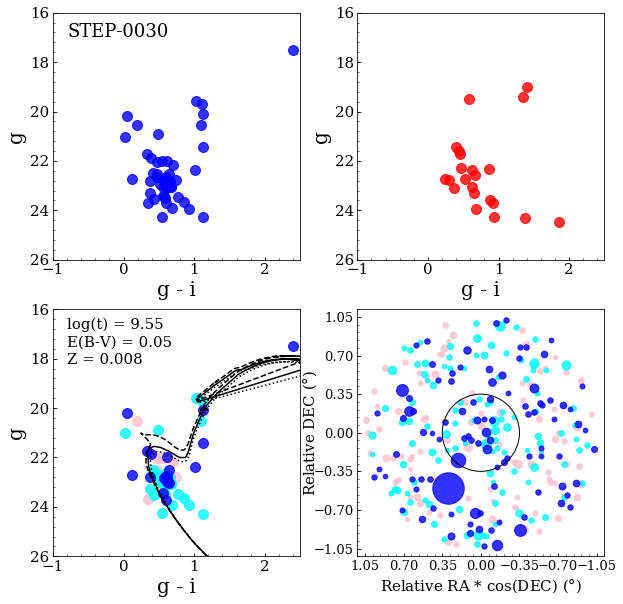}\\
\hspace{-1.2cm}
    \includegraphics[width=0.35\textwidth]{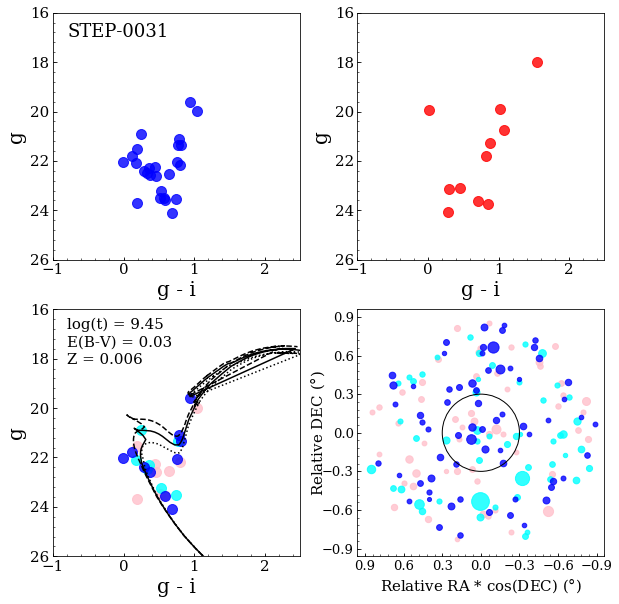}
    \includegraphics[width=0.35\textwidth]{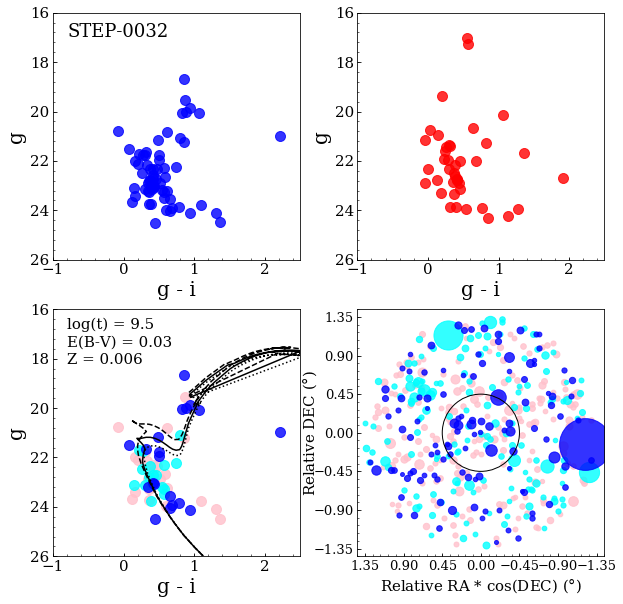}
    \includegraphics[width=0.35\textwidth]{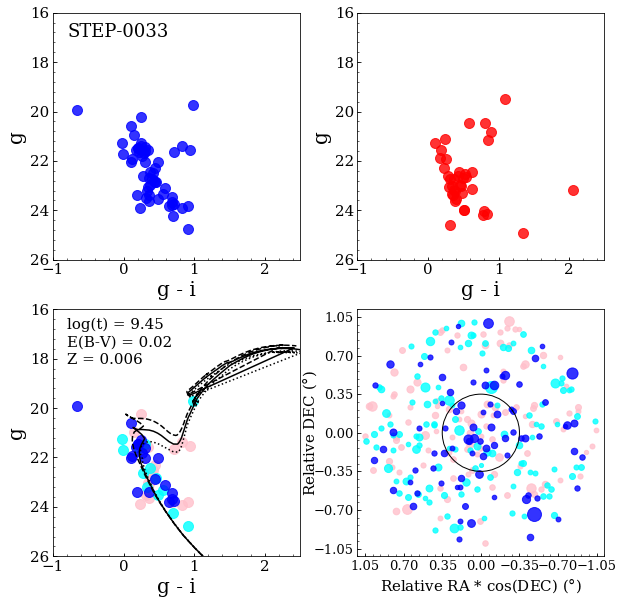}\\
\hspace{-1.2cm}
    \includegraphics[width=0.35\textwidth]{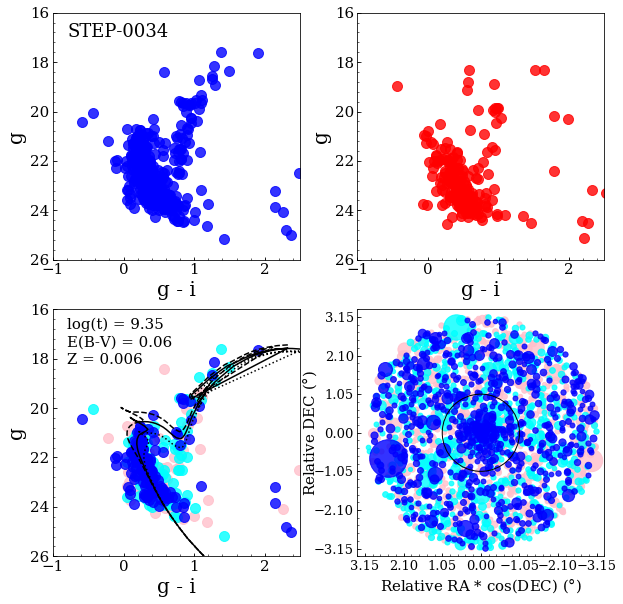}
    \includegraphics[width=0.35\textwidth]{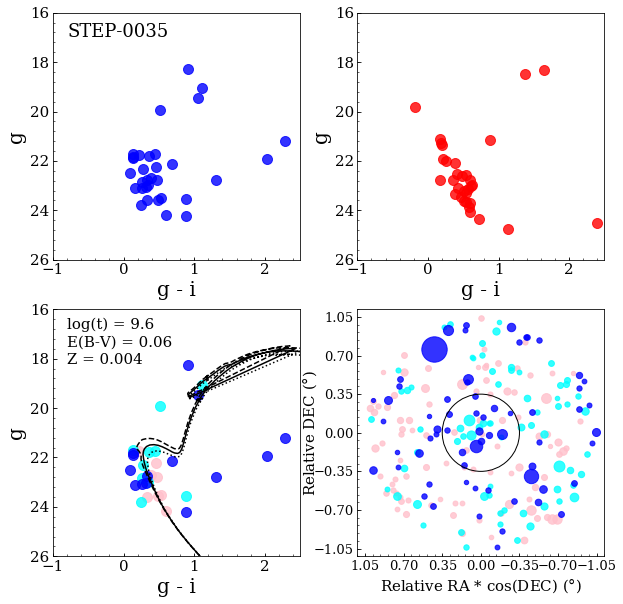}
    \includegraphics[width=0.35\textwidth]{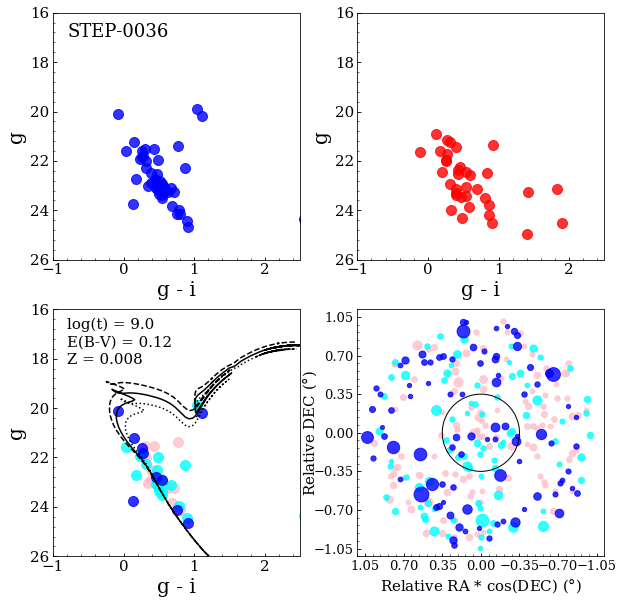}\\
    \contcaption{}
\end{figure*}

\begin{figure*}
\hspace{-12.2cm}
    \includegraphics[width=0.35\textwidth]{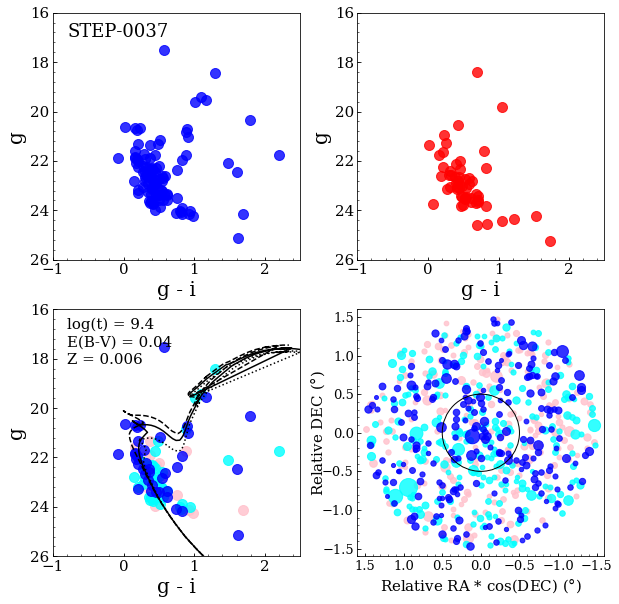}
    \\
    \contcaption{}
\end{figure*}

\section{EFF estimated parameters}

Table \ref{tab:EFF_parameters} lists the core and slope parameter ($\alpha$ and $\gamma$) of the EFF profile obtained via a MCMC, along with the ratio between the central surface density and the estimated background.
Figure \ref{fig:EFF_profiles} shows the RDPs with overlapped the EFF profile.

\begin{table*}
    \centering
    \caption{Elson, Fall \& Freeman's fitting parameters obtained by using all stars with P$\geq$75\%. In the first and fourth column are listed the ID of the SCs, while the core parameter is in the second and fifth column and the slope parameter in the third and seventh column. The fourth and last column indicates the ratio between the estimated central density and the estimated background.}
    \label{tab:EFF_parameters}
    \begin{tabular}{l|c|c|c|l|c|c|c}
    \hline
    ID & $\alpha$ & $\gamma$ & $n_0$/$\phi$ &  ID & $\alpha$ & $\gamma$ & $n_0$/$\phi$\\
     & (arcmin) & & & & (arcmin) & &\\
    \hline

step-0001 & $0.09^{+0.04}_{-0.04}$ & $3.64^{+0.92}_{-1.01}$ & 14.41 &
step-0002 & $0.06^{+0.02}_{-0.03}$ & $3.49^{+1.01}_{-0.99}$ & 9.83 \\
step-0004 & $0.10^{+0.03}_{-0.04}$ & $3.60^{+0.94}_{-0.99}$ & 18.30 &
step-0006 & $0.08^{+0.04}_{-0.03}$ & $3.39^{+1.08}_{-0.96}$ & 12.96 \\
step-0008 & $0.13^{+0.06}_{-0.02}$ & $3.28^{+0.51}_{-0.73}$ & 4.39 &
step-0009 & $0.25^{+0.06}_{-0.06}$ & $3.01^{+0.66}_{-0.67}$ & 4.49 \\
step-0010 & $0.25^{+0.07}_{-0.07}$ & $2.90^{+0.71}_{-0.63}$ & 3.26 &
step-0011 & $0.15^{+0.07}_{-0.04}$ & $3.28^{+0.51}_{-0.72}$ & 4.03 \\
step-0012 & $0.15^{+0.01}_{-0.00}$ & $3.85^{+0.11}_{-0.23}$ & 7.04 &
step-0013 & $0.11^{+0.01}_{-0.01}$ & $3.70^{+0.22}_{-0.39}$ & 7.59 \\
step-0014 & $0.44^{+0.15}_{-0.14}$ & $14.51^{+3.87}_{-5.28}$ & 3.70 &
step-0015 & $0.09^{+0.04}_{-0.04}$ & $3.36^{+1.07}_{-0.93}$ & 11.70 \\
step-0016 & $0.06^{+0.03}_{-0.03}$ & $3.63^{+0.93}_{-1.03}$ & 4.08 &
step-0017 & $0.24^{+0.04}_{-0.03}$ & $8.39^{+1.15}_{-1.76}$ & 6.87 \\
step-0018 & $0.11^{+0.03}_{-0.03}$ & $3.50^{+0.95}_{-0.91}$ & 6.25 &
step-0023 & $0.10^{+0.04}_{-0.04}$ & $2.88^{+1.26}_{-0.67}$ & 11.16 \\
step-0024 & $0.10^{+0.04}_{-0.03}$ & $3.39^{+1.03}_{-0.90}$ & 10.82 &
step-0025 & $0.10^{+0.03}_{-0.04}$ & $3.46^{+1.01}_{-0.93}$ & 8.04 \\
step-0026 & $0.15^{+0.07}_{-0.06}$ & $3.80^{+0.84}_{-1.05}$ & 4.31 &
step-0030 & $0.13^{+0.03}_{-0.02}$ & $4.26^{+0.53}_{-0.81}$ & 5.70 \\
step-0031 & $0.08^{+0.04}_{-0.03}$ & $3.41^{+1.07}_{-1.00}$ & 10.69 &
step-0032 & $0.12^{+0.02}_{-0.04}$ & $2.81^{+1.27}_{-0.62}$ & 6.07 \\
step-0033 & $0.12^{+0.02}_{-0.03}$ & $2.89^{+1.02}_{-0.62}$ & 5.81 &
step-0034 & $0.27^{+0.05}_{-0.07}$ & $3.44^{+0.77}_{-0.79}$ & 12.40 \\
step-0035 & $0.21^{+0.02}_{-0.01}$ & $4.61^{+0.29}_{-0.55}$ & 4.06 &
step-0036 & $0.08^{+0.04}_{-0.02}$ & $3.81^{+0.82}_{-1.01}$ & 6.15 \\
step-0037 & $0.19^{+0.04}_{-0.03}$ & $3.77^{+0.85}_{-1.03}$ & 3.12 &
ymca-0001 & $0.16^{+0.07}_{-0.04}$ & $3.92^{+0.76}_{-1.01}$ & 7.13 \\
ymca-0002 & $0.22^{+0.03}_{-0.02}$ & $4.48^{+0.37}_{-0.65}$ & 16.36 &
ymca-0003 & $0.08^{+0.04}_{-0.04}$ & $3.62^{+0.94}_{-1.04}$ & 17.39 \\
ymca-0004 & $0.20^{+0.06}_{-0.06}$ & $3.64^{+0.92}_{-1.00}$ & 13.92 &
ymca-0005 & $0.35^{+0.07}_{-0.04}$ & $12.26^{+1.96}_{-2.99}$ & 6.50 \\
ymca-0007 & $0.37^{+0.07}_{-0.05}$ & $11.83^{+2.21}_{-3.04}$ & 5.48 &
ymca-0008 & $0.08^{+0.04}_{-0.04}$ & $3.32^{+1.11}_{-0.93}$ & 26.83 \\
ymca-0009 & $0.10^{+0.03}_{-0.04}$ & $3.53^{+0.97}_{-0.98}$ & 16.19 &
ymca-0010 & $0.10^{+0.03}_{-0.04}$ & $3.35^{+1.03}_{-0.89}$ & 18.34 \\
ymca-0012 & $0.11^{+0.03}_{-0.04}$ & $3.28^{+1.07}_{-0.86}$ & 24.48 &
ymca-0013 & $0.09^{+0.03}_{-0.02}$ & $3.86^{+0.79}_{-1.03}$ & 8.91 \\
ymca-0014 & $0.40^{+0.07}_{-0.07}$ & $9.74^{+3.37}_{-2.97}$ & 9.37 &
ymca-0015 & $0.18^{+0.04}_{-0.02}$ & $3.95^{+0.73}_{-1.01}$ & 7.06 \\
ymca-0018 & $0.11^{+0.03}_{-0.03}$ & $3.27^{+1.07}_{-0.86}$ & 13.58 &
ymca-0019 & $0.10^{+0.03}_{-0.04}$ & $3.10^{+1.13}_{-0.79}$ & 16.04 \\
ymca-0020 & $0.09^{+0.04}_{-0.04}$ & $3.49^{+1.01}_{-0.98}$ & 13.15 &
ymca-0021 & $0.08^{+0.04}_{-0.04}$ & $3.47^{+1.04}_{-1.01}$ & 13.75 \\
ymca-0023 & $0.27^{+0.09}_{-0.11}$ & $5.43^{+3.01}_{-2.05}$ & 1.50 &
ymca-0024 & $0.09^{+0.03}_{-0.03}$ & $3.66^{+0.91}_{-1.01}$ & 7.68 \\
ymca-0025 & $0.24^{+0.09}_{-0.07}$ & $2.86^{+0.94}_{-0.59}$ & 56.85 &
ymca-0026 & $0.07^{+0.04}_{-0.04}$ & $3.65^{+0.93}_{-1.07}$ & 5.87 \\
ymca-0028 & $0.10^{+0.03}_{-0.04}$ & $3.10^{+1.16}_{-0.79}$ & 10.44 &
ymca-0029 & $0.09^{+0.03}_{-0.03}$ & $3.54^{+0.98}_{-0.98}$ & 11.16 \\
ymca-0030 & $0.10^{+0.03}_{-0.04}$ & $3.22^{+1.13}_{-0.87}$ & 11.36 &
ymca-0031 & $0.11^{+0.03}_{-0.04}$ & $3.12^{+1.10}_{-0.79}$ & 14.71 \\
ymca-0034 & $0.09^{+0.04}_{-0.04}$ & $3.21^{+1.14}_{-0.88}$ & 15.75 &
ymca-0035 & $0.22^{+0.04}_{-0.02}$ & $8.67^{+0.96}_{-1.66}$ & 13.63 \\
ymca-0036 & $0.08^{+0.04}_{-0.04}$ & $3.43^{+1.05}_{-0.98}$ & 19.52 &
ymca-0037 & $0.72^{+0.22}_{-0.18}$ & $5.59^{+2.31}_{-1.50}$ & 31.1 \\
ymca-0038 & $0.07^{+0.04}_{-0.04}$ & $3.58^{+0.96}_{-1.02}$ & 19.87 &
ymca-0039 & $0.09^{+0.04}_{-0.04}$ & $3.47^{+1.01}_{-0.97}$ & 17.23 \\
ymca-0040 & $0.08^{+0.04}_{-0.04}$ & $3.64^{+0.93}_{-1.02}$ & 14.27 &
ymca-0041 & $0.11^{+0.03}_{-0.04}$ & $3.49^{+0.94}_{-0.89}$ & 10.85 \\
ymca-0042 & $0.08^{+0.03}_{-0.02}$ & $3.87^{+0.79}_{-1.04}$ & 4.76 &
ymca-0043 & $0.08^{+0.04}_{-0.04}$ & $3.68^{+0.92}_{-1.04}$ & 10.20 \\
ymca-0044 & $0.10^{+0.03}_{-0.04}$ & $3.51^{+0.99}_{-0.99}$ & 9.99 &
ymca-0045 & $0.20^{+0.07}_{-0.07}$ & $5.91^{+2.73}_{-2.53}$ & 3.33 \\
ymca-0046 & $0.12^{+0.02}_{-0.03}$ & $3.25^{+0.87}_{-0.72}$ & 14.75 &
ymca-0047 & $0.07^{+0.04}_{-0.04}$ & $3.71^{+0.89}_{-1.05}$ & 6.57 \\
ymca-0048 & $0.08^{+0.03}_{-0.02}$ & $3.92^{+0.75}_{-1.00}$ & 6.57 & & & &\\
\hline   
\end{tabular}  
\end{table*}

\begin{table*}
    \centering
    \caption{Same of Tab\ref{tab:EFF_parameters} but with EFF fitting parameters obtained by using all stars with P$\geq$50\%.}
    \label{tab:EFF_parameters_50}
    \begin{tabular}{l|c|c|c|l|c|c|c}
    \hline
    ID & $\alpha$ & $\gamma$ & $n_0$/$\phi$ &  ID & $\alpha$ & $\gamma$ & $n_0$/$\phi$\\
     & (arcmin) & & & & (arcmin) & &\\
    \hline
step-0003 & $0.17^{+0.05}_{-0.04}$ & $3.12^{+0.59}_{-0.67}$ & 4.55 &
step-0005 & $0.21^{+0.05}_{-0.06}$ & $3.00^{+0.67}_{-0.66}$ & 3.63 \\
step-0007 & $0.06^{+0.03}_{-0.03}$ & $3.52^{+0.99}_{-0.97}$ & 5.02 &
step-0019 & $0.14^{+0.04}_{-0.04}$ & $2.65^{+0.54}_{-0.45}$ & 4.73 \\
step-0020 & $0.05^{+0.03}_{-0.03}$ & $3.63^{+0.93}_{-1.05}$ & 2.43 &
step-0021 & $0.11^{+0.03}_{-0.03}$ & $3.26^{+1.08}_{-0.86}$ & 4.03 \\
step-0022 & $0.06^{+0.03}_{-0.03}$ & $3.60^{+0.95}_{-1.02}$ & 2.89 &
step-0027 & $0.77^{+0.15}_{-0.17}$ & $3.39^{+1.04}_{-0.93}$ & 2.63 \\
step-0028 & $0.10^{+0.03}_{-0.03}$ & $3.52^{+1.00}_{-1.02}$ & 4.03 &
step-0029 & $0.08^{+0.04}_{-0.03}$ & $3.71^{+0.88}_{-1.04}$ & 3.91 \\
ymca-0006 & $0.57^{+0.09}_{-0.05}$ & $8.71^{+0.93}_{-1.39}$ & 2.20 &
ymca-0011 & $0.14^{+0.04}_{-0.03}$ & $3.24^{+0.53}_{-0.71}$ & 5.45 \\
ymca-0016 & $0.54^{+0.08}_{-0.03}$ & $8.90^{+0.82}_{-1.57}$ & 1.66 &
ymca-0017 & $0.16^{+0.05}_{-0.04}$ & $2.81^{+0.47}_{-0.51}$ & 4.99 \\
ymca-0022 & $0.13^{+0.04}_{-0.02}$ & $3.10^{+0.29}_{-0.46}$ & 4.04 &
ymca-0027 & $0.14^{+0.05}_{-0.03}$ & $2.90^{+0.42}_{-0.54}$ & 4.04 \\
ymca-0032 & $0.06^{+0.03}_{-0.03}$ & $3.57^{+0.97}_{-1.03}$ & 4.07 &
ymca-0033 & $0.06^{+0.03}_{-0.03}$ & $3.63^{+0.93}_{-0.99}$ & 4.56 \\
\hline   
\end{tabular}  
\end{table*}

\begin{figure*}
    \vspace{-0.5cm}
    \includegraphics[width=0.23\textwidth]{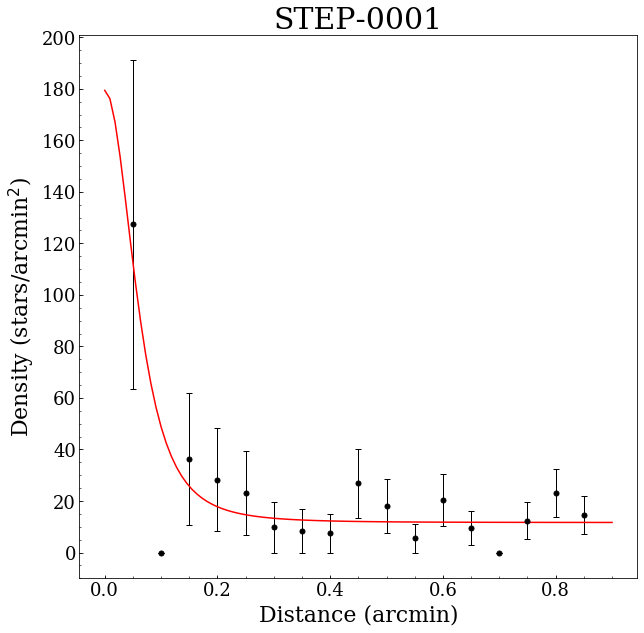}
    \includegraphics[width=0.23\textwidth]{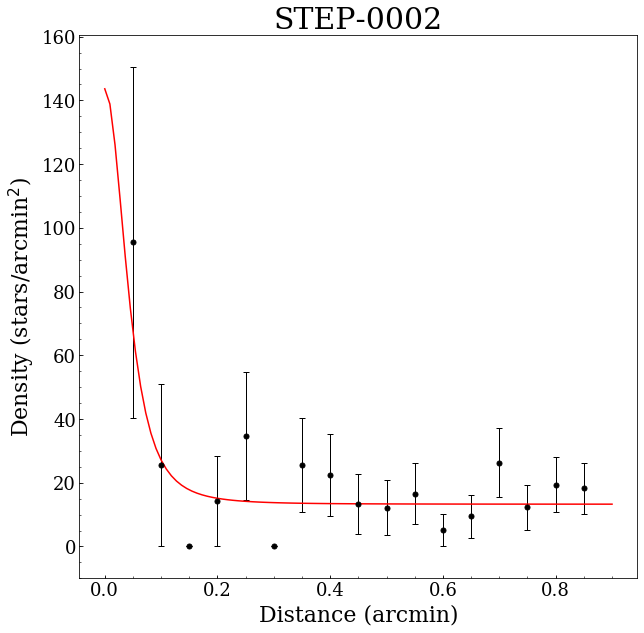}
    \includegraphics[width=0.23\textwidth]{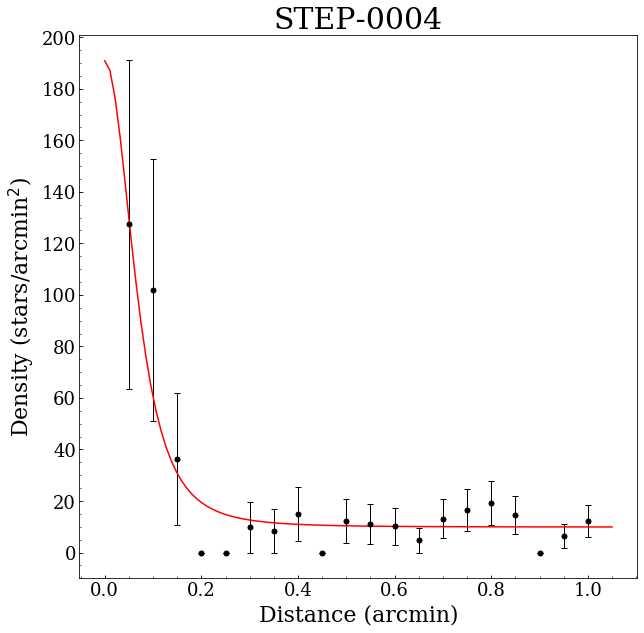}
    \includegraphics[width=0.23\textwidth]{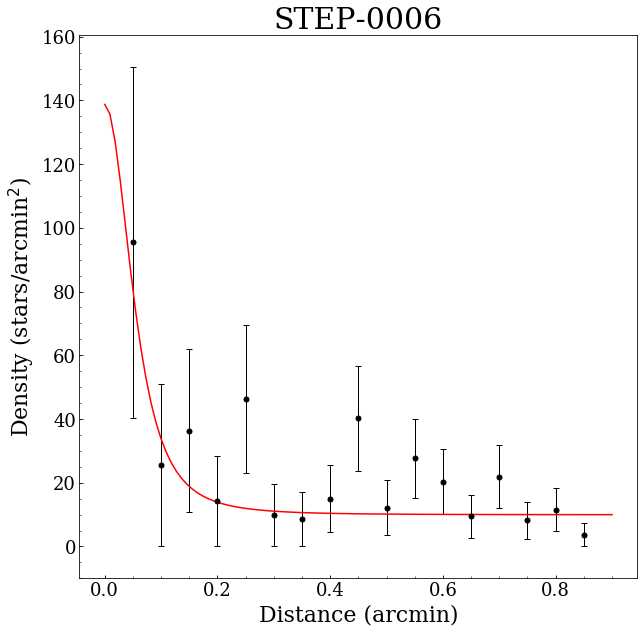}\\
    \includegraphics[width=0.23\textwidth]{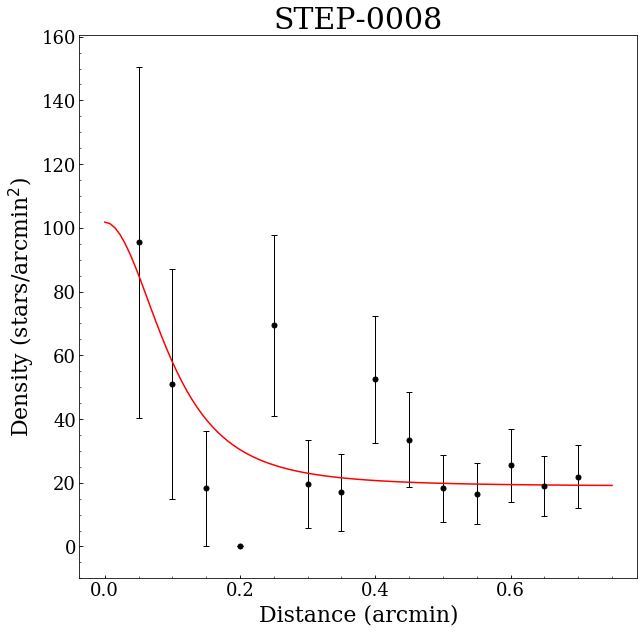}
    \includegraphics[width=0.23\textwidth]{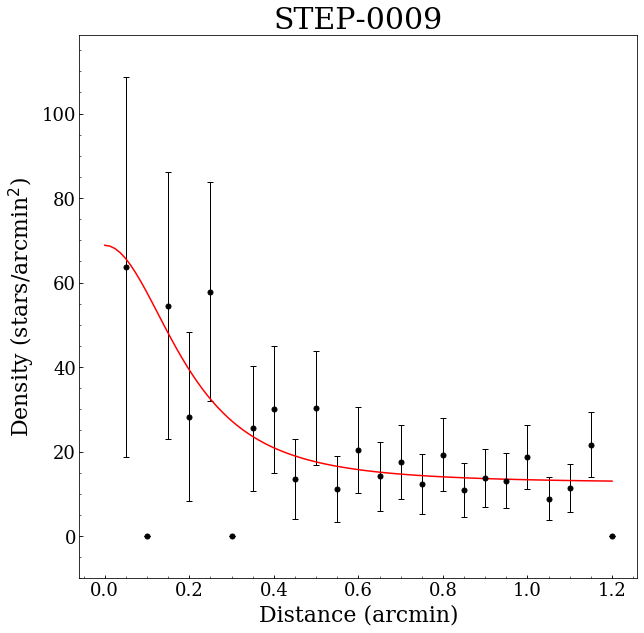}
    \includegraphics[width=0.23\textwidth]{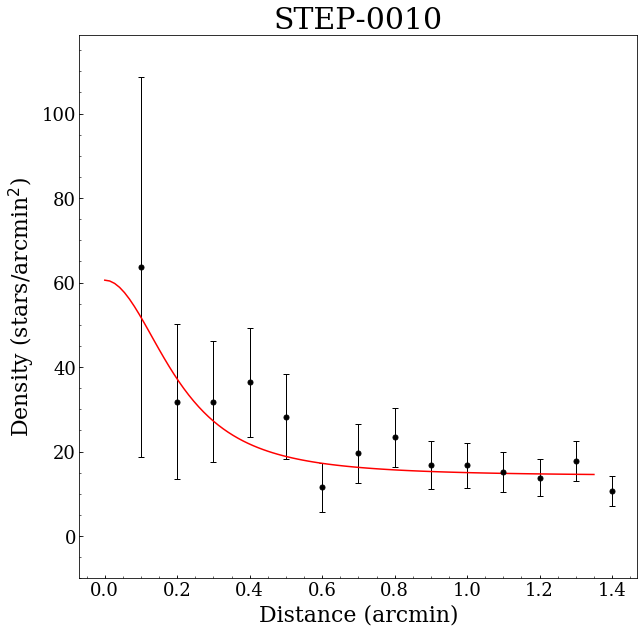}
    \includegraphics[width=0.23\textwidth]{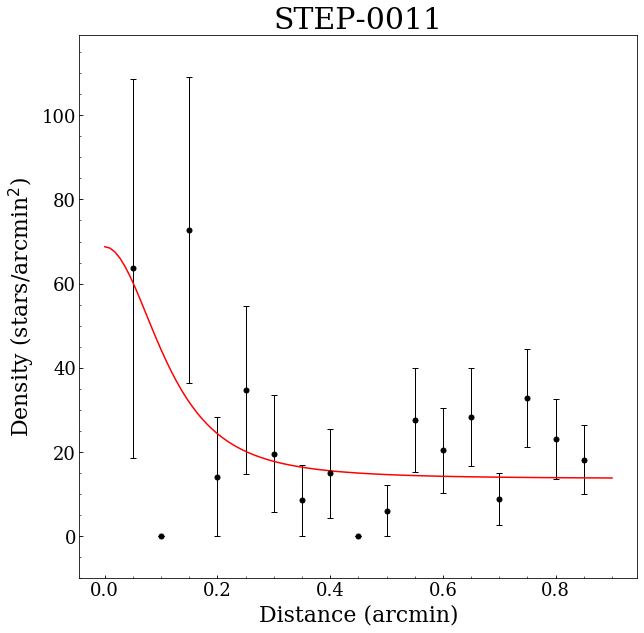}\\
    \includegraphics[width=0.23\textwidth]{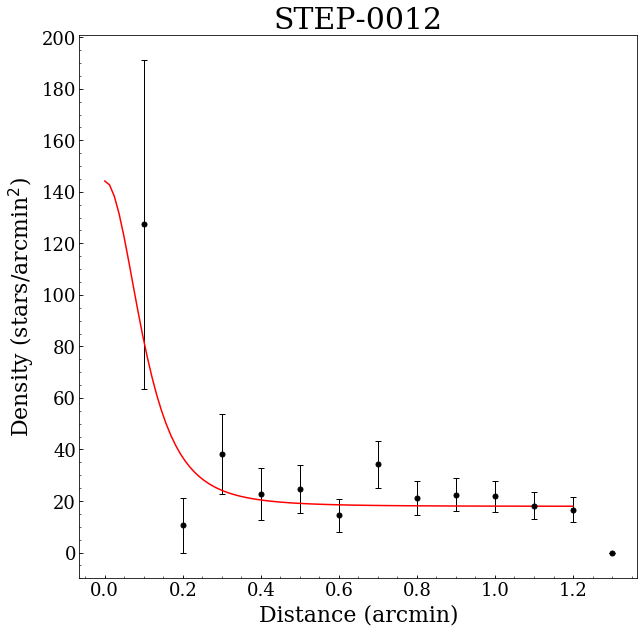}
    \includegraphics[width=0.23\textwidth]{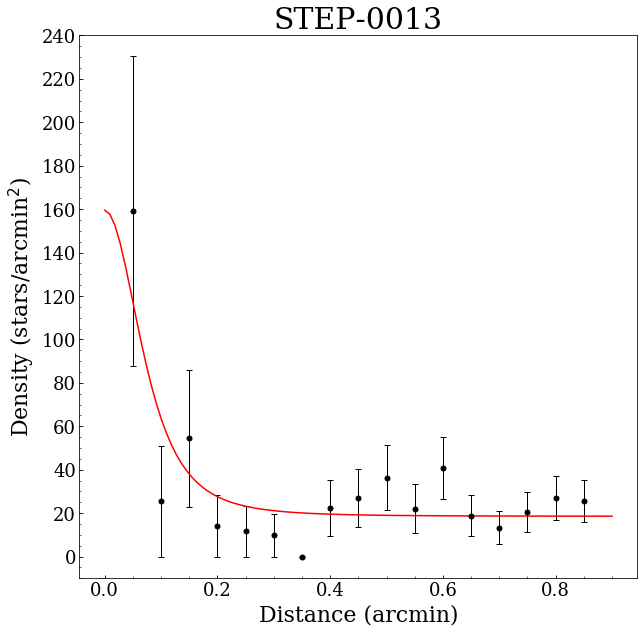}
    \includegraphics[width=0.23\textwidth]{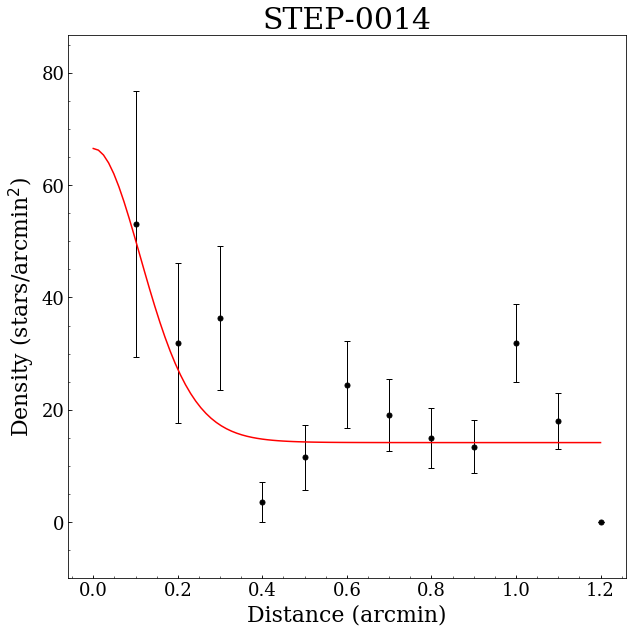}
    \includegraphics[width=0.23\textwidth]{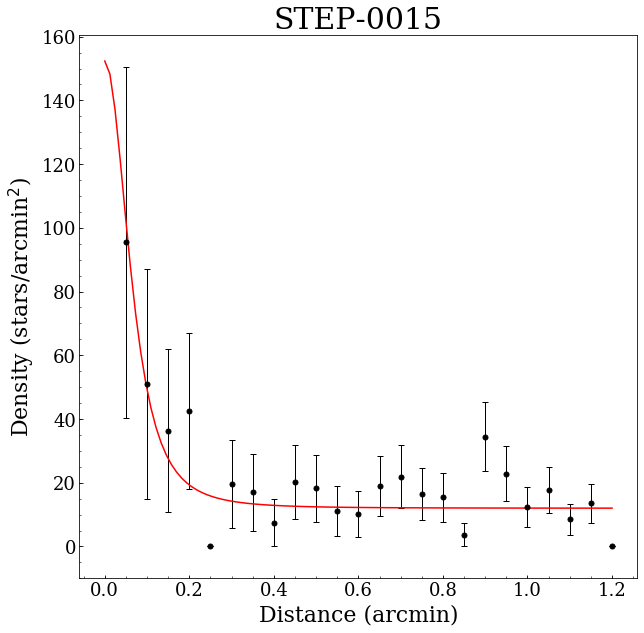}\\
    \includegraphics[width=0.23\textwidth]{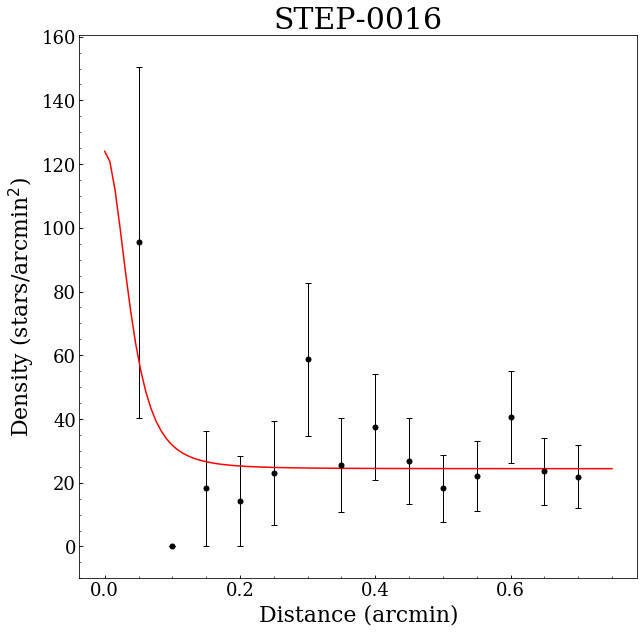}
    \includegraphics[width=0.23\textwidth]{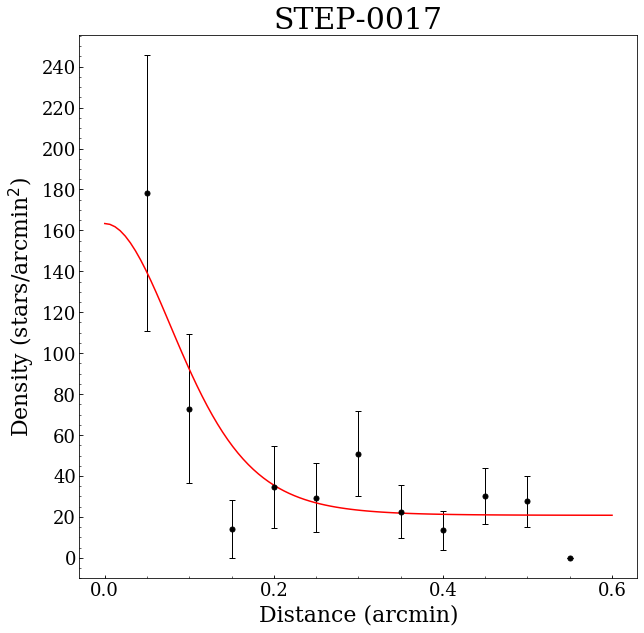}
    \includegraphics[width=0.23\textwidth]{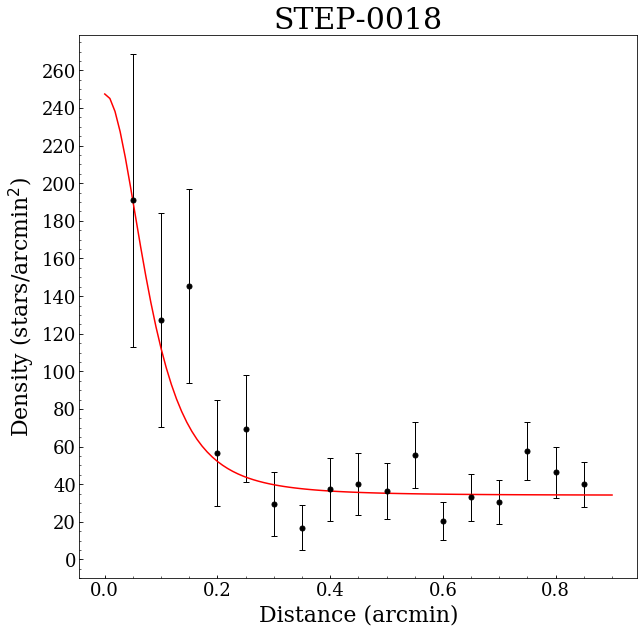}
    \includegraphics[width=0.23\textwidth]{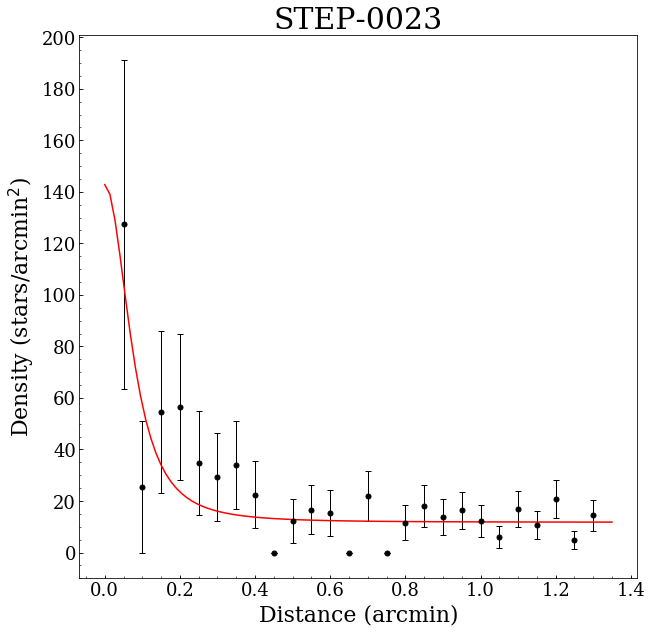}\\
    \includegraphics[width=0.23\textwidth]{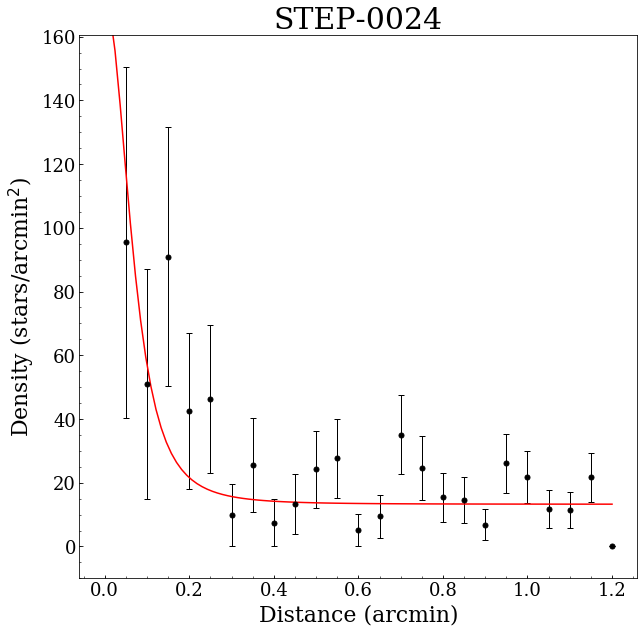}
    \includegraphics[width=0.23\textwidth]{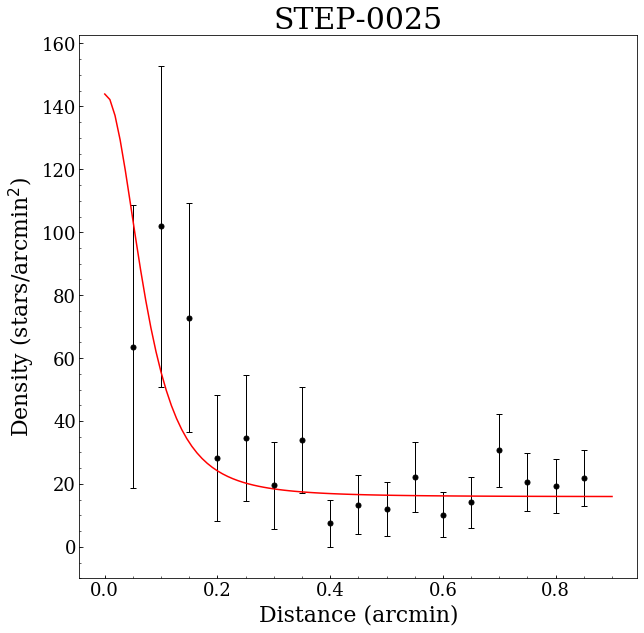}
    \includegraphics[width=0.23\textwidth]{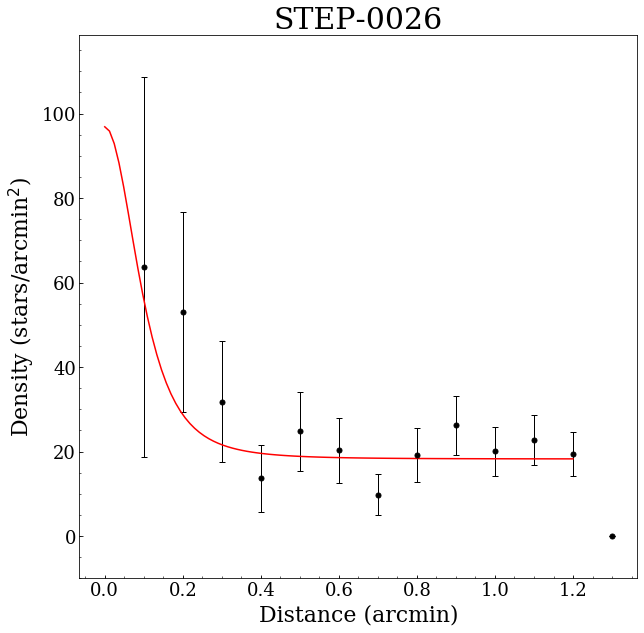}
    \includegraphics[width=0.23\textwidth]{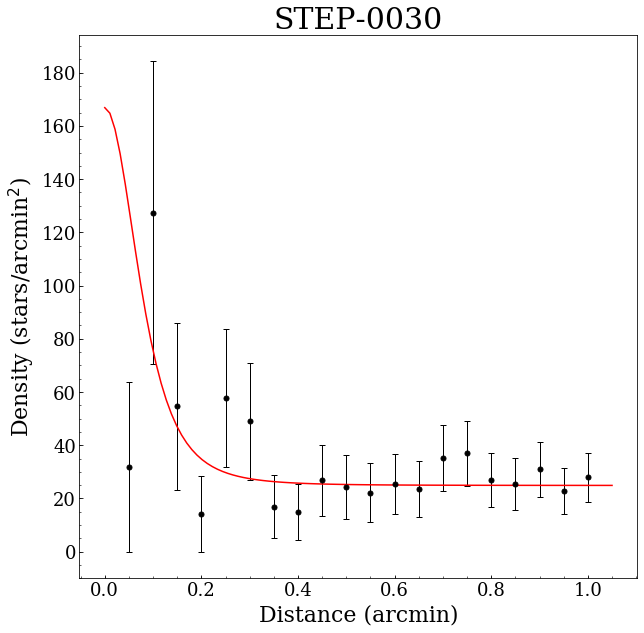}\\
    \includegraphics[width=0.23\textwidth]{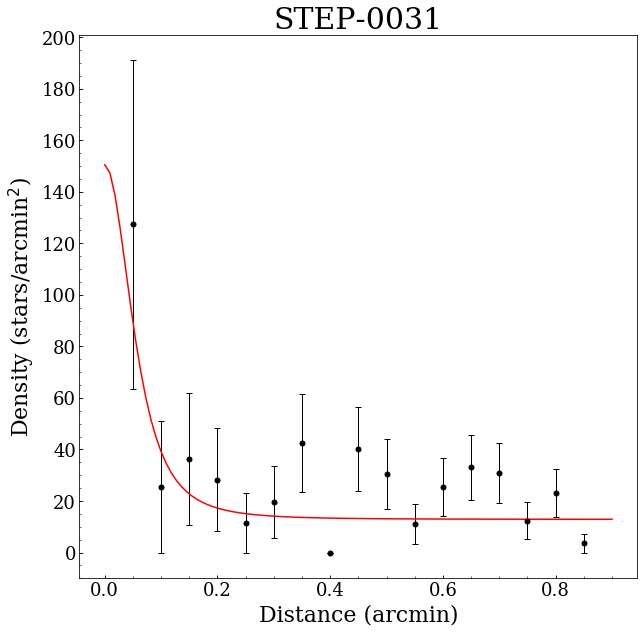}
    \includegraphics[width=0.23\textwidth]{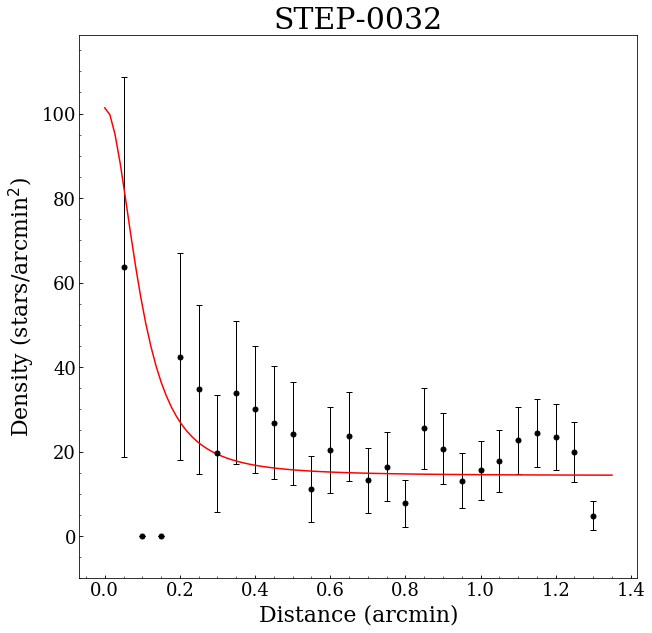}
    \includegraphics[width=0.23\textwidth]{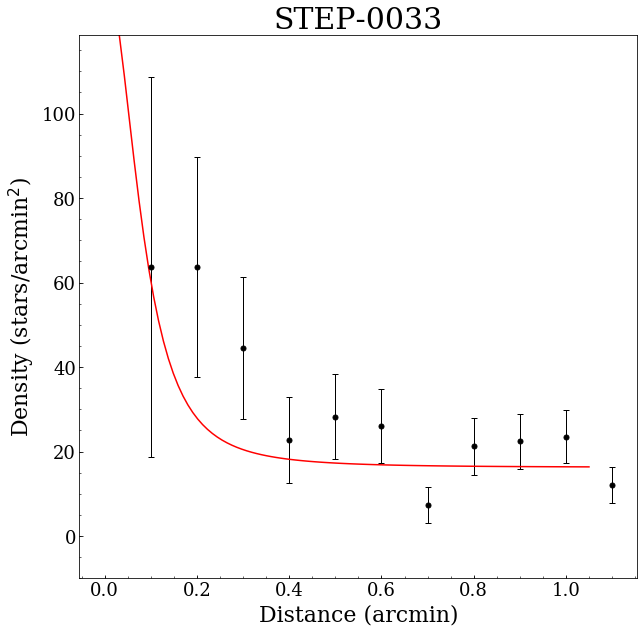}
    \includegraphics[width=0.23\textwidth]{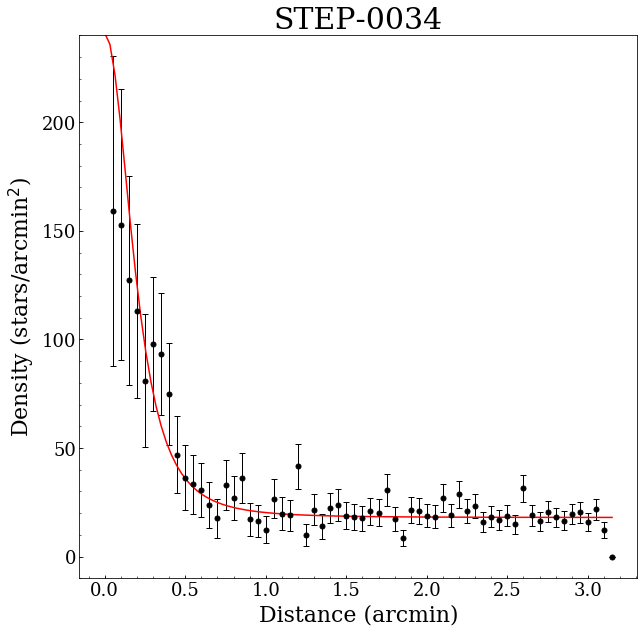}\\
    
\caption{RDP obtained using only stars with P $\geq$75\%. The red line represents the best fit with an EFF profile.}
    \label{fig:EFF_profiles}
\end{figure*}{}
\begin{figure*}
    \centering
    \includegraphics[width=0.23\textwidth]{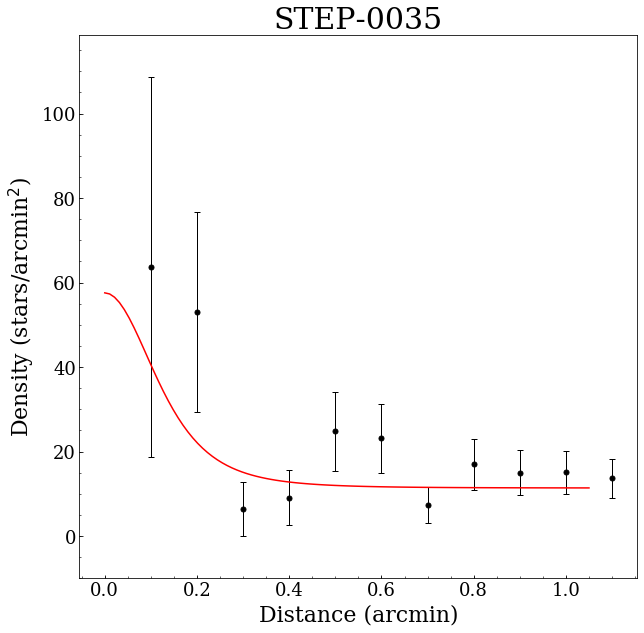}
    \includegraphics[width=0.23\textwidth]{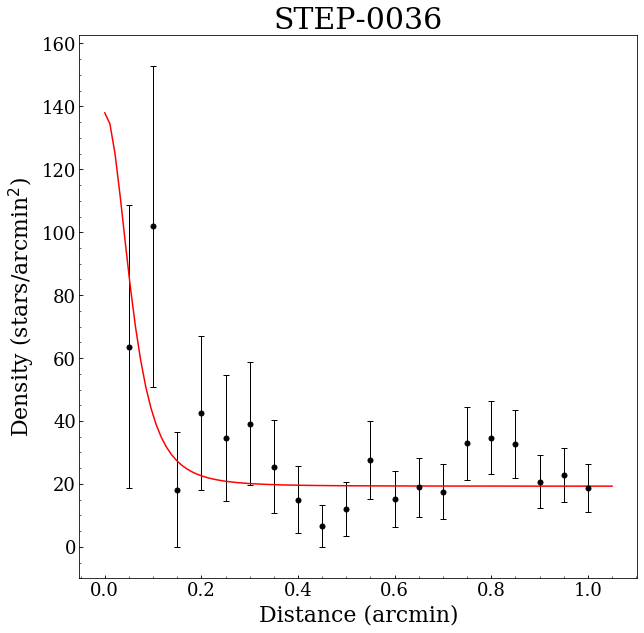}
    \includegraphics[width=0.23\textwidth]{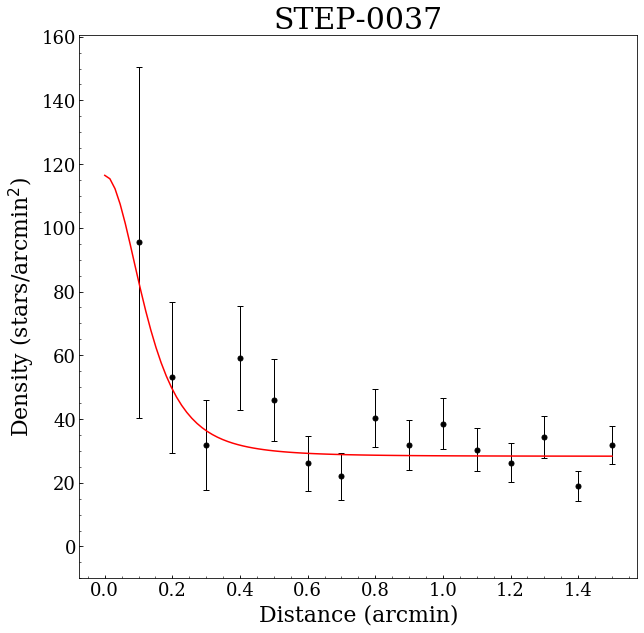}
    \includegraphics[width=0.23\textwidth]{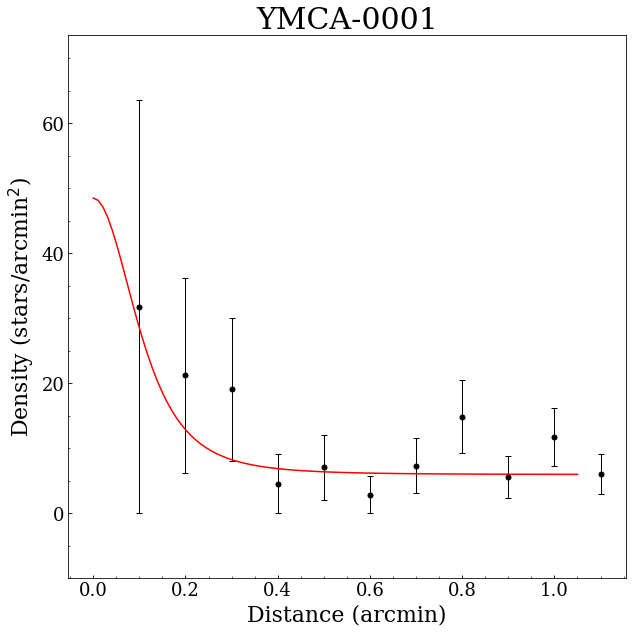}\\
    \includegraphics[width=0.23\textwidth]{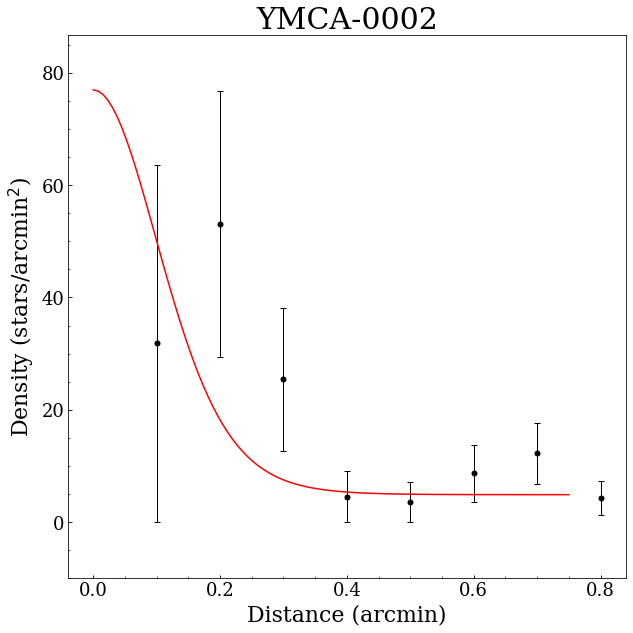}
    \includegraphics[width=0.23\textwidth]{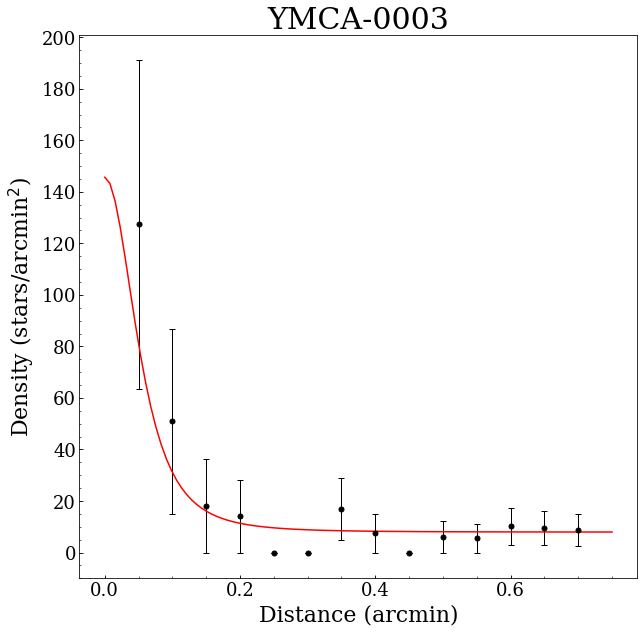}
    \includegraphics[width=0.23\textwidth]{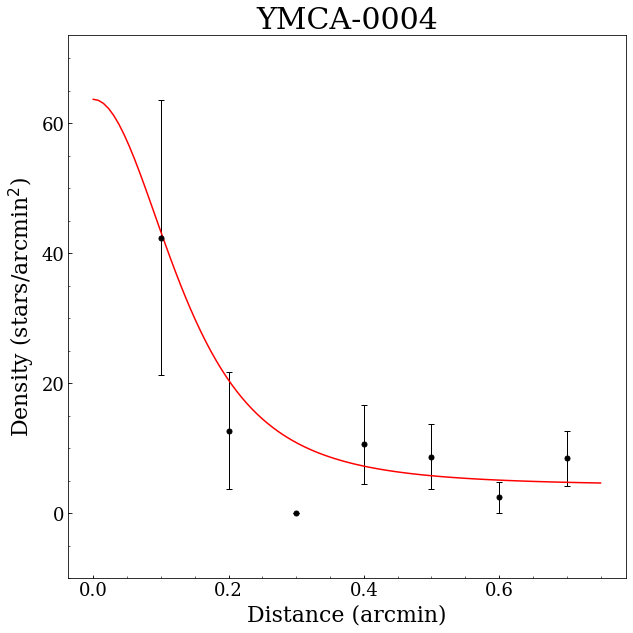}
    \includegraphics[width=0.23\textwidth]{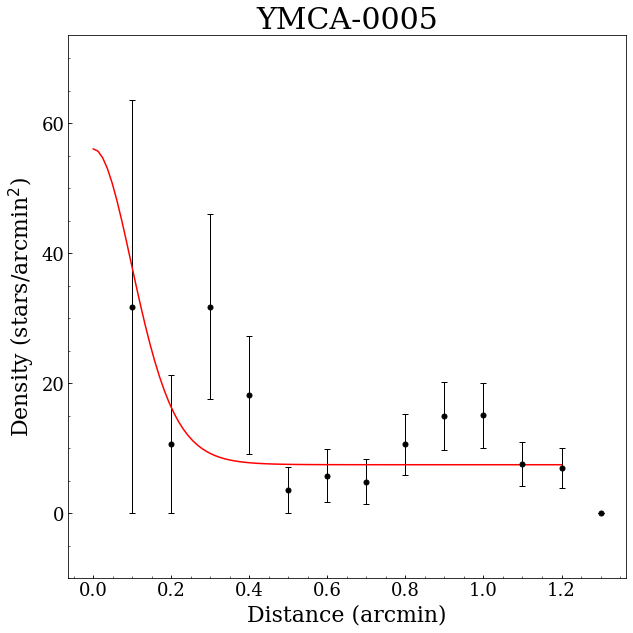}\\
    \includegraphics[width=0.23\textwidth]{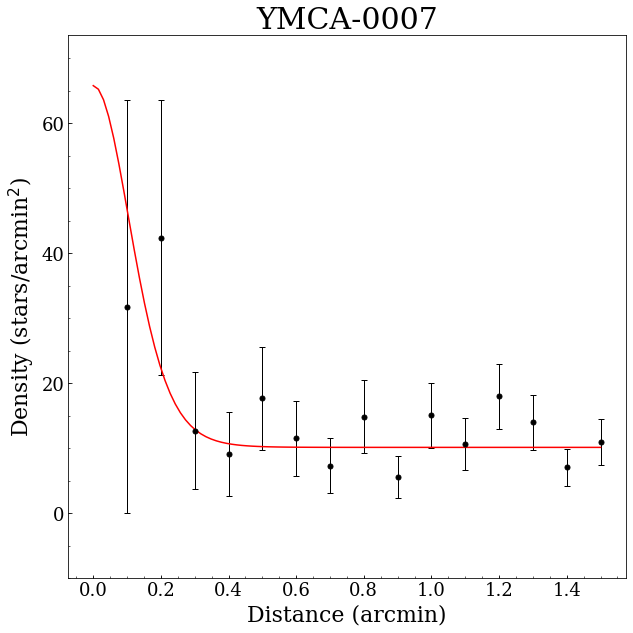}
    \includegraphics[width=0.23\textwidth]{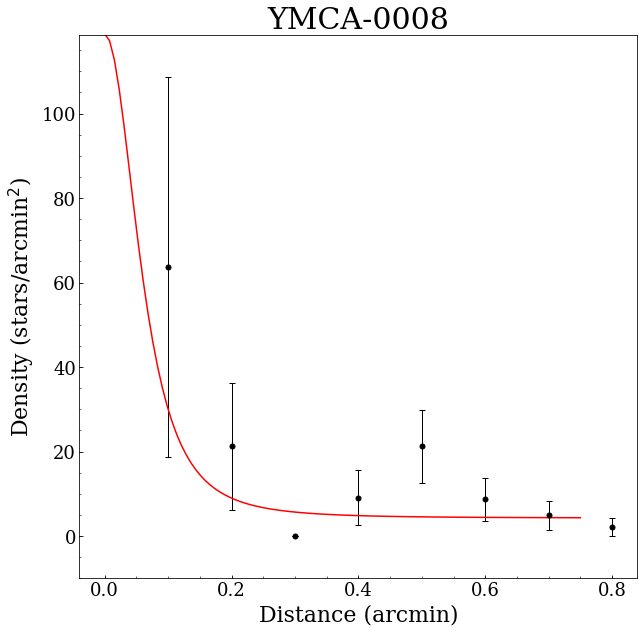}
    \includegraphics[width=0.23\textwidth]{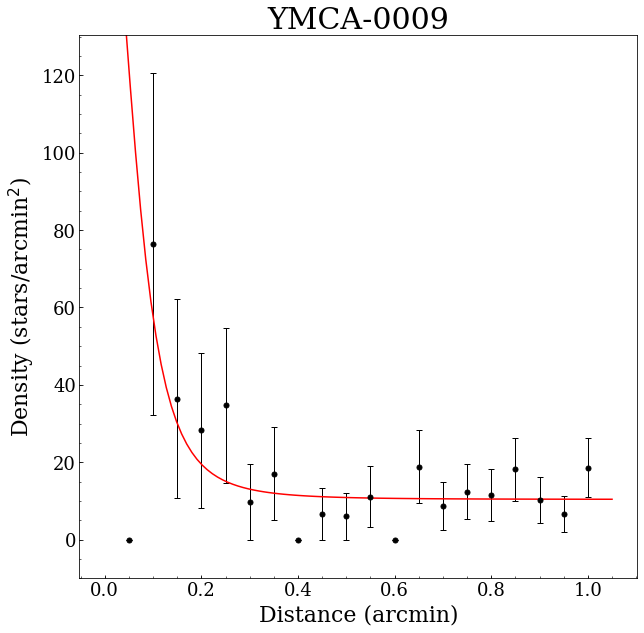}
    \includegraphics[width=0.23\textwidth]{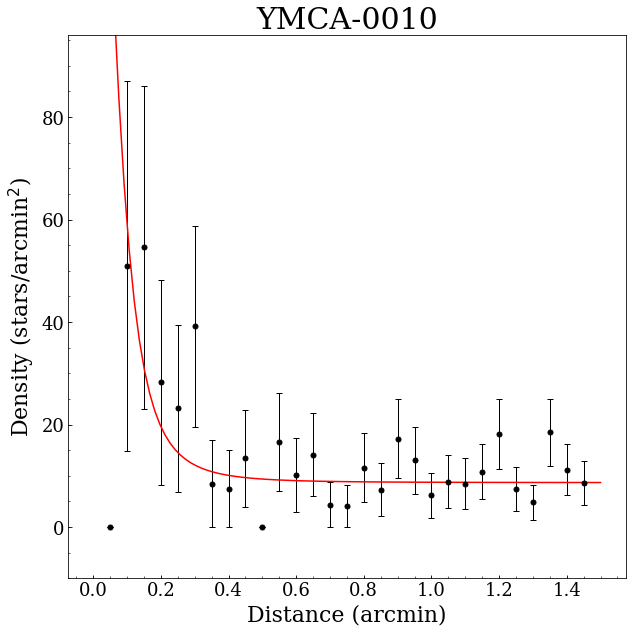}\\
    \includegraphics[width=0.23\textwidth]{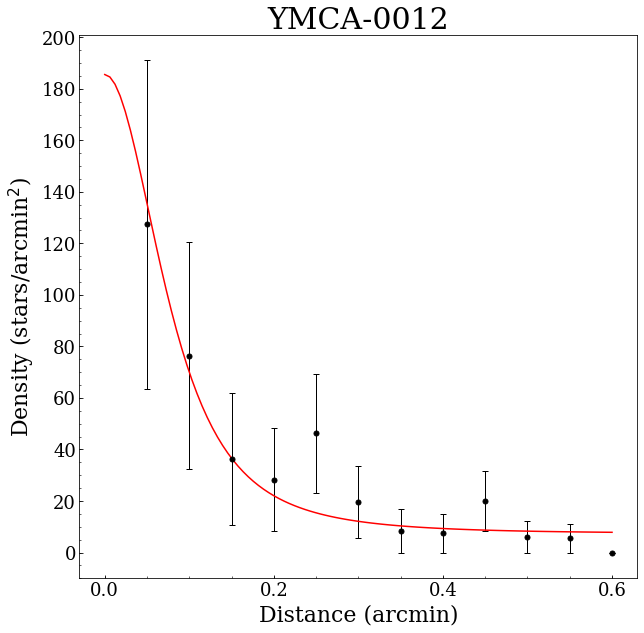}
    \includegraphics[width=0.23\textwidth]{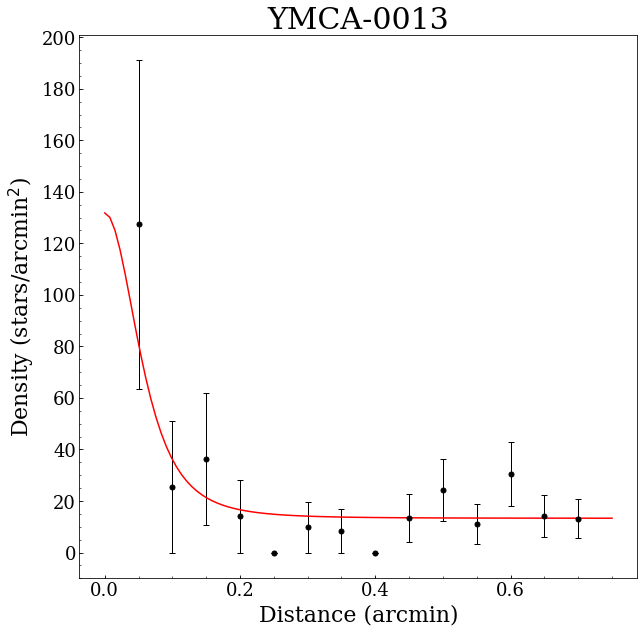}
    \includegraphics[width=0.23\textwidth]{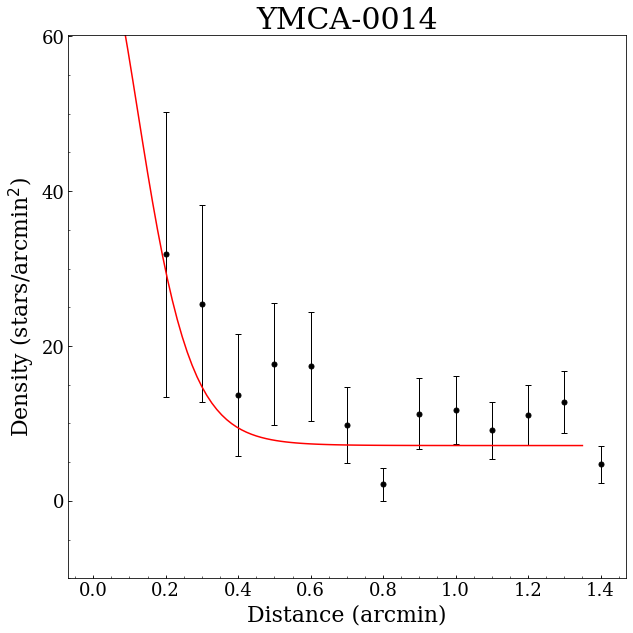}
    \includegraphics[width=0.23\textwidth]{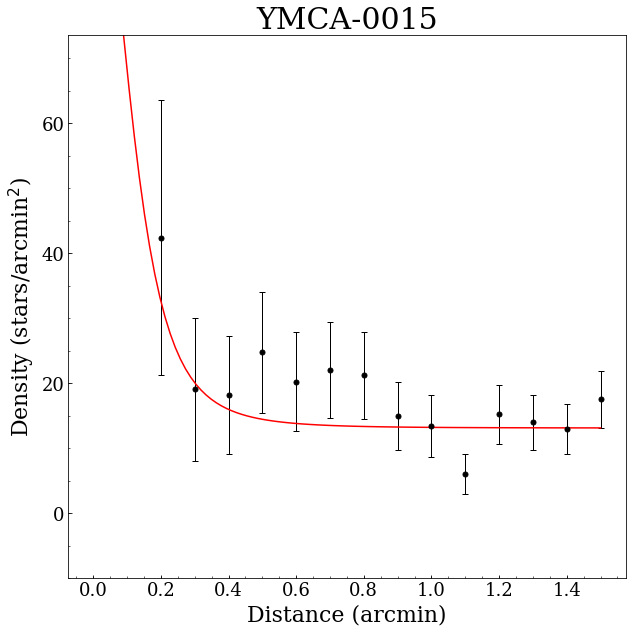}\\
    \includegraphics[width=0.23\textwidth]{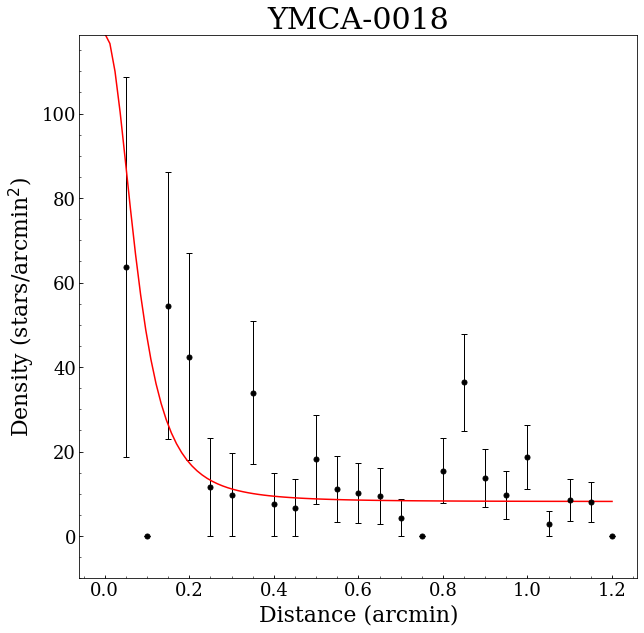}
    \includegraphics[width=0.23\textwidth]{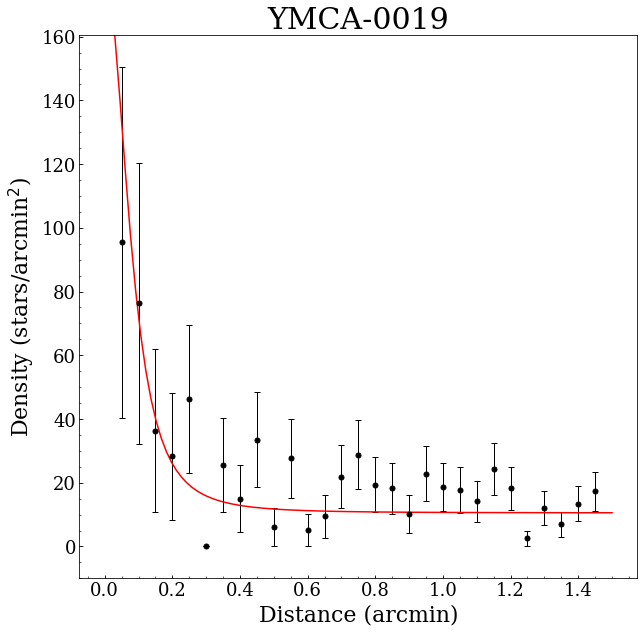}
    \includegraphics[width=0.23\textwidth]{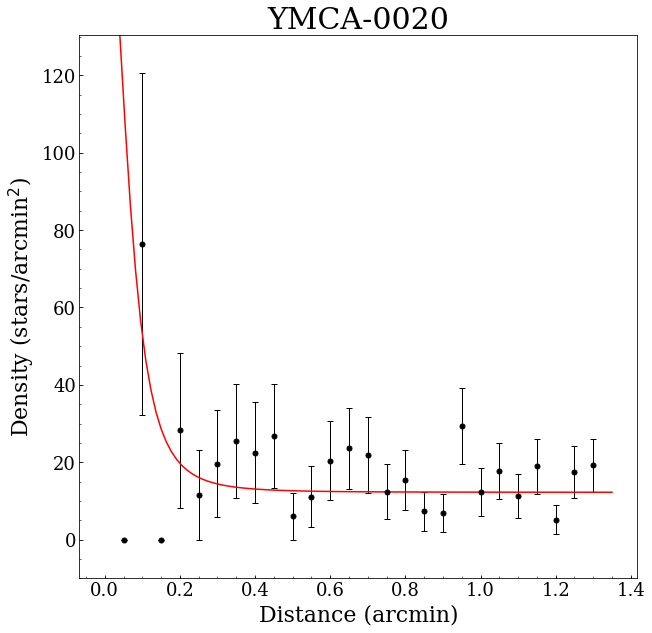}
    \includegraphics[width=0.23\textwidth]{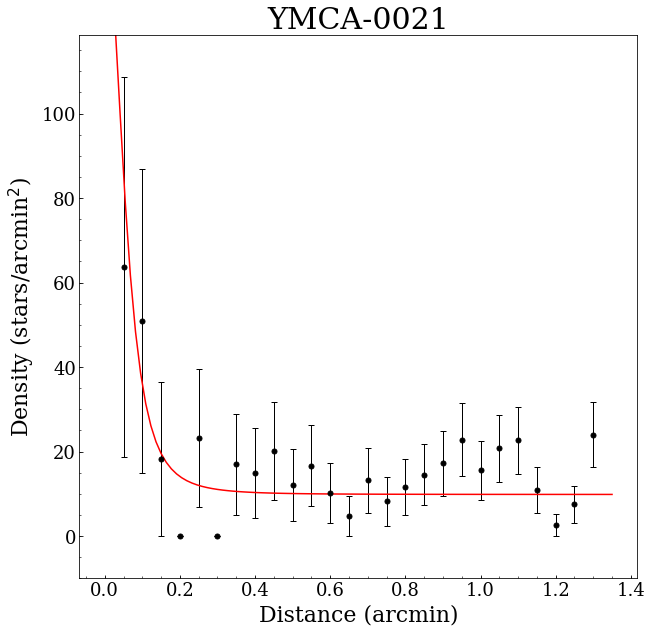}\\
    \includegraphics[width=0.23\textwidth]{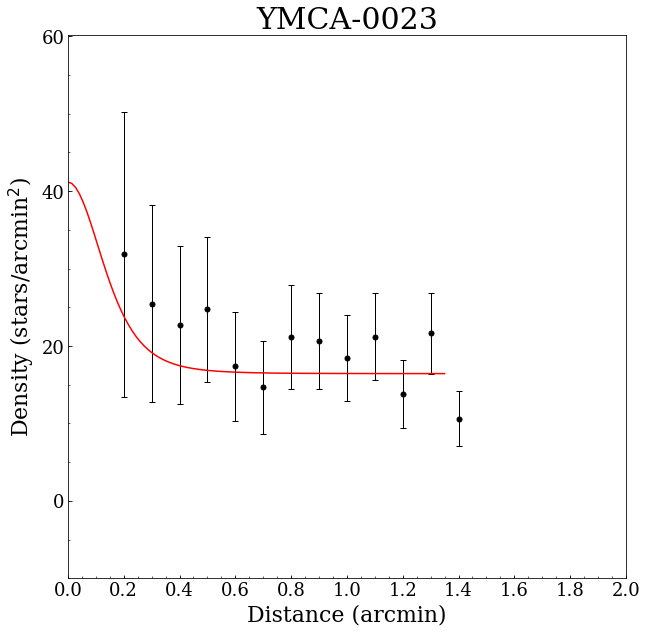}
    \includegraphics[width=0.23\textwidth]{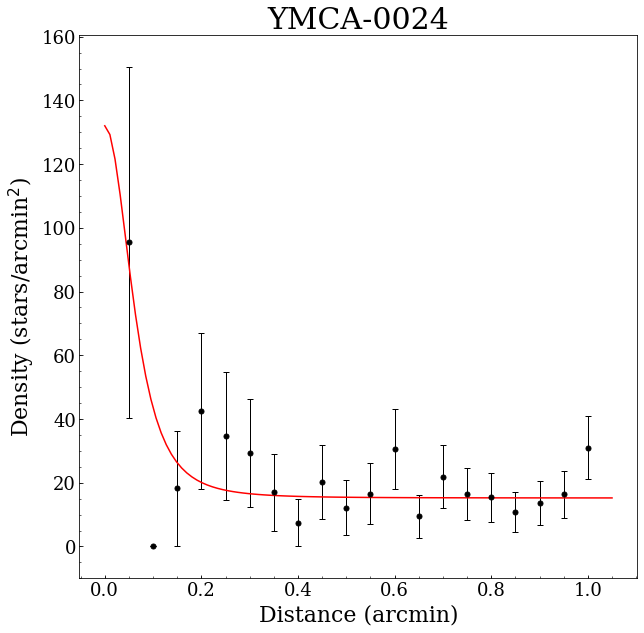}
    \includegraphics[width=0.23\textwidth]{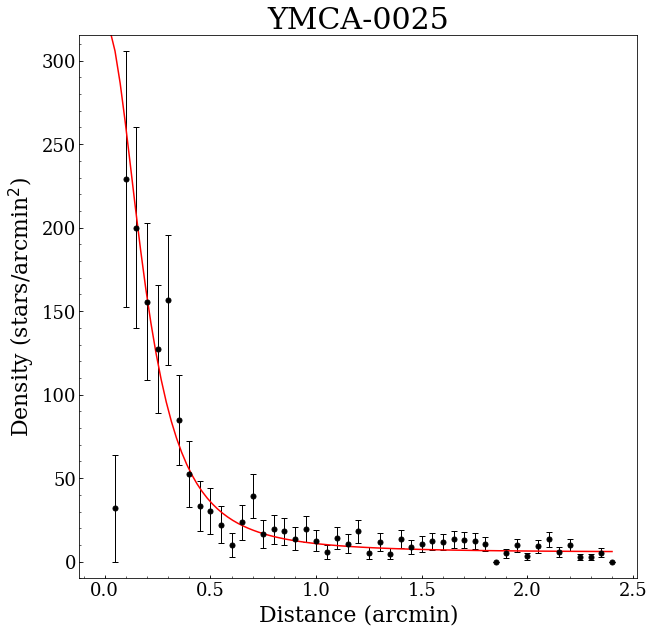}
    \includegraphics[width=0.23\textwidth]{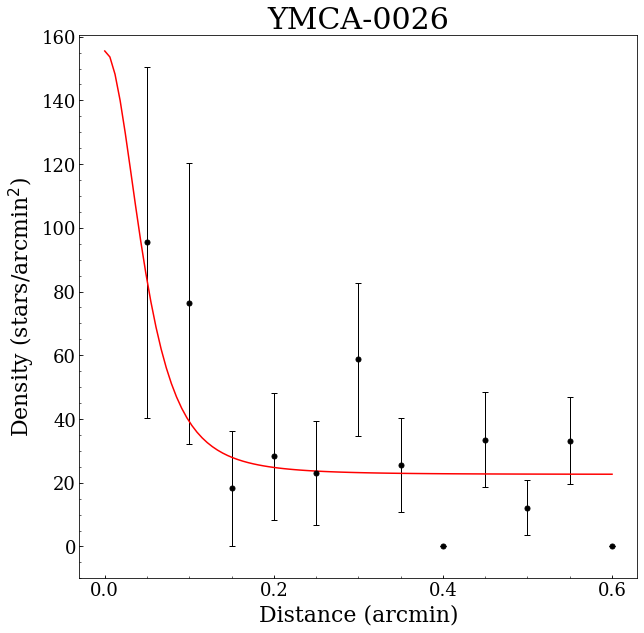}\\
    \contcaption{}
\end{figure*}{}
\begin{figure*}
    \centering
    \includegraphics[width=0.23\textwidth]{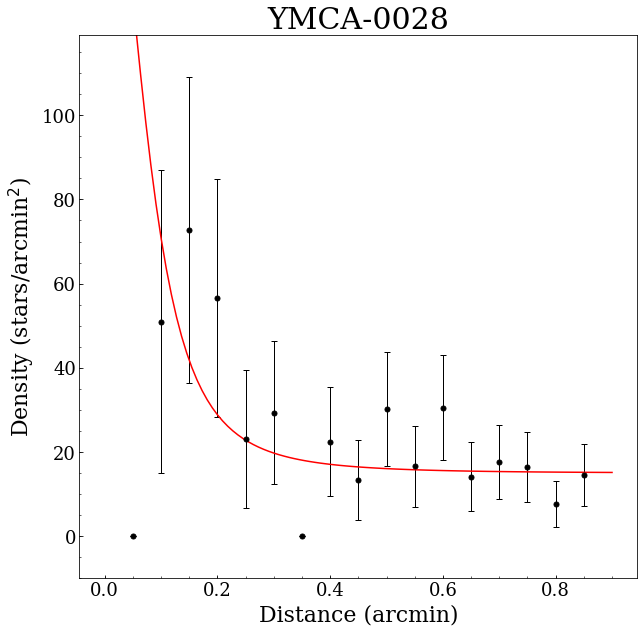}
    \includegraphics[width=0.23\textwidth]{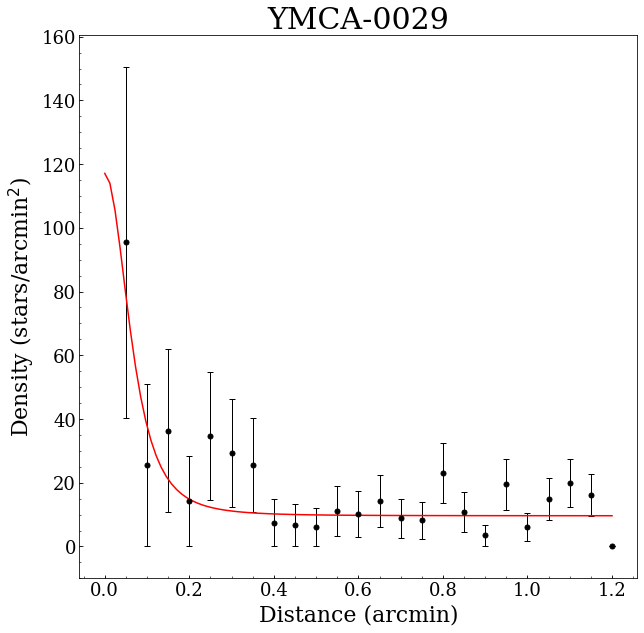}
    \includegraphics[width=0.23\textwidth]{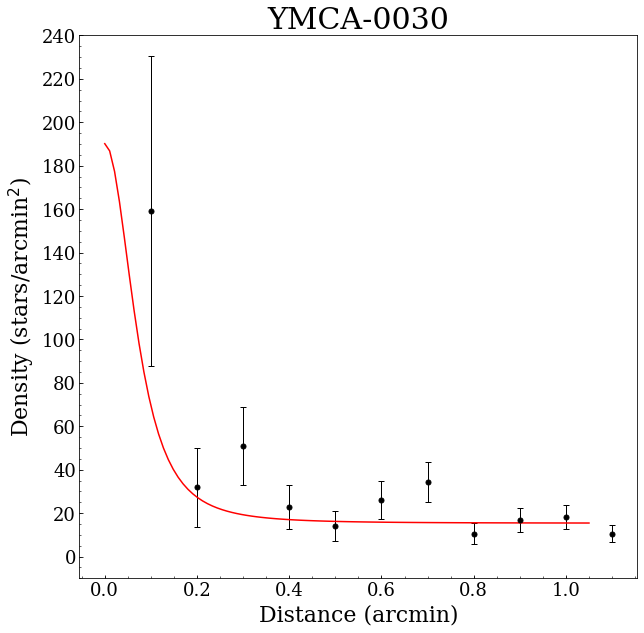}
    \includegraphics[width=0.23\textwidth]{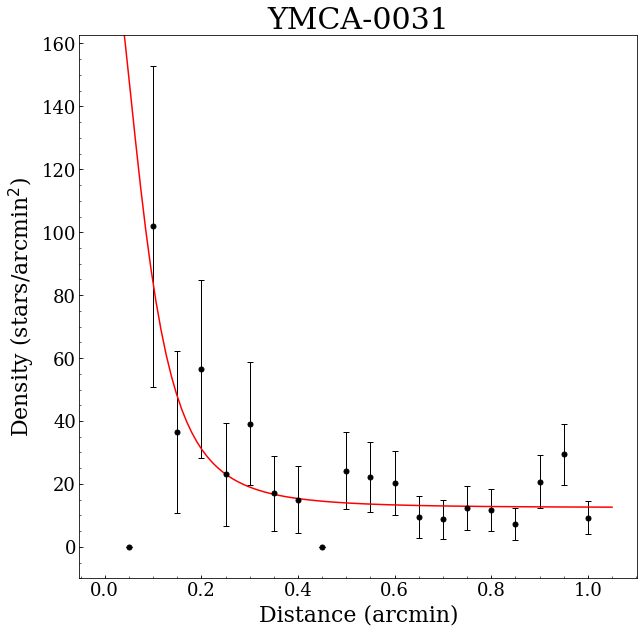}\\
    \includegraphics[width=0.23\textwidth]{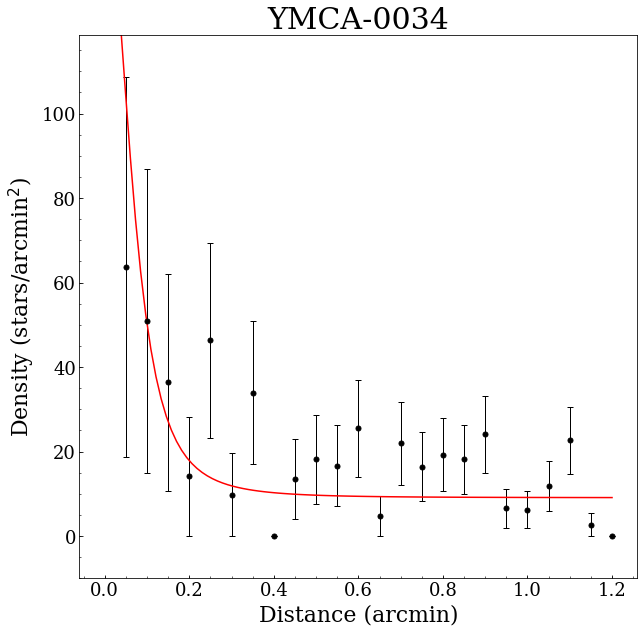}
    \includegraphics[width=0.23\textwidth]{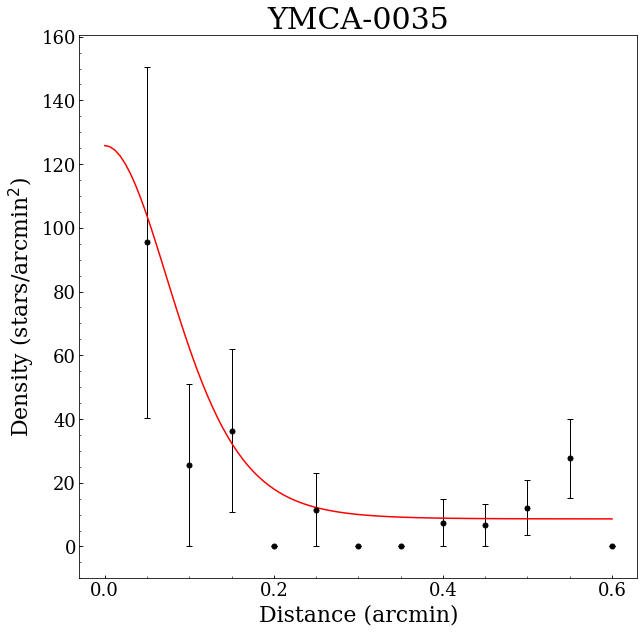}
    \includegraphics[width=0.23\textwidth]{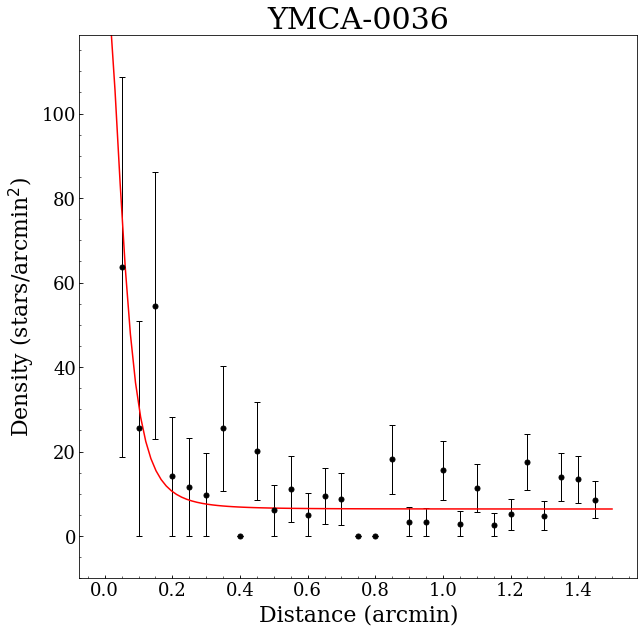}
    \includegraphics[width=0.23\textwidth]{YMCA-0037_rdp_75.png}\\
    \includegraphics[width=0.23\textwidth]{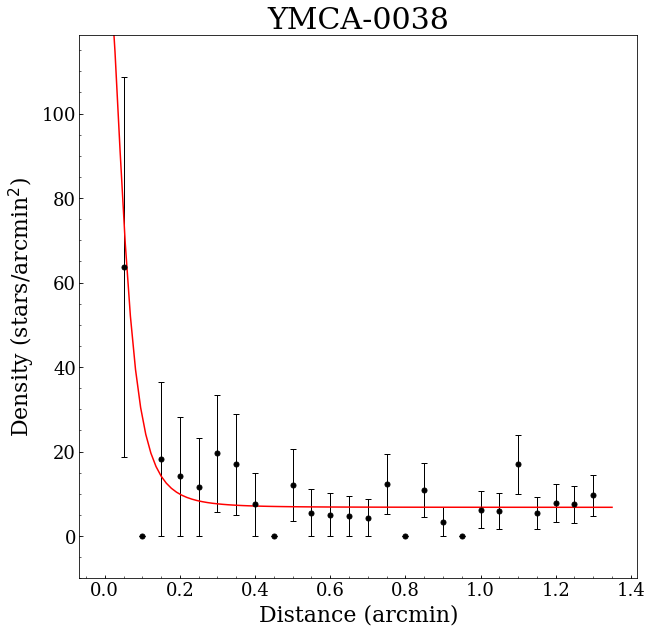}
    \includegraphics[width=0.23\textwidth]{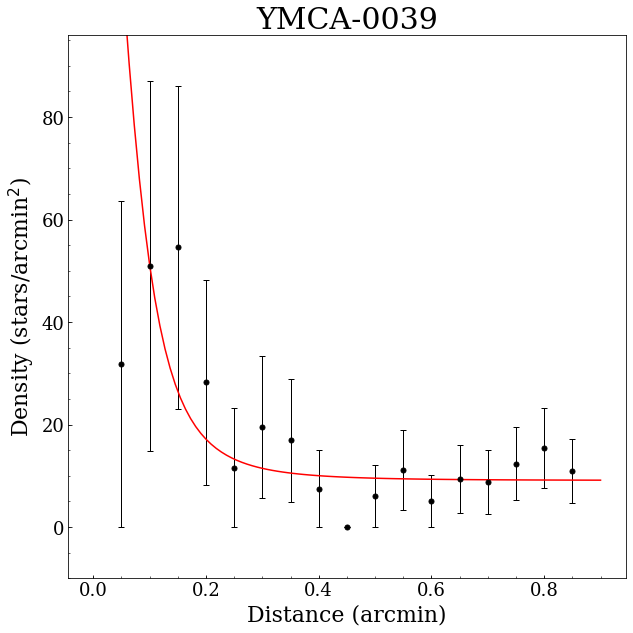}
    \includegraphics[width=0.23\textwidth]{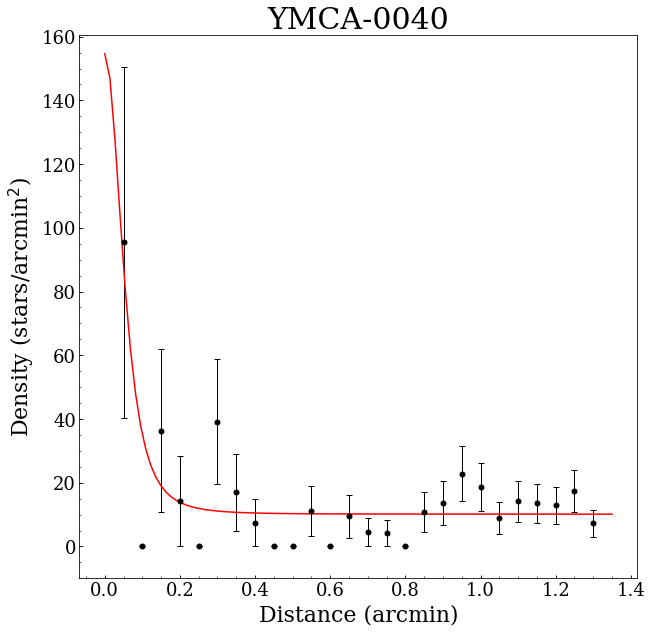}
    \includegraphics[width=0.23\textwidth]{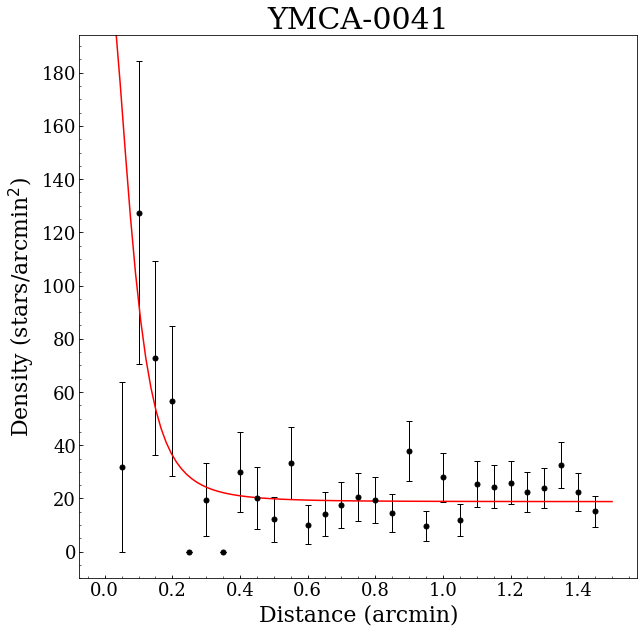}\\
    \includegraphics[width=0.23\textwidth]{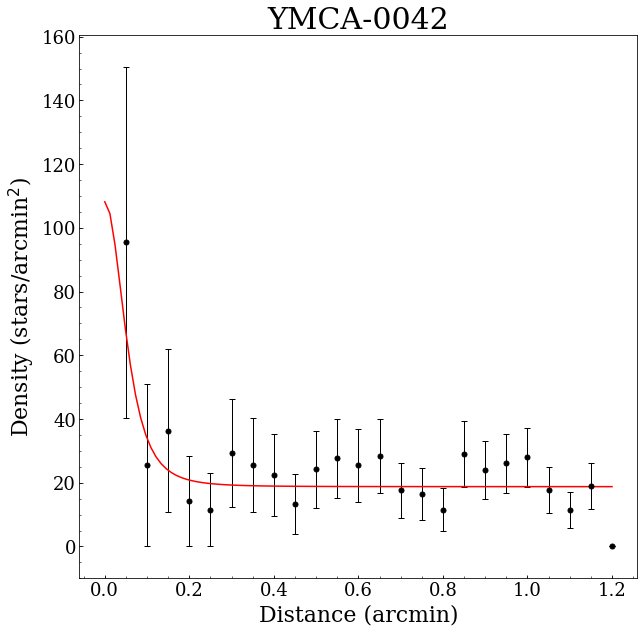}
    \includegraphics[width=0.23\textwidth]{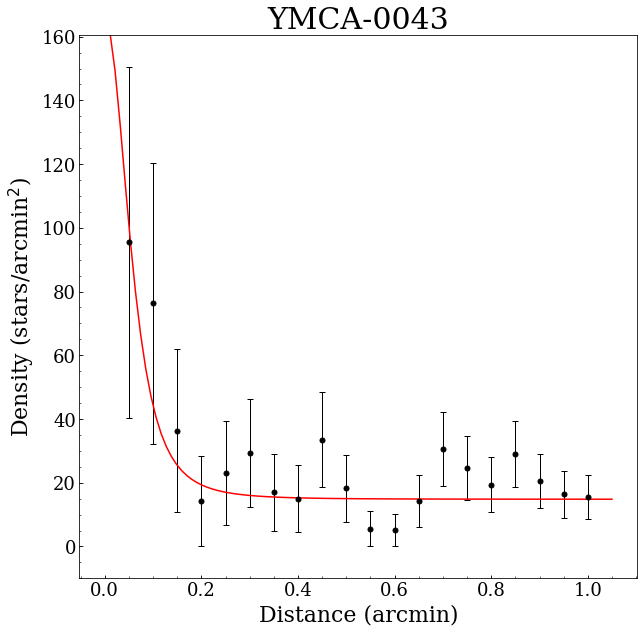}
    \includegraphics[width=0.23\textwidth]{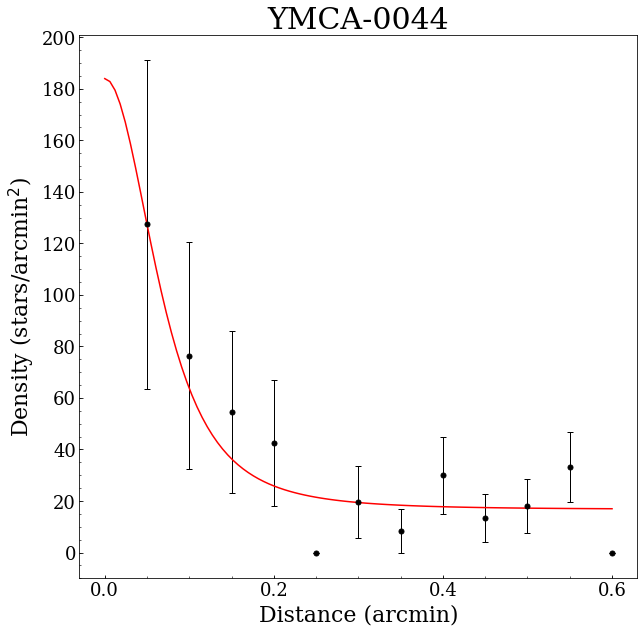}
    \includegraphics[width=0.23\textwidth]{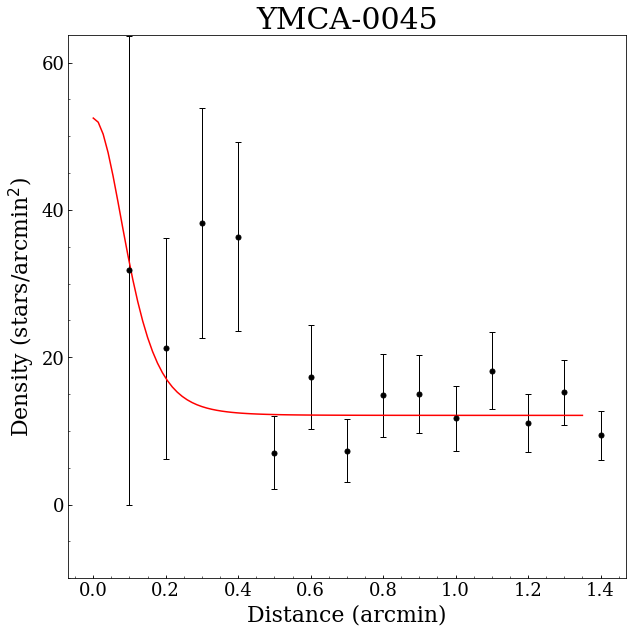}\\
    \includegraphics[width=0.23\textwidth]{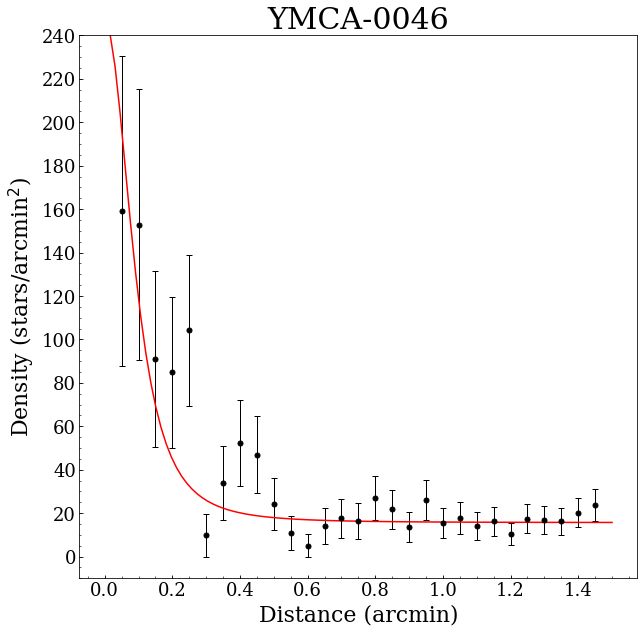}
    \includegraphics[width=0.23\textwidth]{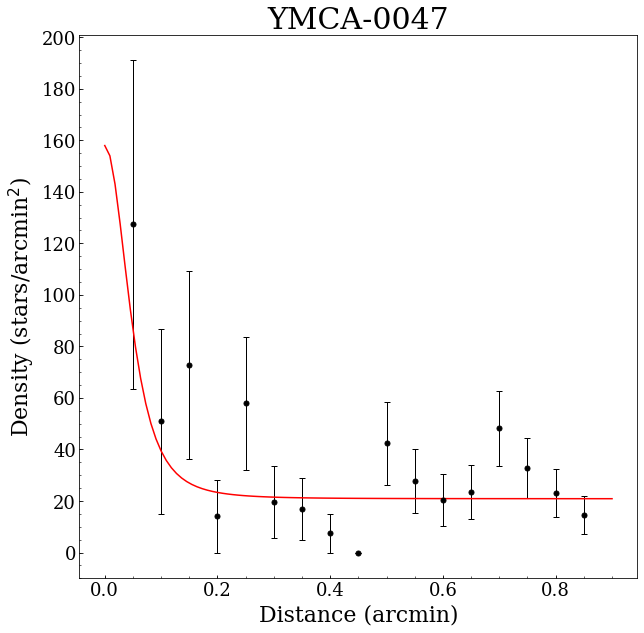}
    \includegraphics[width=0.23\textwidth]{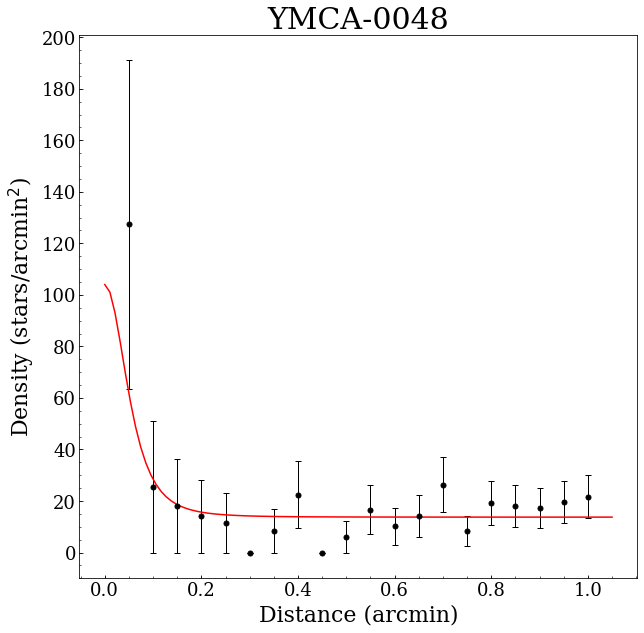}\\
    \contcaption{}
\end{figure*}{}

\begin{figure*}
    \vspace{-0.5cm}
    \includegraphics[width=0.23\textwidth]{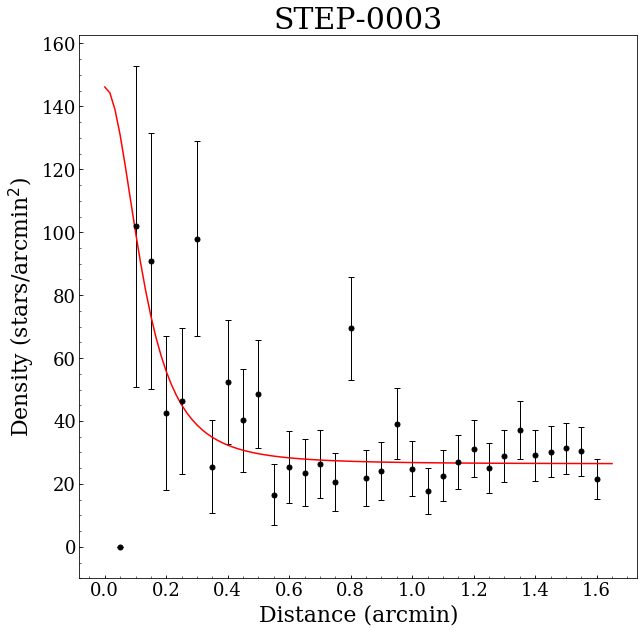}
    \includegraphics[width=0.23\textwidth]{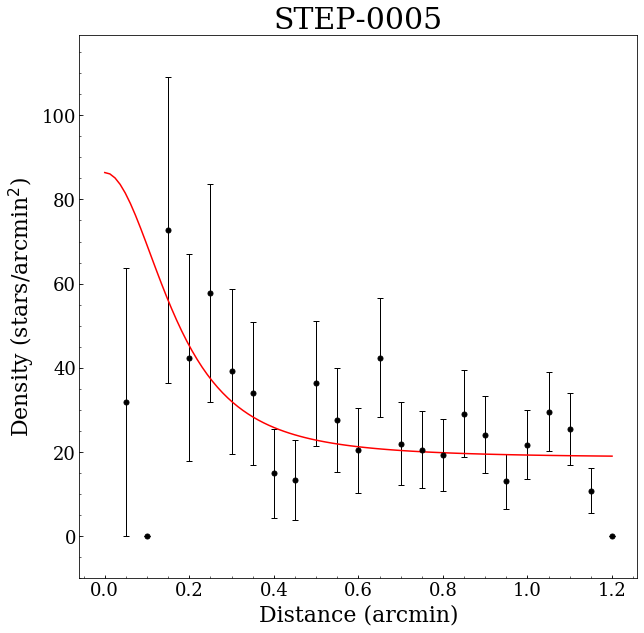}
    \includegraphics[width=0.23\textwidth]{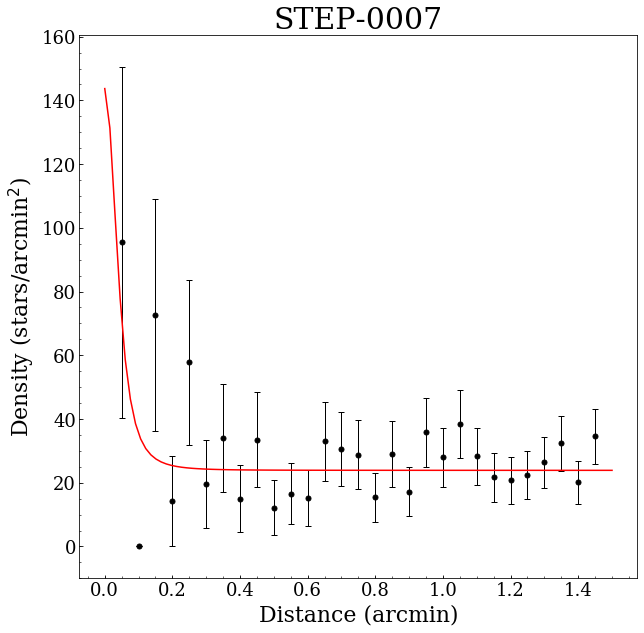}
    \includegraphics[width=0.23\textwidth]{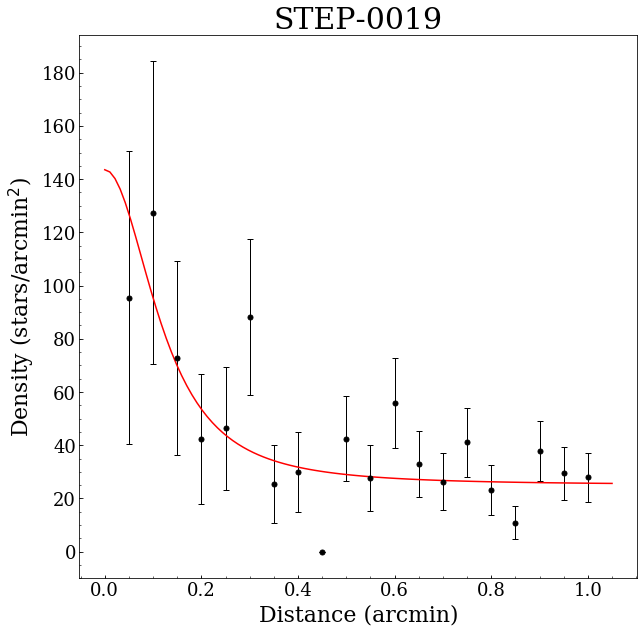}\\
    \includegraphics[width=0.23\textwidth]{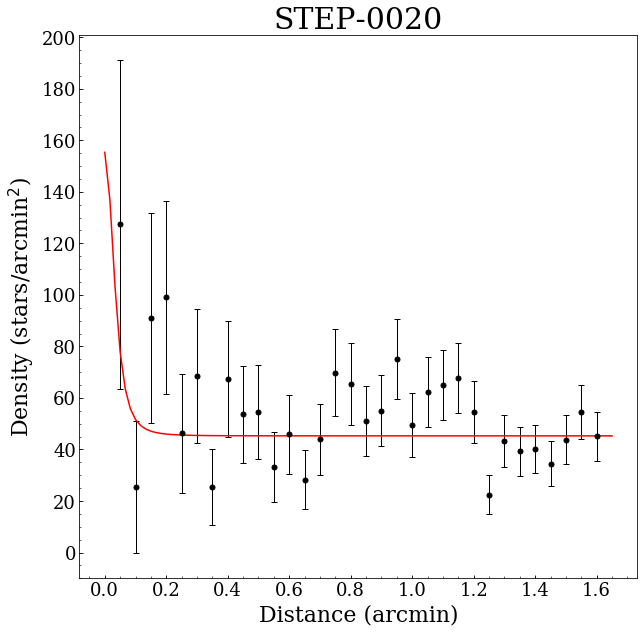}
    \includegraphics[width=0.23\textwidth]{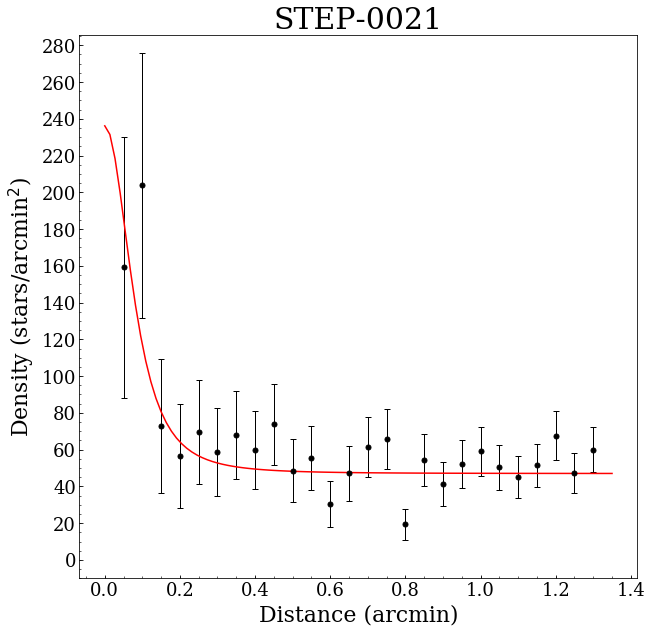}
    \includegraphics[width=0.23\textwidth]{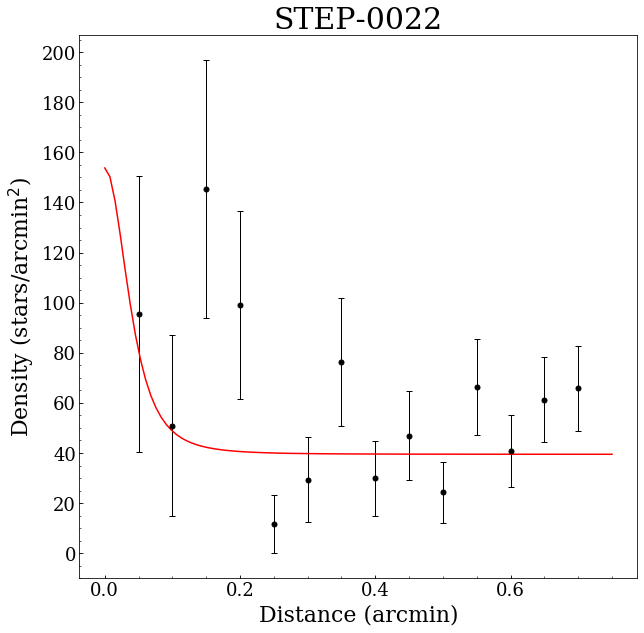}
    \includegraphics[width=0.23\textwidth]{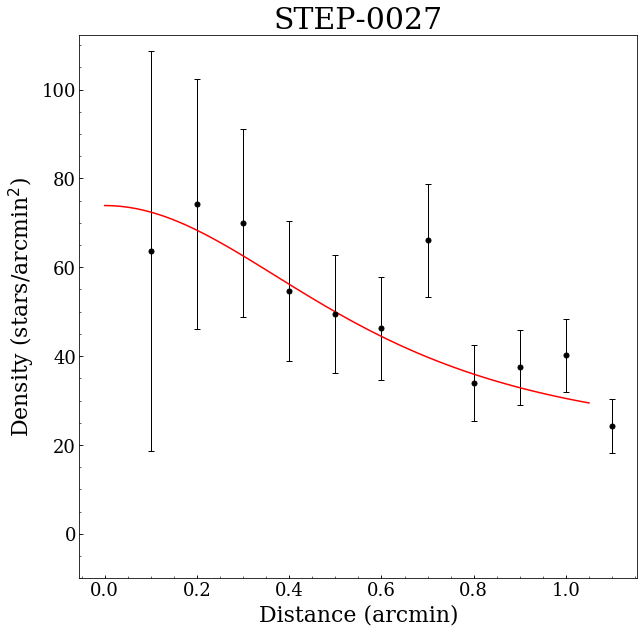}\\
    \includegraphics[width=0.23\textwidth]{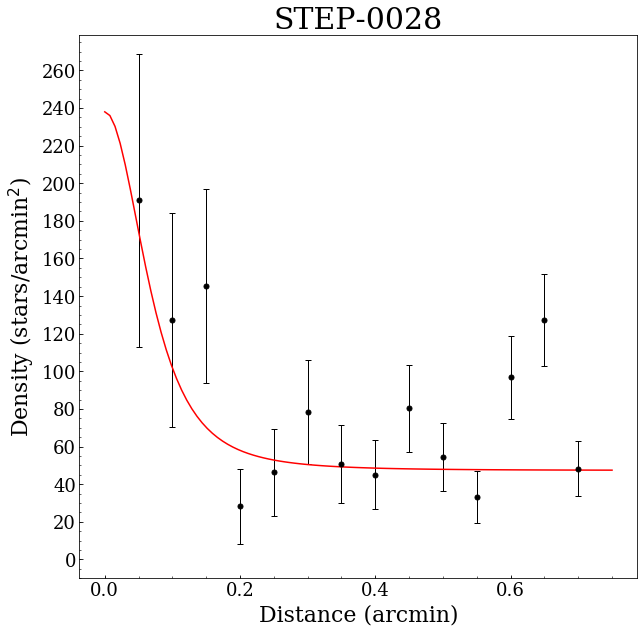}
    \includegraphics[width=0.23\textwidth]{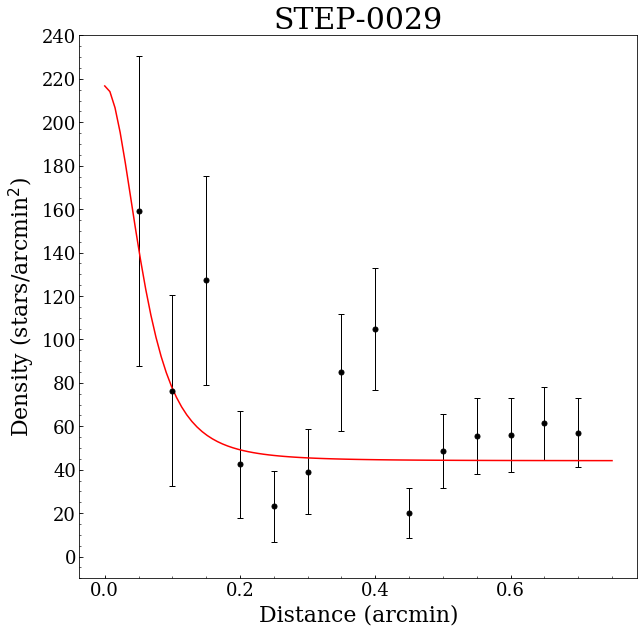}
    \includegraphics[width=0.23\textwidth]{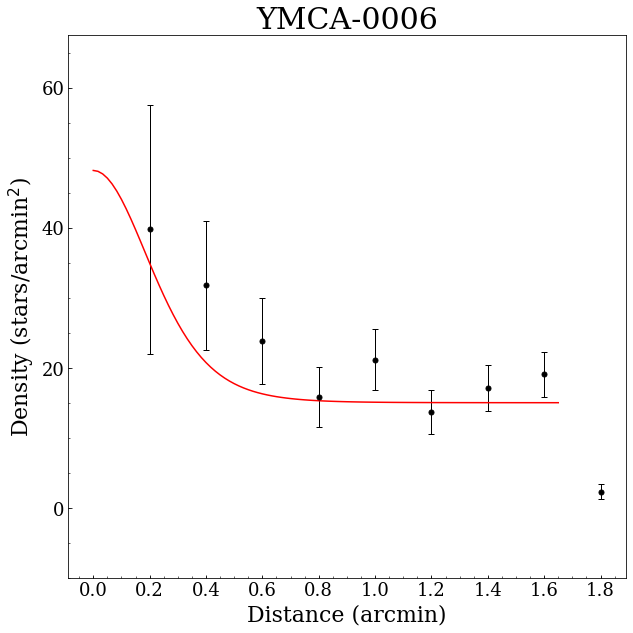}
    \includegraphics[width=0.23\textwidth]{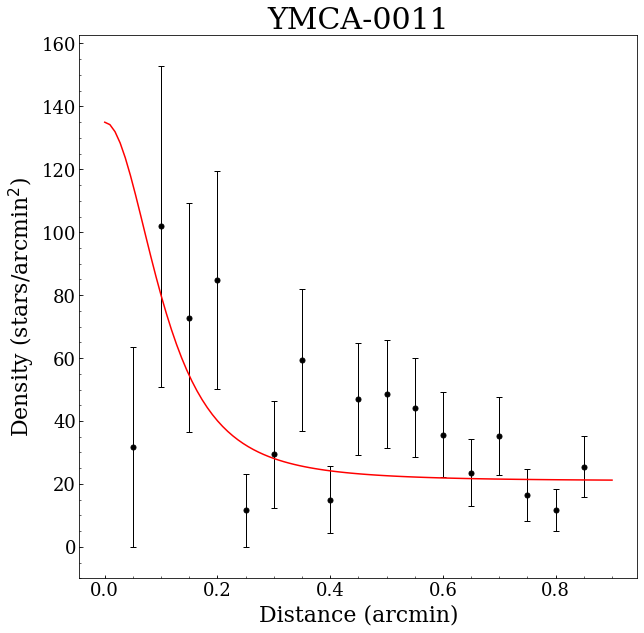}\\
    \includegraphics[width=0.23\textwidth]{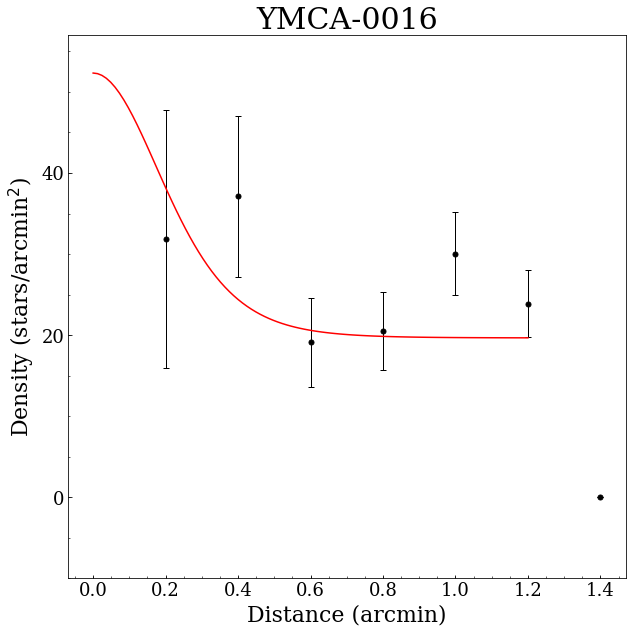}
    \includegraphics[width=0.23\textwidth]{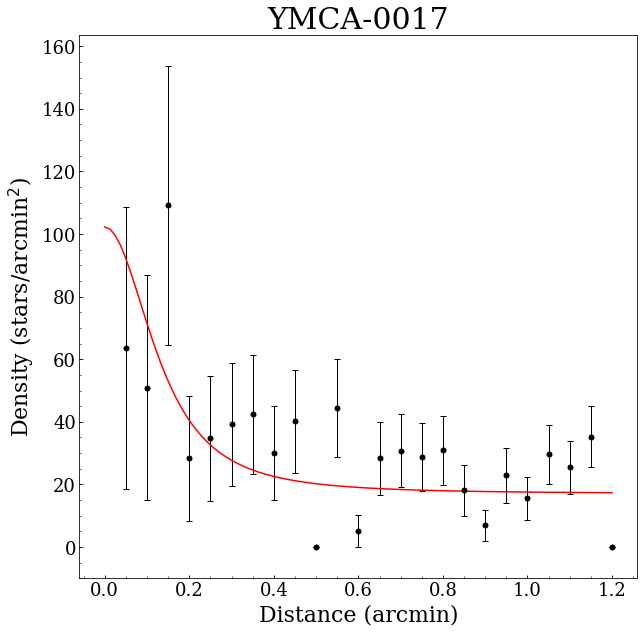}
    \includegraphics[width=0.23\textwidth]{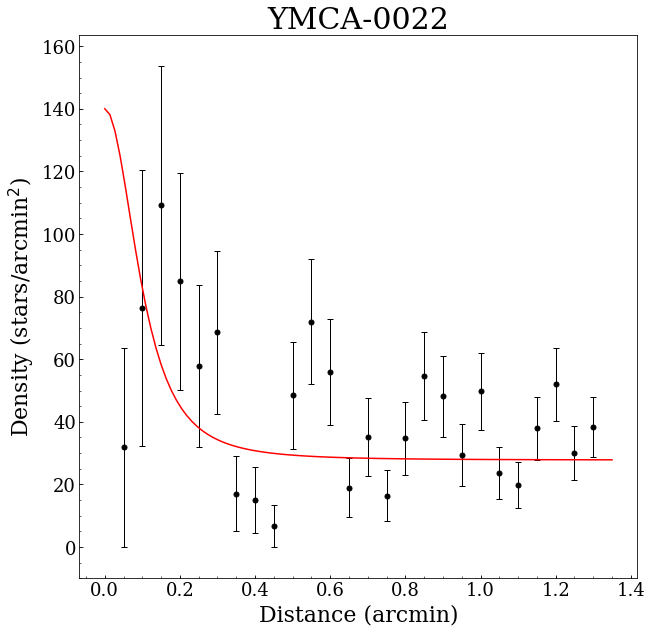}
    \includegraphics[width=0.23\textwidth]{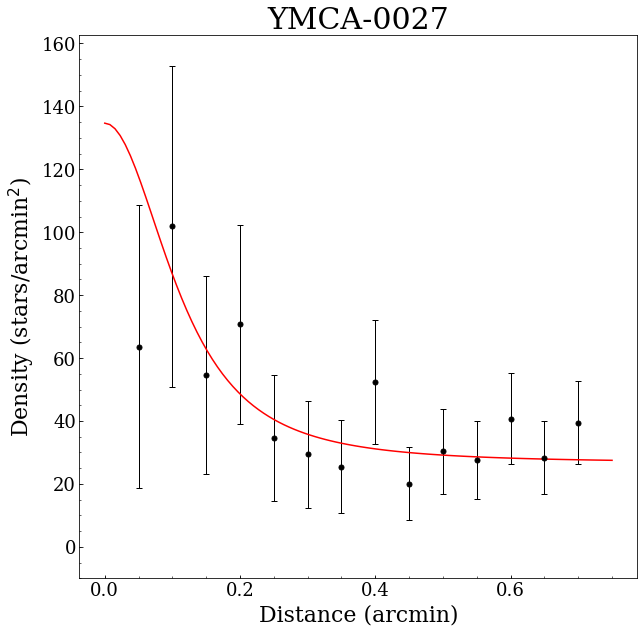}\\
    \includegraphics[width=0.23\textwidth]{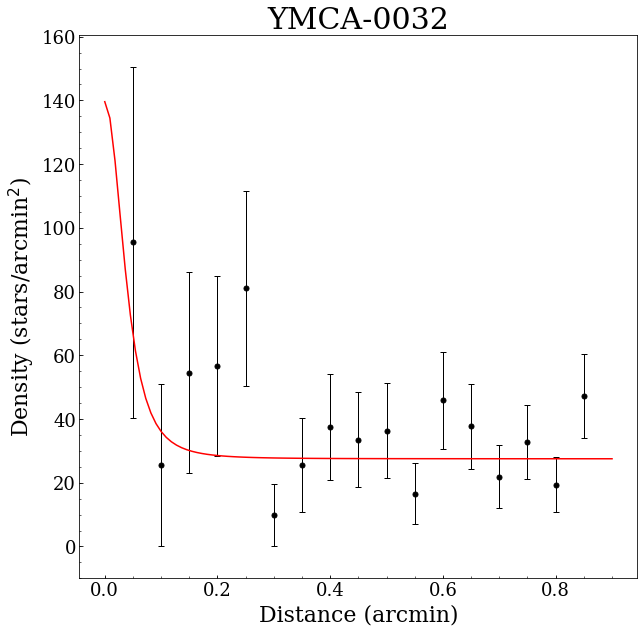}
    \includegraphics[width=0.23\textwidth]{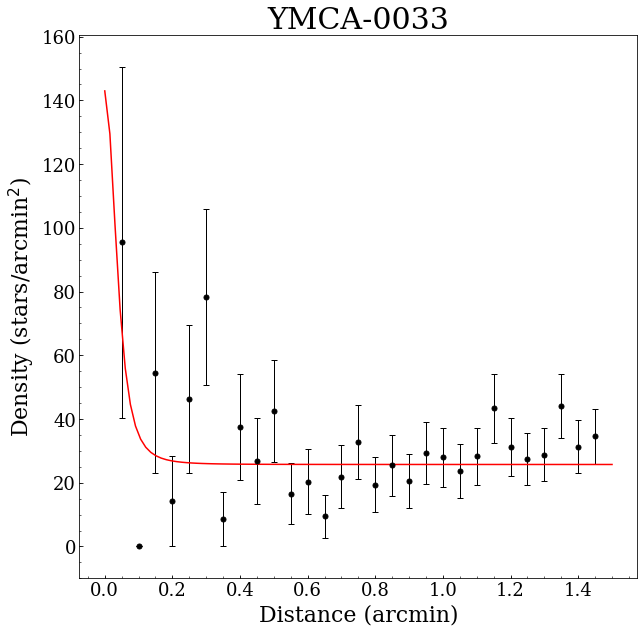}\\
    
\caption{RDP obtained using only stars with P $\geq$50\%. The red line represents the best fit with an EFF profile.}
    \label{fig:EFF_profiles_50}
\end{figure*}{}


\bsp	
\label{lastpage}
\end{document}